\documentclass{jfm}
\usepackage{fix-cm}
\usepackage{graphicx}
\usepackage{newtxtext}
\usepackage{hyperref}
\usepackage{xcolor}
\usepackage{newtxmath}
\usepackage[justification=justified]{caption}
\usepackage{natbib}
\usepackage{hyperref}
\hypersetup{
    colorlinks = true,
    urlcolor   = blue,
    citecolor  = black,
}

\newcommand{\RomanNumeralCaps}[1] 

\usepackage{graphicx}

\title{On the generation of free-surface waves by instabilities in quadratic shear flows}

\author{Harishankar Muppirala\aff{1},
  Ramana Patibandla\footnote{Currently at University of Massachusetts Dartmouth, North Dartmouth, Massachusetts, US}\aff{1}
 \and Anubhab Roy\aff{1}\corresp{\email{anubhab@iitm.ac.in}}}

\affiliation{\aff{1}Department of Applied Mechanics and Biomedical Engineering, Indian Institute of Technology Madras, Chennai, India}

\begin{document}
\maketitle

\begin{abstract}
This paper investigates the generation of free-surface waves in a liquid layer driven by linear instabilities in Couette–Poiseuille (quadratic) shear flows. The base velocity profiles are characterized by a curvature parameter, and two-dimensional viscous and inviscid perturbations are analyzed across a wide parameter space of curvature, wavenumber, and Reynolds number, for fixed Froude and Bond numbers. In the inviscid limit, analytical solutions of the Rayleigh equation reveal that velocity profiles ranging from Nusselt to linear flows remain stable against the rippling instability, with long-wave growth occurring only under strong interfacial forcing, whereas weaker forcing produces well-defined stability boundaries. For the viscous problem, Orr–Sommerfeld computations and asymptotic analyses reveal that a slight convex curvature of the shear flow suppresses long-wave instabilities, while a slight concave curvature suppresses short-wave instabilities, so even small deviations from a linear profile produce qualitatively different behaviors. Furthermore, we observe that strongly forced long waves are more unstable at large $Re$ than the inviscid value they latch on to as $Re \to \infty$. Growth-rate maps highlight smooth transitions between long-wave and rippling modes and reveal an additional shear instability near the linear profile at high Reynolds numbers. Based on energy transfers and eigenfunction structures, five distinct instability types are identified: shear, rippling, long-wave interfacial, short-wave interfacial, and a composite mode that combines features of shear, rippling and long-wave interfacial instabilities at large Reynolds numbers.
\end{abstract}


{\bf MSC Codes } {\it(Optional)} Please enter your MSC Codes here.

\section{Introduction}
\label{sec:Introduction}

Generation of waves through linear instabilities of parallel two-phase shear flows is a canonical problem with wide ranging applications. Such instabilities underline phenomena as diverse as the generation of surface waves on the ocean \citep{sullivan2010dynamics}, spray-generation in combustion applications \citep{lefebvre2017atomization}, oil transportation \citep{joseph2013fundamentals} and coating technology \citep{weinstein2004coating}. In geophysical settings, shear driven instabilities are recognized as the fundamental mechanism of wave generation for wind-driven waves on lakes and the ocean. These waves in turn interact and modulate large-scale winds, currents, and mixing in the ocean \citep{young1993review,constantin2019equatorial,dai2010experiment}.

\begin{figure}
    \centering
    \hspace{-1.6cm}
    \begin{minipage}[t]{0.4\textwidth}
      \centering
      \includegraphics[width=\textwidth]{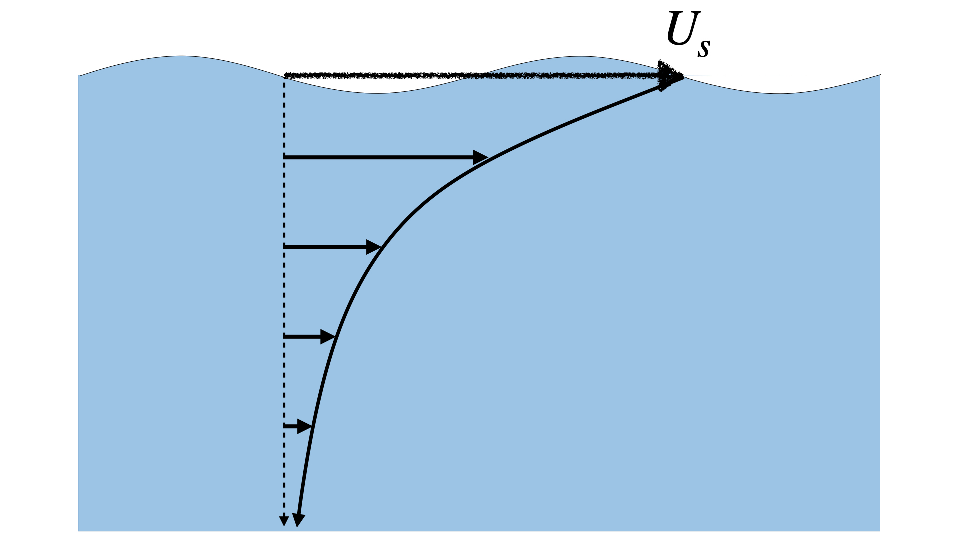}
      \caption*{(a)}
    \end{minipage}
     \hspace{-0.75cm}
    \begin{minipage}[t]{0.4\textwidth}
      \centering
      \includegraphics[width=\textwidth]{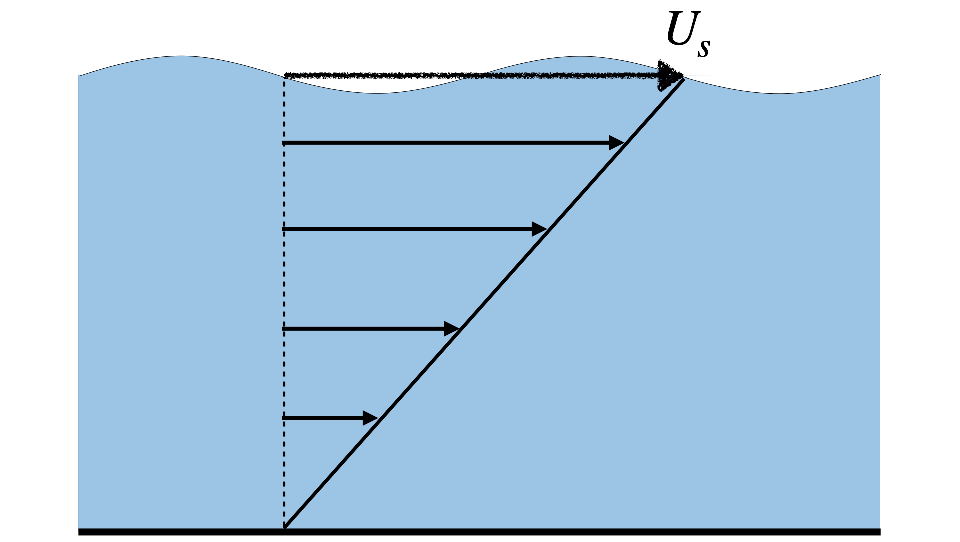}
      \caption*{(b)}
    \end{minipage}
    \hspace{-0.75cm}
    \begin{minipage}[t]{0.4\textwidth}
      \centering
      \includegraphics[width=\textwidth]{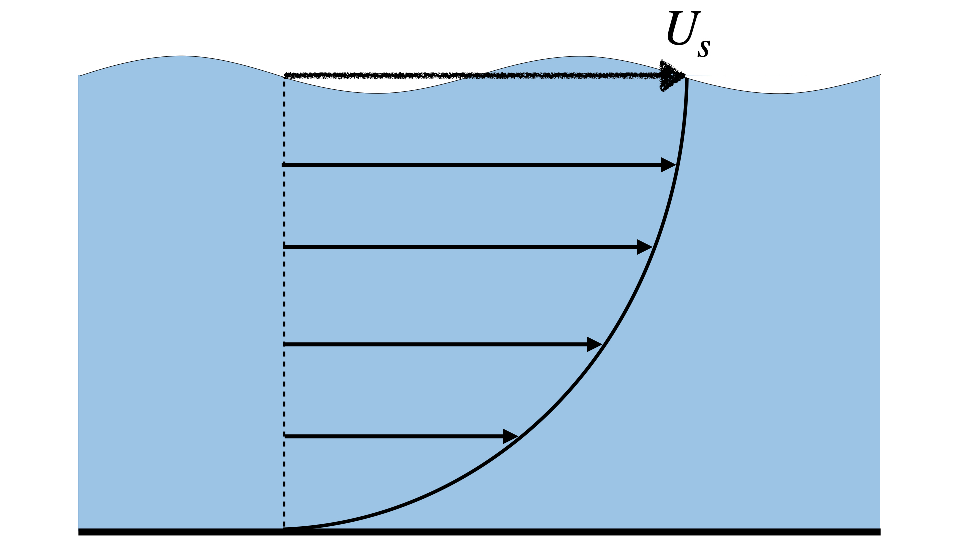}
      \caption*{(c)}
    \end{minipage}
    \hspace{-1.6cm}
\caption{A schematic showing some free-surface flow configurations and their respective instabilities as described in the literature: (a) An exponential velocity profile in a semi-infinite domain - rippling instability \citep{morland1991waves}, (b) A linear velocity profile in a finite depth domain - shear mode instability \citep{miles1960hydrodynamic}, and (c) A Nusselt profile in a finite depth domain - long wave interfacial instability\citep{benjamin1957wave,10.1063/1.1706737}.}
\label{Schematics}
\end{figure}

Depending on the application, various flow configurations have been considered in the literature, such as two-phase Couette flows, two-phase Poiseuille flows, core-annular flows and free-surface flows (\cite{govindarajan2014instabilities}, see Figure 2). Multiple instabilities were found in different regions of the parameter space and they were broadly classified depending on the dominant energy transfer mechanism leading to a growth in perturbation kinetic energy \citep{boomkamp1996classification}, and the eigenfunction structure \citep{charru2000phase}. Among the various mechanisms proposed for wave generation on the ocean surface \citep{ayet2022dynamical}, the Miles instability \citep{miles1957generation} remains the most prominent and experimentally supported model \citep{plant1982relationship}. This inviscid instability occurs due to a transfer of energy, through wave Reynolds stress, from a critical layer ($z_c$, defined as the vertical location where the wave phase speed ($c_r$) matches the background flow velocity i.e $c_r = U(z_c)$) in the air phase to the interface \citep{janssen2004interaction}. The asymptotic calculation of \citet{miles1957generation} and \citet{miles1959generation} shows that this energy transfer is proportional to the air-layer velocity profile curvature ($U''(z_c)$) implying that Miles instability will not exist in a constant shear flow (such as in the viscous sub-layer). Evidence of a critical layer in the air-layer was provided in the experiments of \cite{buckley2016structure} and \cite{carpenter2022evidence}. A similar mechanism of instability but with the critical layer in the water phase was studied by \citet{stern1973capillary} and \citet{morland1991waves}, indicating wave generation due to shear currents. The presence of wind stress driven currents in the ocean has been documented both in classical near equatorial wind-drift theories \citep{stommel1959wind} and more recent exact solutions for azimuthal equatorial flows with a free surface \citep{constantin2016exact}. These studies highlight the variety of velocity profile curvatures \citep[including flow reversals; see][]{constantin2019ekman} that can arise in the water layer and are crucial for the onset of instability. In particular, asymptotic analysis of \citet{shrira1993surface} and \citet{bonfils2023flow} show that a negative velocity profile curvature in the water layer is necessary for wave growth. A combined two-phase inviscid problem comparing the growth rates of both instabilities was studied by \citet{young2014generation}, who found that the latter instability is important at smaller wavelengths (naming it the ``rippling instability'') and can attain higher growth rates compared to Miles instability. A combined two-phase viscous problem was also studied by \citet{zeisel2008viscous}, who identified a second instability at high friction velocities and found, counter-intuitively, that viscosity enhances Miles instability growth rates. More recently, \citet{kadam2023wind} showed that this enhancement is due to a fundamental shift in the instability energy source from Reynolds stress at the critical layer to tangential stress at the interface. They also showed that viscosity reduces rippling instability growth rates, although the underlying instability mechanism is altered. The present study is motivated by the non-trivial dependence of instability mechanisms on both viscosity and background flow curvature. To this end, we focus on quadratic (Couette--Poiseuille) velocity profiles, which not only reproduce the forward and backward bulges observed in field and laboratory flows, but also remain tractable for both inviscid and viscous stability analyses. This framework enables us to disentangle the respective roles of curvature and viscosity on free surface instabilities. Some of the well-known free-surface instabilities associated with different shear flows, as discussed above, are schematically illustrated in figure~\ref{Schematics}.

Viscosity-stratified instabilities were first identified by \citet{yih1967instability} for Couette–Poiseuille base-state flows, where the velocity profile is quadratic with coefficients depending on viscosity ratio ($m = \mu_2/\mu_1$), density ratio ($r = \rho_2/\rho_1$), and height ratio ($n = h_2/h_1$). Subsequent long-wave asymptotic analyses revealed instabilities at arbitrarily small Reynolds numbers ($Re = \rho_2 U_s h_2/\mu_2$). A similar instability had been noted earlier by \citet{kapitza1948wave} and studied rigorously by \citep{benjamin1957wave,10.1063/1.1706737} in the context of falling films, called the Kapitza instability. The former is viscosity-induced while the latter is gravity-induced, with combined effects also studied in the literature (see \citet{boomkamp1996classification}). Long- and short-wave instabilities were later detailed by \citet{smith1990mechanism}, \citet{charru2000phase}, and \citet{kalliadasis2011falling}, with \citet{hooper1983shear} further identifying a short-wave instability in constant-shear flows due to tangential stress at the interface, a mechanism later described by \citet{hinch1984note}. In both cases, growth rates are proportional to $|m-1|$ for $r=1$, while density stratification ($r \neq 1$) can stabilize the flow under certain conditions \citep{hooper1983shear}. These instabilities, dominated by interfacial mechanisms, occur in a range of natural and technological contexts from magma and glaciers to oil transport and biological flows \citep{govindarajan2014instabilities}.  

Apart from inviscid and interfacial instabilities, shear-mode instabilities have also been identified. For a single-phase free-surface constant shear flow, \citet{miles1960hydrodynamic} derived the critical Reynolds number ($Re_c \approx 203$), showing that a viscous instability occurs at high $Re$ for modes which are inviscidly neutral with a critical layer, consistent with Heisenberg’s criterion \citep{lin1946stability}. Later, \citet{smith1982instability} considered surface-tension-gradient-driven shear flows and found much lower $Re_c$, while \citet{hooper1987shear} used asymptotic theory to study shear-mode instabilities in two-phase flows. Further works such as \citet{yiantsios1988linear}, \citet{hooper1989stability}, and \citet{miesen1995hydrodynamic} established that shear modes can coexist with interfacial modes while mode coalescence at high $Re$ is reported by \citet{timoshin1997instabilities}, \citet{ozgen1998two}, \citet{timoshin2000mode}, and \citet{kaffel2015eigenspectra}. A recent review by \citet{mohammadi2016stability} emphasized the importance of detailed studies of shear–interfacial mode interactions and their role across parameter regimes.

\begin{table}
  \small
  \def~{\hphantom{0}}
  \begin{tabular}{lccccc}
    \bfseries Paper & \bfseries $U_s$ (mm/s) & \bfseries $h$ (mm) & \bfseries $Re$ & \bfseries $G$ & \bfseries $Bo$ \\[6pt]
    \cite{cohen1965generation}\textsuperscript{\textdagger} & 40 -- 220 & 1 -- 7 & $O(10)$ -- $O(10^3)$ & $O(10^{-1})$ -- $O(10)$ & $\sim O(1)$ \\[6pt]
    \cite{craik1966wind} & 3 -- 68 & 0.1 -- 1.5 & $O(10^{-1})-O(10^2)$ & $O(1)-O(10^2)$ & $O(10^{-3})-O(10^{-1})$ \\[6pt]
    \cite{hidy1966wind}\textsuperscript{*} & 16 -- 296 & 25 -- 100 & $O(10^2)-O(10^4)$ & $O(1)-O(10^3)$ & $O(10^2)-O(10^3)$ \\[6pt]
    \cite{buckley2016structure}\textsuperscript{*} & 10 -- 330 & 700 & $\sim O(10^5)$ & $O(10^2)-O(10^4)$  & $\sim O(10^6)$ \\[6pt]
    \cite{paquier2016viscosity}\textsuperscript{\textdagger *} & 60 -- 168 & 35 & $O(10)-O(10^3)$ & $O(10)-O(10^2)$ & $\sim O(10^2)$ \\[6pt]
    \cite{liu2017experimental} & 44 -- 387 & 0.195 -- 1.09 & $O(10)-O(10^2)$ & $O(10^{-2})-O(1)$ & $O(10^{-3})-O(10^{-1})$  \\
  \end{tabular}
  \caption{Range of parameters reported in various experiments on air--water or air--water+glycerol two-phase flows. Key non-dimensional parameters are Reynolds number ($Re = \rho_2 U_s h_2 / \mu_2$), inverse squared Froude number ($G = g h_2 / U_s^2$), and Bond number ($Bo = \rho_2 g h_2^2 / T$). Here, $U_s$ is the interfacial velocity, $g$ is the acceleration due to gravity, $T$ is the surface tension, $\rho_2$ is the density, $\mu_2$ is the dynamic viscosity, and  $h_2$ is the depth of the bottom layer. References marked (*) estimate $U_s$ as $2\%$ of free-stream velocity, while those marked (\textdagger) use surface tension values from \citet{takamura2012physical}.}
  \label{Experiments}
\end{table}

In addition to theoretical studies, numerous experiments have investigated two-phase instabilities across configurations \citep[see][for details]{govindarajan2014instabilities}. Viscosity-stratified instabilities in oil–water systems have been studied in core-annular \citep{bai1992lubricated, kouris2001core}, circular Couette \citep{sangalli1995finite, barthelet1995experimental}, and planar parallel flows \citep{charles1965experimental, kao1972experimental}. For air–water like systems, several experiments in planar channel flows are summarized in Table~\ref{Experiments}. These cover a broad range of $Re$, $G$, and $Bo$ and report distinct instabilities such as ``fast’’ and ``slow’’ waves \citep{cohen1965generation, craik1966wind}, viscous modifications of Miles instability \citep{hidy1966wind, buckley2016structure}, transitions between wrinkle and wave regimes \citep{paquier2016viscosity}, and glaze-ice-related film instabilities \citep{liu2017experimental}. Interpretations vary across studies, with energy analyses attributing most of these waves to interfacial instabilities \citep{boomkamp1996classification}, though Reynolds stress contributions are sometimes non-negligible \citep{naraigh2011interfacial}. These works collectively demonstrate the complex interplay of viscosity, shear, and interfacial mechanisms in air–water instabilities.  

The experiments and theory above emphasize that the two-phase stability problem is governed by six non-dimensional parameters—$Re$, $G$, $Bo$, $m$, $r$, and $n$—making comprehensive characterization difficult. Fixing the phases (e.g., air–water) reduces this to three: $Re$, $G$, and $Bo$. Further simplification arises from the free-surface approximation, where the air layer is treated as passive. This approximation neglects air-side instabilities like Miles’ instability but remains justified due to the large density ratio ($\rho_2/\rho_1 \gg 1$) and small viscosity ratio ($\nu_2/\nu_1 \ll 1$). \citet{young2014generation} have shown that this approximation preserves key features of the rippling instability. In the present work, we adopt this framework and consider a family of quadratic Couette–Poiseuille velocity profiles parameterized by curvature $a$, allowing a systematic exploration of profile curvature effects without explicitly specifying $m$, $r$, and $n$.  

To the best of our knowledge, a comprehensive study of how viscosity and velocity profile curvature affect rippling instability and its interaction with other modes has not been carried out. The present work addresses this gap. Section~\ref{sec:Problem_formulation} describes the governing equations and base-state velocity profiles. Inviscid stability results, including analytical solutions, asymptotic expressions, stability boundaries, and complete parameter-space behavior, are discussed in \S\ref{invlsa}. Viscous stability analyses, combining asymptotic and numerical approaches, are presented in \S\ref{vislsa}. Section \ref{instabilityFamily} characterizes instability families using energy and eigenfunction structures, and conclusions are summarized in \S\ref{conclusion}.

\section{Problem formulation}\label{sec:Problem_formulation}

We consider a two-dimensional, incompressible, parallel flow in a water layer, represented by a family of quadratic base-state velocity profiles, i.e. Couette–Poiseuille flows. The domain is unbounded in the horizontal direction and bounded vertically by a free surface at the top and a rigid wall at the bottom, located at a dimensional depth $h$ below the interface. The base-state shear flow velocity at the free surface, $U_s$, is chosen as the characteristic velocity scale, while $h$ is chosen as the characteristic length scale, and $h/U_s$ as the characteristic time scale, to nondimensionalize all variables. Henceforth, all quantities in the analysis are nondimensional unless otherwise specified.  

Assuming small-amplitude perturbations, the governing set of nondimensional linearized equations for the perturbation fields can be written as
\begin{equation}\label{eq1}
    \frac{\partial u}{\partial x} + \frac{\partial w}{\partial z} = 0,
\end{equation}
\begin{equation}\label{eq2}
    \frac{\partial u}{\partial t} + U \frac{\partial u}{\partial x} + u \frac{dU}{dz} = - \frac{\partial p}{\partial x} + \frac{1}{Re}\nabla^2 u,
\end{equation}
\begin{equation}\label{eq3}
    \frac{\partial w}{\partial t} + U \frac{\partial w}{\partial x} = - \frac{\partial p}{\partial z} + \frac{1}{Re}\nabla^2 w,
\end{equation}
where $x$ and $z$ are the horizontal and vertical coordinates, respectively, $t$ denotes time, $u(x,z,t)$ and $w(x,z,t)$ are the perturbation velocity components, $p(x,z,t)$ is the perturbation pressure, and $U(z)$ is the base-state velocity profile. The operator $\nabla^2(\cdot)$ denotes the two-dimensional Laplacian in the $(x,z)$ coordinates. The Reynolds number is defined as
\[
Re = \frac{\rho U_s h}{\mu},
\]
where $\rho$ and $\mu$ are the density and dynamic viscosity of water, respectively.  

At the rigid wall ($z=-1$), the no-slip and no-penetration boundary conditions are
\begin{subequations}\label{eq:4}
    \begin{equation}
        u = 0, \qquad w = 0.
        \tag{\theequation{a,b}}
    \end{equation}
\end{subequations}
At the free surface ($z=0$), the kinematic, tangential stress, and normal stress boundary conditions, linearized about the base state, take the form
\begin{subequations}\label{eq:5}
\begin{equation}
    \frac{\partial \hat{\eta}}{\partial t} + \frac{\partial \hat{\eta}}{\partial x} = w, 
    \qquad
    \hat{\eta}\frac{d^2U}{dz^2} + \frac{\partial u}{\partial z} + \frac{\partial w}{\partial x} = 0,
    \tag{\theequation{a,b}}
\end{equation}
\end{subequations}
\begin{equation}\label{eq:6}
    p + \frac{2}{Re}\frac{\partial u}{\partial x}
    - G\left(\hat{\eta} - Bo^{-1}\frac{\partial^2 \hat{\eta}}{\partial x^2}\right) = 0,
\end{equation}
where $z = \hat{\eta}(x,t)$ is the free surface displacement, $G$ is the inverse squared Froude number,
\[
G = \frac{gh}{U_s^2},
\]
and $Bo$ is the Bond number,
\[
Bo = \frac{\rho g h^2}{T}.
\]
Here, $g$ is the gravitational acceleration and $T$ is the air–water surface tension. Condition (\ref{eq:5}b) expresses that the total tangential stress vanishes at the free surface, while \eqref{eq:6} represents the balance of perturbation normal stresses with surface tension.  

Introducing a perturbation streamfunction $\psi(x,z,t)$ such that $u = \partial \psi / \partial z$ and $w = - \partial \psi / \partial x$, and adopting a normal-mode decomposition
\[
\{\psi, p, \hat{\eta}\} = \{\phi(z), f(z), \eta\} \, e^{ik(x-ct)},
\]
equations \eqref{eq1}–\eqref{eq3} can be combined into the Orr–Sommerfeld equation \citep{drazin2004hydrodynamic},
\begin{equation}\label{OSE}
    (U - c) \left(\frac{d^2}{dz^2} - k^2\right)\phi - U'' \phi
    = \frac{1}{ikRe}\left(\frac{d^2}{dz^2} - k^2\right)^2 \phi,
\end{equation}
where $k$ is the nondimensional wavenumber and $c$ is the complex phase speed of the perturbation.  

At the rigid wall ($z=-1$), the boundary conditions reduce to
\begin{subequations}\label{no slip and no pen}
    \begin{equation}
        \phi(-1) = 0, \qquad \phi'(-1) = 0.
        \tag{\theequation{a,b}}
    \end{equation}
\end{subequations}
The linearized conditions at the free surface ($z=0$) are
\begin{subequations}\label{kin and tan}
    \begin{equation}
        \eta = \frac{\phi}{c - U}, \qquad \eta U'' + \phi'' + k^2 \phi = 0,
        \tag{\theequation{a,b}}
    \end{equation}
\end{subequations}
\begin{equation}\label{nor}
    \phi''' - 3k^2 \phi' + ikRe \left(U' \phi + (c - U)\phi'\right)
    - ikRe\, G\, \eta\left(1 + \frac{k^2}{Bo}\right) = 0.
\end{equation}
Here, primes denote differentiation with respect to $z$. For a given $k$ and specified nondimensional parameters ($Re, G, Bo$), the Orr–Sommerfeld equation \eqref{OSE}, subject to the boundary conditions \eqref{no slip and no pen}–\eqref{nor}, yields the eigenvalue $c(k) = c_r(k) + i c_i(k)$ and eigenfunction $\phi(z;k)$. The real part $c_r$ represents the phase speed, while $kc_i$ denotes the temporal growth rate. Modes with $c_i>0$ are unstable, whereas those with $c_i<0$ are stable.  

The present study focuses on the limit of strong interfacial forcing ($G=0$), where gravitational effects are negligible compared to shear. In this limit, for any finite $Bo$, equation \eqref{nor} shows that surface tension also does not contribute. This asymptotic regime has been examined in several previous works \citep{miles1960hydrodynamic, smith1982instability, miesen1995hydrodynamic} and is especially relevant for thin-film flows (see Table~\ref{Experiments}). For completeness, cases with nonzero $G$ are also considered; in these, gravity is finite but surface tension is assumed small, corresponding to a large Bond number $Bo \approx 10^5$ throughout the study.

\subsection{Base-state profiles}

To investigate the effect of velocity profile curvature on stability characteristics, we consider a one-parameter family of Couette–Poiseuille base-state flow profiles defined as
\begin{equation}\label{base-state profiles}
    U(z) = a z^2 + (a+1)z + 1,
\end{equation}
where $a$ is a curvature parameter that may take values in the range $-\infty < a < \infty$. Any quadratic velocity distribution can be recast into this nondimensional form, making it a convenient canonical representation. The parameter $a$ thus provides a systematic way to explore the influence of curvature on instability.  

Several notable flow configurations correspond to special values of $a$. For $a=0$, equation \eqref{base-state profiles} reduces to a linear velocity profile, characteristic of a simple shear flow. The case $a=-1$ yields the Nusselt profile, which arises in the classical problem of a liquid film flowing down an inclined plane under gravity. For $a=3$, the profile exhibits a flow reversal near the free surface, a phenomenon observed in laboratory experiments on wind-driven water waves \citep{hidy1966wind, paquier2015surface, paquier2016viscosity}. Interestingly, similar flow reversals occur in large-scale geophysical settings, such as equatorial oceanic currents, though with more complex vertical structures \citep{stommel1959wind, constantin2016exact}. In the asymptotic limits, $a \to -\infty$ yields a rescaled forward Poiseuille-type flow, whereas $a \to +\infty$ corresponds to a backward Poiseuille-type profile.  

\begin{figure}
  \centerline{\includegraphics[scale=0.25]{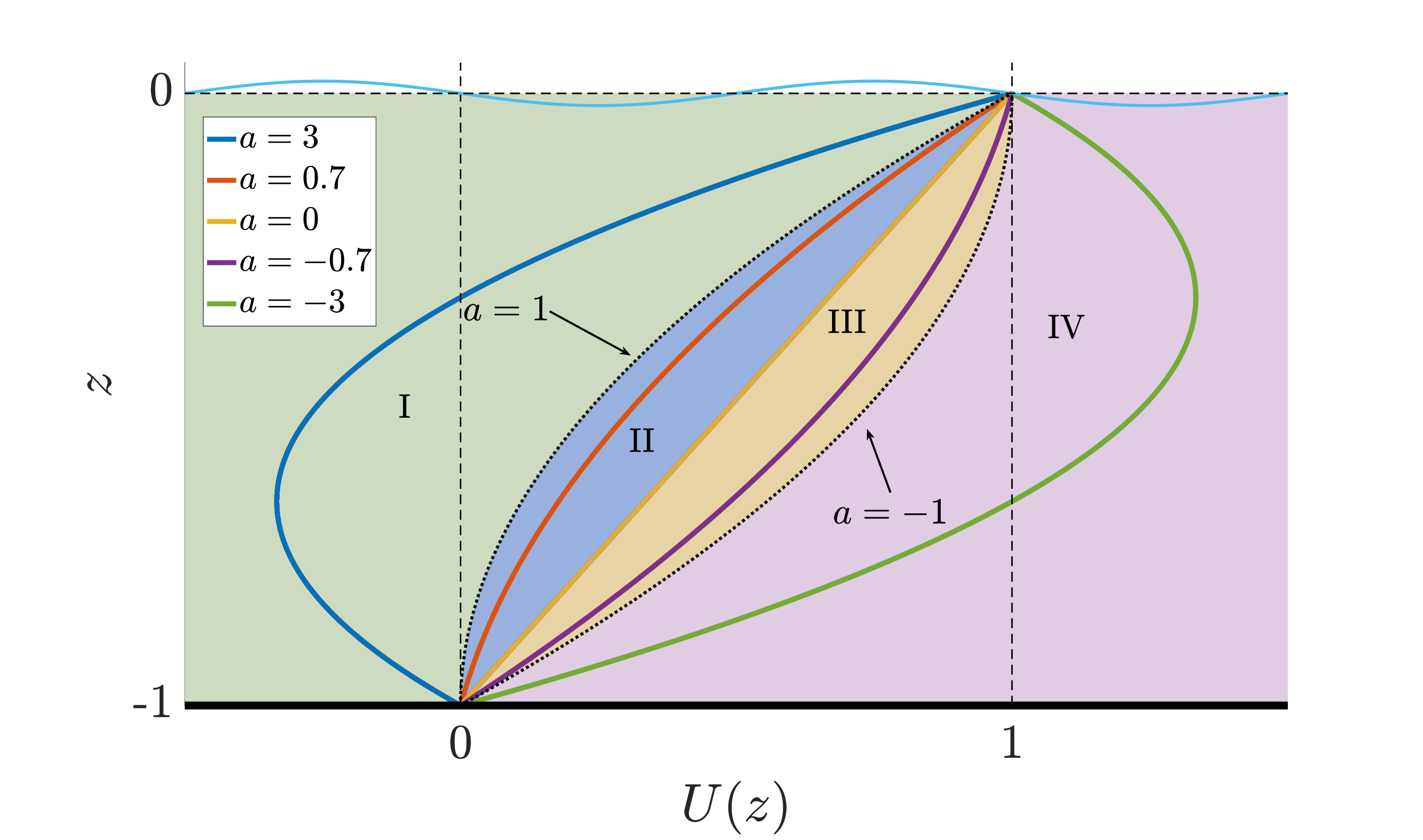}}% Images in 100% size
  \caption{Base-state velocity profiles for different values of the curvature parameter $a$, highlighting the four regimes: I — backward bulging ($a \in [1,\infty)$), II — monotonic convex ($a \in (0,1)$), III — monotonic concave ($a \in [-1,0]$), and IV — forward bulging ($a \in (-\infty,-1]$).}
  \label{base-states fig}
\end{figure}

The parameter space of $a$ can be naturally divided into four distinct regimes, each corresponding to a qualitatively different profile shape: (i) \emph{backward bulging} ($1 \leq a < \infty$), which includes the flow-reversal case; (ii) \emph{monotonic convex} ($0 < a < 1$); (iii) \emph{monotonic concave} ($-1 < a \leq 0$), which contains the linear shear profile; and (iv) \emph{forward bulging} ($-\infty < a \leq -1$), which includes the Nusselt profile. Representative examples of these regimes are illustrated in figure~\ref{base-states fig}.

\section{Inviscid linear stability analysis}\label{invlsa}

In the inviscid limit ($\Rey \to \infty$), the Orr–Sommerfeld equation \eqref{OSE} reduces to the Rayleigh equation \citep{drazin2004hydrodynamic},  
\begin{equation}\label{Rayleigh eqn}
    (U(z)-c)\left(\phi''(z) - k^2\phi(z)\right) - U''(z)\phi(z) = 0,
\end{equation}
which governs the eigenfunction $\phi(z)$ for a given wavenumber $k$ and complex phase speed $c$. At the rigid bottom wall ($z=-1$), only the no-penetration condition remains,  
\begin{equation}\label{inv nop pen}
    \phi(-1) = 0,
\end{equation}
while the number of independent conditions at the free surface $z=0$ reduces from three to two. Specifically, the kinematic and normal-stress conditions can be combined into a single dynamic constraint \citep{young2014generation},  
\begin{equation}\label{inv dynamic bc}
     \Xi_w (c-U)^2 + U'(0)(c-U) - G\left(1 + k^2 Bo^{-1}\right) = 0, \quad \text{where} \quad \Xi_w = \dfrac{\phi'(0)}{\phi(0)} 
\end{equation}
 
\subsection{Analytical solution}
\label{sec:analytical solution}
For the family of base-state velocity profiles considered, the Rayleigh equation \eqref{Rayleigh eqn} can be solved analytically in closed form, generalizing the special case $a=3$ studied in \citet{kadam2023wind}. After a sequence of transformations, the solution to \eqref{Rayleigh eqn} for arbitrary $a$ may be expressed in terms of confluent Heun functions (denoted by H$_c$) as
\begin{equation}
\label{analyticalsol}
\phi(z) = e^{\kappa(\zeta(z)-\zeta_c)}(\zeta_c^2-\zeta(z)^2)\left(C_1 \phi_1(z) + C_2\phi_2(z)\right),
\end{equation}
where
\begin{equation}
\nonumber
\phi_1(z, \zeta_c) = \dfrac{\textrm{H$_c$}\left[2+4\kappa\zeta_c,4\kappa\zeta_c,2,0,4\kappa\zeta_c,\frac{\zeta_c+\zeta(z)}{2\zeta_c}\right]}{\zeta_c-\zeta(z)},\hspace{1cm}\phi_2(z, \zeta_c) = \phi_1(z, -\zeta_c).
\end{equation}
\begin{equation}
\nonumber
\zeta(z) = U'(z) = 2az + (a+1),\hspace{1cm} \zeta_c = \zeta(z_c),\hspace{1cm} \kappa = k/2a,
\end{equation}
with $C_1$ and $C_2$ as integration constants, and $z_c$ denoting the critical layer depth obtained by solving $U(z_c)-c=0$. Additional details on confluent Heun functions and their alternate representations are provided in Appendix \ref{appA}. Here $\zeta(z)$ is the negative of the base-state vorticity and $\zeta_c$ its value at the critical level.

The dispersion relation follows by enforcing boundary conditions \eqref{inv nop pen} and \eqref{inv dynamic bc}, giving
\begin{equation}
\label{dispersionrelation}
\Xi_w(c-1)^2 + (a+1)(c-1) - G(1+k^2Bo^{-1}) = 0,
\end{equation}
where
\begin{equation}
\label{disprel2}
\Xi_w = k + \dfrac{a+1}{1-c} + \dfrac{\phi_1(-1)\phi_2'(0)-\phi_2(-1)\phi_1'(0)}{\phi_1(-1)\phi_2(0)-\phi_2(-1)\phi_1(0)}.
\end{equation}

For prescribed $(a,k,c,Bo)$, equation \eqref{dispersionrelation} yields $G$ explicitly. Conversely, for fixed $(a,G,Bo,k)$, a root-finding procedure is required to determine $c$ and the corresponding growth rate $kc_i$. Thus, despite its compact analytical form, the dispersion relation cannot be analyzed directly without numerical evaluation. In the sections that follow, we examine its implications: Section~\ref{longwave} addresses long-wave asymptotics for the growth rate, while Section~\ref{stabilityboundary} derives explicit expressions for the stability boundary in different regimes of $a$.

\subsection{Long wave asymptotics ($k \ll 1$)}\label{longwave}
\begin{figure}
    \hspace{-1.5cm}
    \begin{minipage}[t]{0.6\textwidth}
      \centering
      \includegraphics[width=\textwidth]{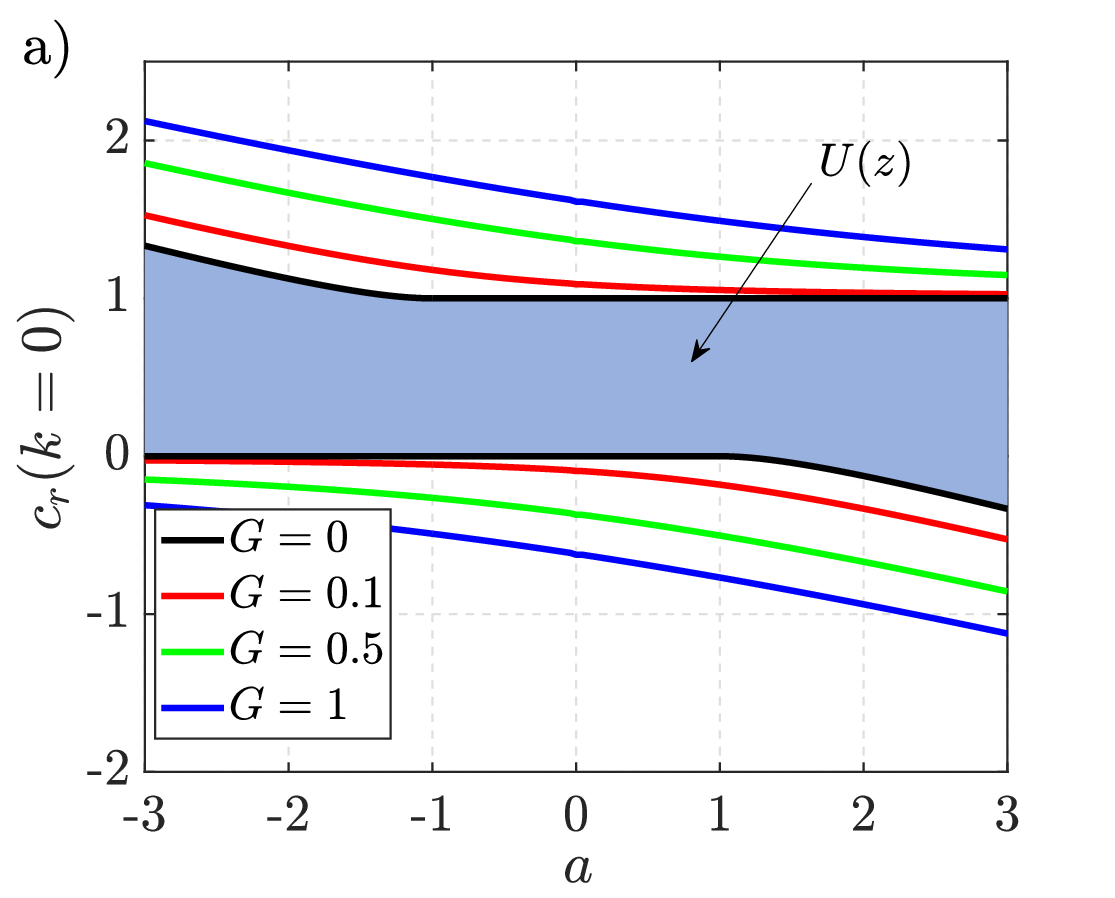}
    \end{minipage}
     \hspace{-0.7cm}
    \begin{minipage}[t]{0.6\textwidth}
      \centering
      \includegraphics[width=\textwidth]{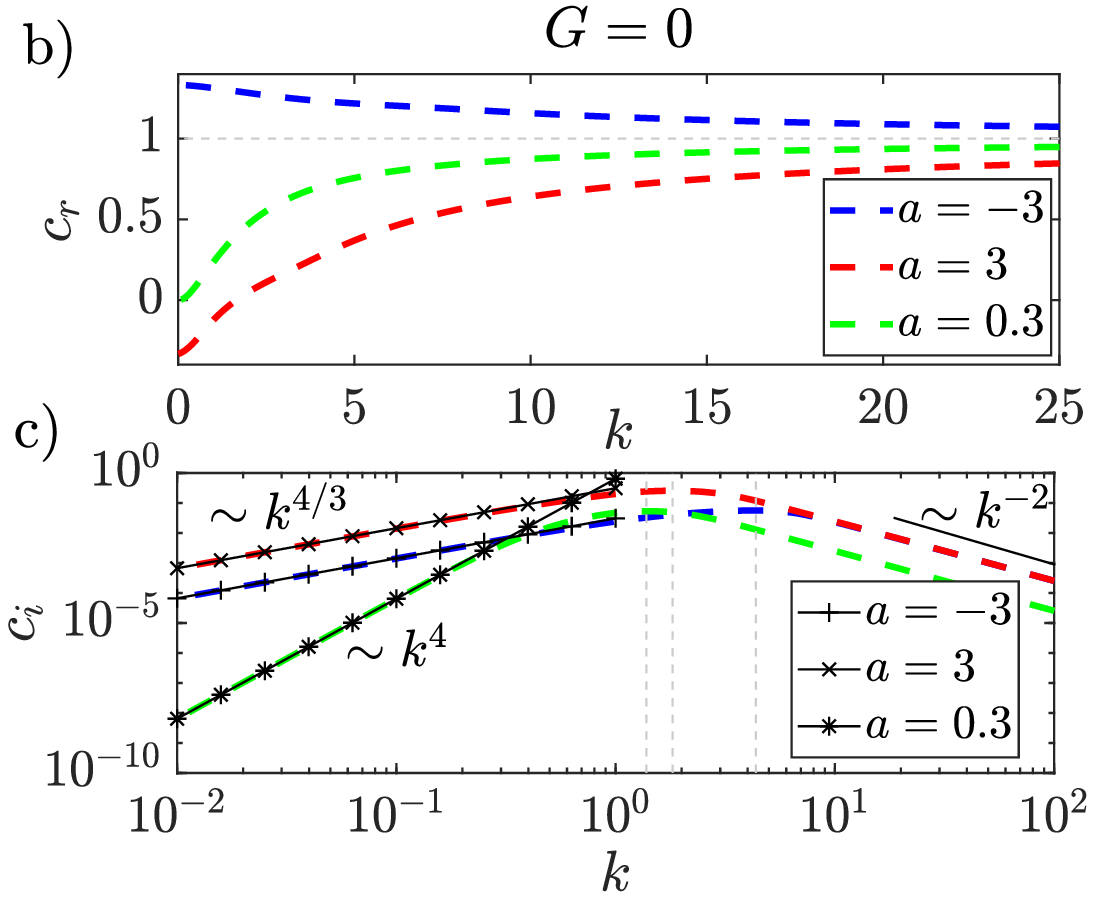}
    \end{minipage}
\caption{(a) Long-wave phase speeds obtained from \eqref{Burns disp rel} for different inverse squared Froude numbers ($G$), as a function of curvature parameter $a$. The shaded blue region corresponds to phase speeds coinciding with the background velocity, i.e. $c_r = U(z)$. For $G=0$, (b) the real part $c_r$, and (c) the imaginary part $c_i$ of the complex phase speed, plotted against wavenumber $k$ for various $a$. Black curves with markers in (c) indicate the asymptotes calculated in Section~\ref{longwave}, while the grey dashed lines mark the location of maximum $c_i$ for each $a$.}
\label{Burns integral cr vs a}
\end{figure}

In order to obtain physical insight without resorting to a numerical evaluation of the complete dispersion relation, we perform asymptotics in the long wave ($k \ll 1$) limit. The leading-order behaviour of long waves can be obtained from the Burns integral relation \citep{burns1953long}, which connects the wave phase speed to the Froude number in the limit $k\to 0$:  
\begin{equation}\label{Burns integral}
    \int_{-1}^{0} \dfrac{1}{(U(z) - c)^2} \, dz - \dfrac{1}{G} = 0.
\end{equation}
For the velocity profiles in \eqref{base-state profiles}, this reduces to
\begin{equation}\label{Burns disp rel}
    G\left(\alpha(1+a(2c-1)) + 2ac(c-1)\log\left(\dfrac{2c-1+a+\alpha}{2c-1+a-\alpha}\right)\right)-c(c-1)\alpha^3 = 0,
\end{equation}
with $\alpha^2 = (a-1)^2 + 4 a c$.  

For $G>0$ and every value of curvature ($a$), equation \eqref{Burns disp rel} yields two neutrally stable modes ($c_i=0$): a prograde branch with speed exceeding the maximum background velocity, and a retrograde branch with speed below the minimum velocity (figure \ref{Burns integral cr vs a}a). Thus, long waves remain neutrally stable for any $G>0$. In the limit of $G =0$, however, the phase speeds $c_r$ for all the curvatures, of the prograde and retrograde modes attain the maximum and minimum values of the base state velocity, respectively  (black curves in figure \ref{Burns integral cr vs a}a). With the maximum given by $\big(-(a-1)^2/4a$ for $a\leq-1$ and 1 for $a>-1\big)$ and minimum $\big(0$ for $a\leq1$ and $-(a-1)^2/4a$ for $a>1\big)$. This coincidence with the background velocity indicates the emergence of a critical layer and thereby the possibility of a rippling-type instability at small but non-zero $k$ \citep{young2014generation}. Notably, when $a<-1$, the prograde mode itself can destabilize due to the forward bulge in the profile, consistent with the observations of \citet{bonfils2023flow}.

Figures \ref{Burns integral cr vs a}b,c illustrate this behaviour at finite $k$ for $G=0$ and representative curvature parameters $a=-3$ (blue), $3$ (red), and $0.3$ (green). The real part $c_r$ (panel b), computed from the dispersion relation \eqref{dispersionrelation}, begins at the extremal velocity and approaches the surface speed $U_s$ as $k$ increases, confirming the existence of a critical layer for all $k$. The imaginary part $c_i$ (panel c) demonstrates finite growth rates across the $k$ range, with maxima occurring at $O(1)$ wavenumbers, a typical feature of rippling instability \citep{young2014generation}. The logarithmic axis for $k$ highlights the distinct small- and large-$k$ asymptotic behaviour.

To capture the small-$k$ growth rates for $G=0$, we adopt the perturbative approach of \citet{renardy2013stability}, using the leading-order pressure perturbation solution as input to the pressure pertubation equation at the next order. The governing equation is  
\begin{equation}\label{Pressure pert Rayleigh}
    \left( \frac{p'}{(U(z) - c)^2}\right)' - k^2 \frac{p}{(U(z) -c)^2} = 0,
\end{equation}
subject to
\refstepcounter{equation}
    $$
    k^2 (U(0)-c)^2 p(0) - G(1 + Bo^{-1} k^2)p'(0) = 0, \quad p'(-1) = 0,
    \eqno{(\theequation{\mathit{a},\mathit{b}})}\label{complete pressure pert bcs}
    $$
which reduce to $p(0)=0$ and $p'(-1)=0$ for $G=0$.  

\begin{enumerate}
\item \textbf{Case I ($|a|>1$).}  
Substituting the phase speeds from \eqref{Burns disp rel} into \eqref{Pressure pert Rayleigh}, the $k=0$ solution reads
\begin{equation}
\label{pres. sol.}
    p_0(z)= 
\begin{cases}
    1 - \left(\dfrac{\zeta(z)}{1+a}\right)^5,& 0 \geq z > z_c,\\
    1, & z_c > z \geq -1, 
\end{cases}
\end{equation}
with the critical layer at $z_c = -(a+1)/2a$ where $c=U_{\min/\max}$. This piecewise form reflects the neutral continuous spectrum. For finite $k$, integration of \eqref{Pressure pert Rayleigh} yields the relation
\begin{equation}
\label{integ. rel.}
    \dfrac{p'(0)}{(1-c)^2} = k^2 \int_{-1}^{0} \dfrac{p}{(U -c)^2} \, dz.
\end{equation}
Using $p(z)\approx p_0(z)$ and setting $c=U_{\min/\max}+c_1$ with $|c_1|\ll 1$ gives
\begin{equation}
\label{eq:3.13}
    -\frac{160 a^3}{(a+1)^5} \sim -k^2 \frac{i \pi }{2 \sqrt{a} (c_1)^{3/2}},
\end{equation}
leading to
\begin{equation}
\label{growth scale inviscid}
    c_1 \sim \left(\dfrac{\pi (a+1)^5}{320 a^3\sqrt{|a|}}\right)^{2/3}k^{4/3}e^{i\pi/3}.
\end{equation}
The non-integer power law ($c_i \sim k^{4/3}$) reflects the singular role of the critical layer. This generalises earlier results of \citet{kadam2023wind} and recovers \citet{renardy2013stability} in the $|a|\to\infty$ limit.

\item \textbf{Case II ($|a|<1$).}  
Here the $k=0$ solutions correspond to $c=1$ (prograde) and $c=0$ (retrograde). Since the Doppler shift prevents the prograde mode phase speed from matching with the background velocity profile, only the retrograde branch can destabilize. Following \citet{renardy2013stability}, we obtain a neutral solution at $c=0$, leading to the asymptotic relation
\begin{equation}
\label{eq:3.17}
    \bar{c} \sim \dfrac{\bar{k}^2p_0(-1)}{2} + \dfrac{\bar{k}^4p_0^2(-1)}{8}\left(\pi i - \ln\left(\dfrac{\bar{k}^2p_0(-1)}{4}\right)\right) + \mathcal{O}(\bar{k}^6),
\end{equation}
with $\bar{k}=(1-a)k/2a$, $\bar{c}=4ac/(1-a)^2$ and
\begin{equation}
    \label{eq:3.17a}
   p_0(-1) = \left(\dfrac{\zeta(0)^5-\zeta(-1)^5}{5(1-a)^5}-\dfrac{2\zeta(0)^3-\zeta(-1)^3}{3(1-a)^3} +\dfrac{\zeta(0)-\zeta(-1)}{1-a}\right).
\end{equation}
This gives $c_i\sim k^4$ for $0<a<1$, while $a<0$ yields stability. Figure \ref{Burns integral cr vs a}c (black stars) confirms the agreement with the full solution.
\end{enumerate}

For large $k$, the growth rate scales as $c_i\sim k^{-2}$ in the $G=0$ limit, consistent with \citet{kaffel2011surface}. More generally, studies by \citet{shrira1993surface} and \citet{bonfils2023flow} show exponential decay for $G>0$, with the instability mechanism tied to curvature at the critical layer. A complementary small-curvature asymptotic analysis is given in Appendix~\ref{appB}.  

Thus, long waves are neutrally stable for any non-zero $G$, while for $G=0$ the presence of a critical layer leads to the rippling instability with growth rates that depend on curvature $a$. In the next section, we derive explicit expressions for the stability boundary.

\subsection{Stability boundary}\label{stabilityboundary}
\begin{figure}
    \hspace{-1.5cm}
    \begin{minipage}[t]{0.6\textwidth}
      \centering
      \includegraphics[width=\textwidth]{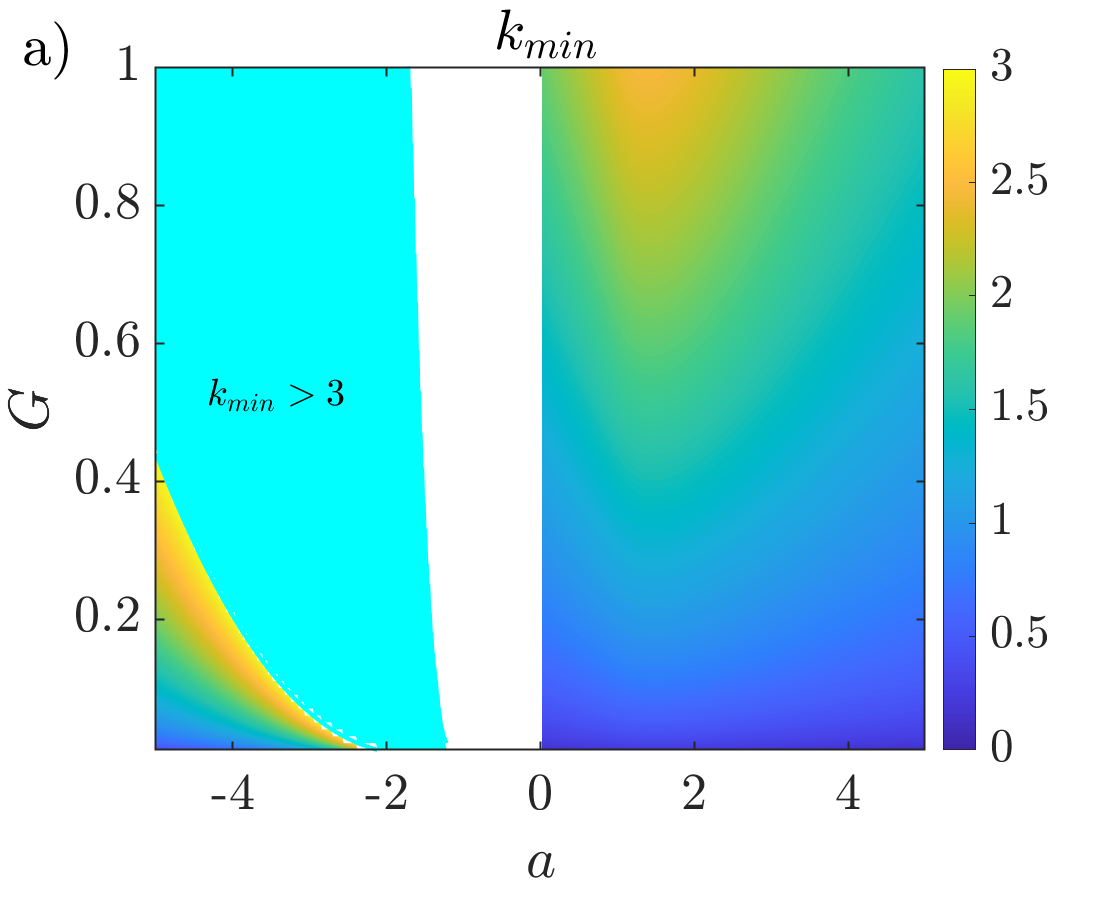}
    \end{minipage}
     \hspace{-0.5cm}
    \begin{minipage}[t]{0.6\textwidth}
      \centering
      \includegraphics[width=\textwidth]{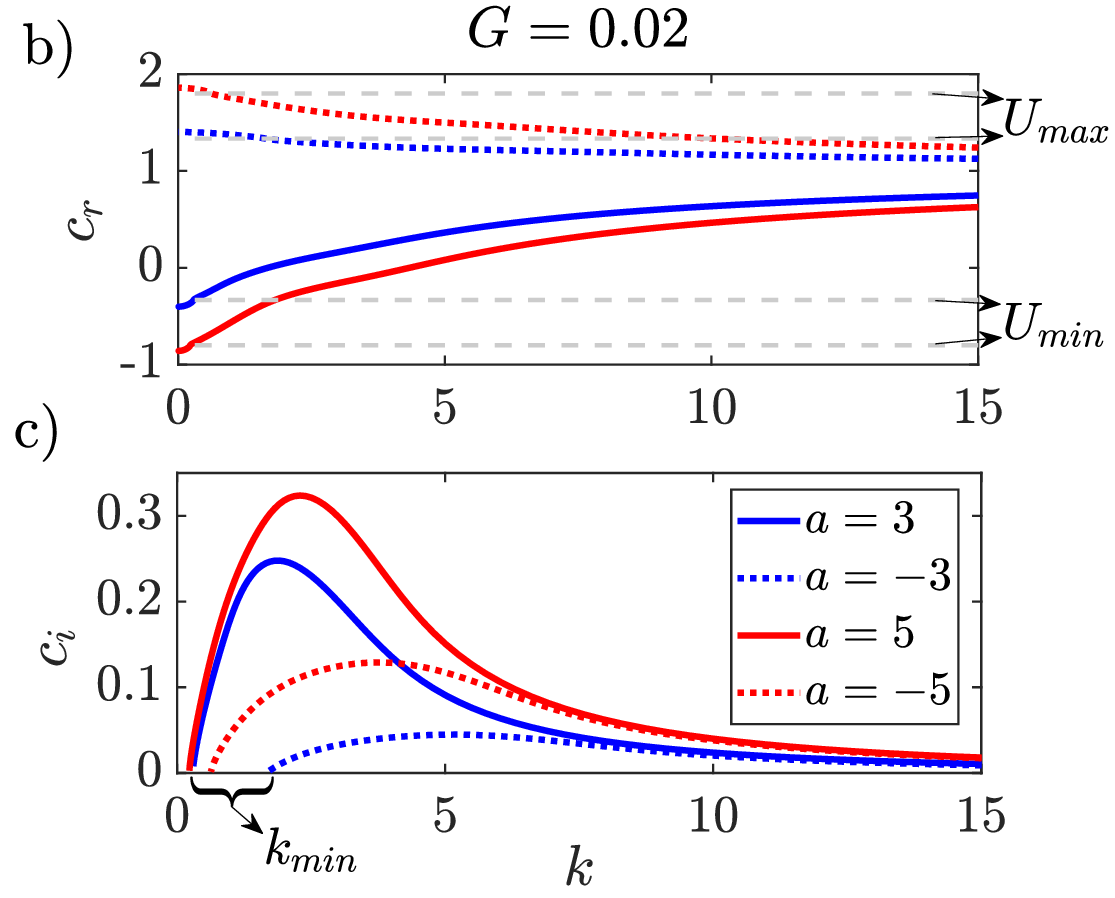}
    \end{minipage}
\caption{(a) Contour plot of the long-wave cutoff $k_{min}$ in the $(G,a)$ plane. The region with $k_{min}>3$ is shown in cyan to highlight the $k_{min}<3$ region. For $G=0.02$, (b) the real part $c_r$ and (c) the imaginary part $c_i$ of the complex phase speed, plotted against wavenumber $k$ for different $a$. Gray dashed lines in (b) indicate the extrema of the base-state velocity for each $a$, and the crossings with solid lines mark the emergence of critical layers.}
\label{Stability boundary contour}
\end{figure}

A closed-form analytical solution, as outlined in \S \ref{sec:analytical solution}, enables the numerical construction of the full stability contour. However, such an approach is computationally intensive and does not provide an immediate answer to the simpler question of whether instability arises for a given set of parameters. In this regard, a stability boundary that demarcates the stable and unstable regions of the parameter space serves as a valuable diagnostic tool. We construct this boundary by invoking the extension of Howard’s semicircle theorem to free-surface shear flows, due to \citet{yih1972surface}, which restricts the phase speed $c_r$ of any unstable mode to lie between the minimum and maximum values of the base-state velocity profile. This property has been widely used \citep{morland1991waves, engevik2000note, renardy2013stability, young2014generation, kadam2023wind} to identify neutral modes at $c=U_{min}$ or $c=U_{max}$, which define the boundary. Here we apply this approach to obtain explicit expressions for the stability boundary in terms of $G$, $k$, and $a$. Substituting the extremal values of the velocity profile into the governing equations provides conditions for neutral stability.

\begin{enumerate}
\item \textbf{Case I ($|a|>1$):}  
In this regime, where the flow is either forward or backward bulging, the velocity extrema are $U_{max/min}=-(a-1)^2/4a$, occurring at $z_c=-(a+1)/2a$. Setting $c=U_{min/max}$ in \eqref{Rayleigh eqn} yields the neutral eigenfunction
\begin{equation}
\label{pres. sol.}
    \phi(z)= 
\begin{cases}
    \dfrac{(1+a)\left(\kappa \zeta(z) \cosh [\kappa \zeta(z)] - \sinh[\kappa \zeta(z)]\right)}{\zeta(z) \left(\kappa (1+a) \cosh[\kappa (1+a)] - \sinh[\kappa (1+a)]\right)}, & 0 \geq z > z_c,\\
    0, & z_c > z \geq -1,
\end{cases}
\end{equation}
with $\zeta(z)$ and $\kappa$ defined in \eqref{analyticalsol}. Enforcing the surface condition at $z=0$ gives the stability boundary
\begin{equation}\label{stability boundary dispersion 1}
     G = \dfrac{(1+a)^3}{16a^2(1+k^2Bo^{-1})}\left(-6a + \dfrac{(1+a)^2 k^2}{k(1+a)\coth[k(1+a)/2a]-2a}\right).
 \end{equation}

\item \textbf{Case II ($0<a<1$):}  
Here only the retrograde branch can destabilize. Taking $c=U_{min}=0$ does not yield a simpler eigenfunction, but substitution into the dispersion relation \eqref{dispersionrelation} gives
\begin{equation}\label{stability boundary dispersion 2}
    G = \dfrac{1}{1+k^2 Bo^{-1}} \left(\dfrac{a \textrm{H}_c'\left[2+4\kappa(a-1), 4\kappa(a-1), 2, 0, 4\kappa(a-1), \tfrac{a}{a-1}\right]}{(a-1)\textrm{H}_c\left[2+4\kappa(a-1), 4\kappa(a-1), 2, 0, 4\kappa(a-1), \tfrac{a}{a-1}\right]}+k-a\right),
\end{equation}
where $\textrm{H}_c'$ denotes the derivative of the confluent Heun function.

\item \textbf{Case III ($-1<a<0$):}  
In this range, the prograde mode remains stable. Substituting $c=1$ into the boundary condition \eqref{inv dynamic bc} simplifies to
\begin{equation}\label{stability boundary dispersion 3}
    G \left(1 + \dfrac{k^2}{Bo}\right)= 0 \quad \Rightarrow \quad G = 0,
\end{equation}
demonstrating that no neutral mode exists for finite $G$.
\end{enumerate}

In summary, \eqref{stability boundary dispersion 1}–\eqref{stability boundary dispersion 2} can be solved to obtain the minimum unstable wavenumber $k_{min}$ as a function of $G$ and $a$. The resulting contour map is shown in figure \ref{Stability boundary contour}a. For $G=0$, $k_{min}=0$ across all $a$, consistent with the long-wave analysis. For $G>0$ and $a>0$, $k_{min}$ generally grows with $G$ but varies non-monotonically with $a$, peaking near $a\approx1.41$. For $a<0$, $k_{min}$ increases rapidly with $G$, reaching $O(10^2)$ at $G=1$. Since such large cutoffs are beyond our focus, we shade $k_{min}>3$ uniformly in cyan. The white band at $0<a<1$ reflects the absence of instability, in agreement with \eqref{stability boundary dispersion 3}. For finite $G$, the same analysis can also provide the short-wave cutoff $k_{max}$, while for $G=0$ no cutoffs exist, as seen earlier in figure \ref{Burns integral cr vs a}c.

\begin{figure}
    \hspace{-1cm}
    \begin{minipage}[t]{0.57\textwidth}
      \centering
      \includegraphics[width=\textwidth]{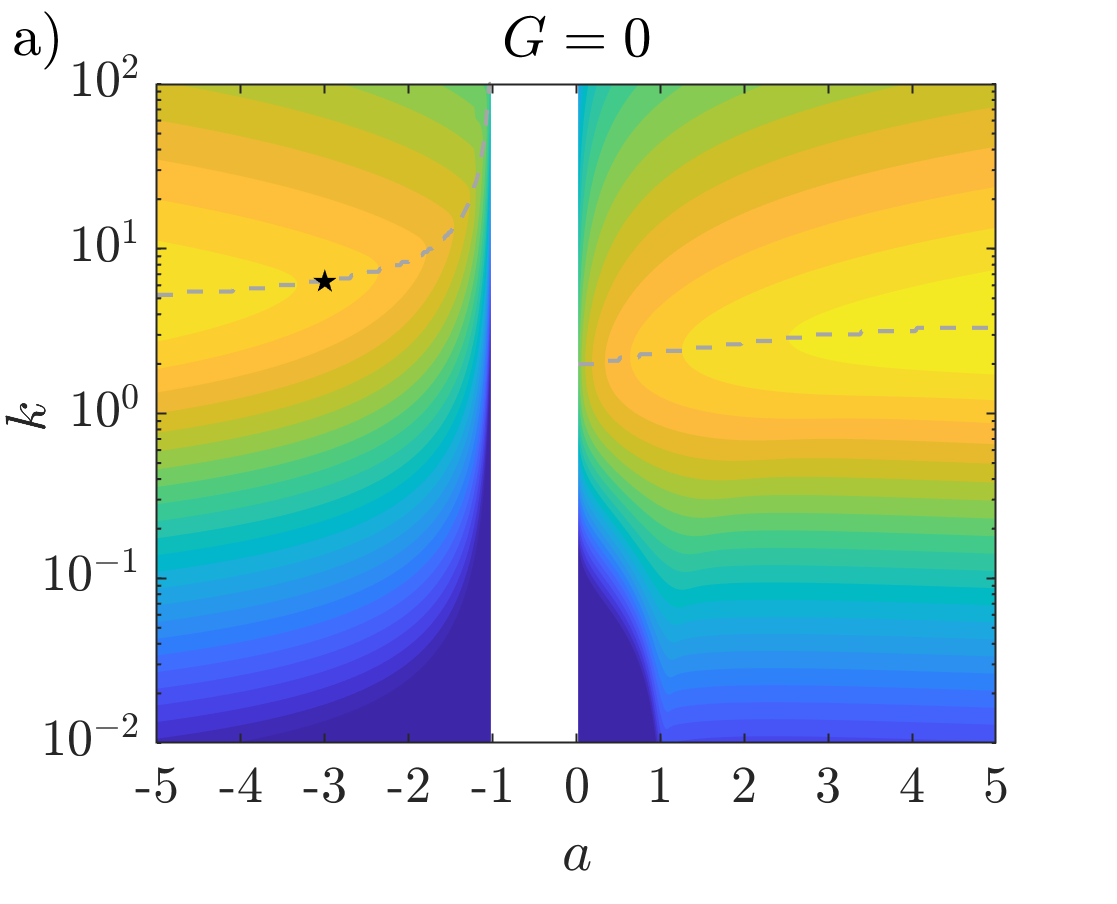}
    \end{minipage}
     \hspace{-0.4cm}
    \begin{minipage}[t]{0.57\textwidth}
      \centering
      \includegraphics[width=\textwidth]{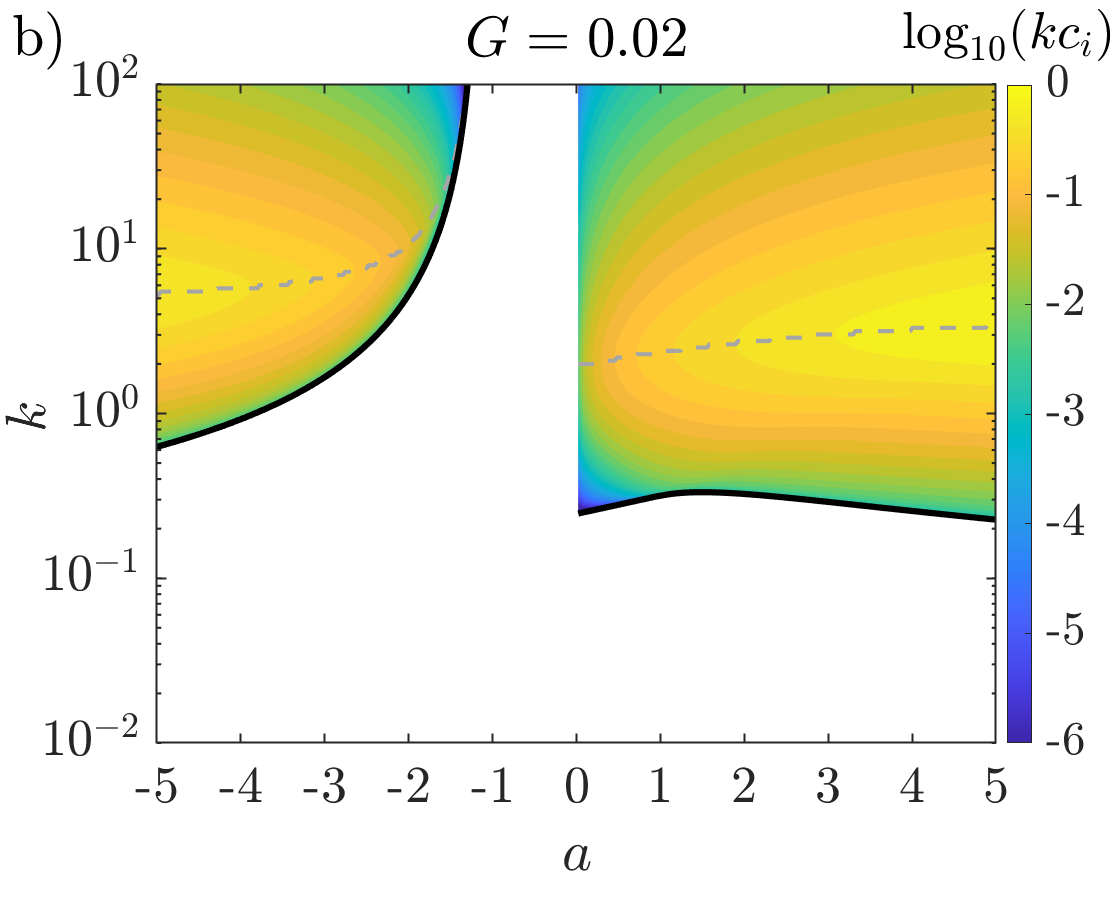}
    \end{minipage}

    \hspace{-1cm}
    \begin{minipage}[t]{0.57\textwidth}
      \centering
      \includegraphics[width=\textwidth]{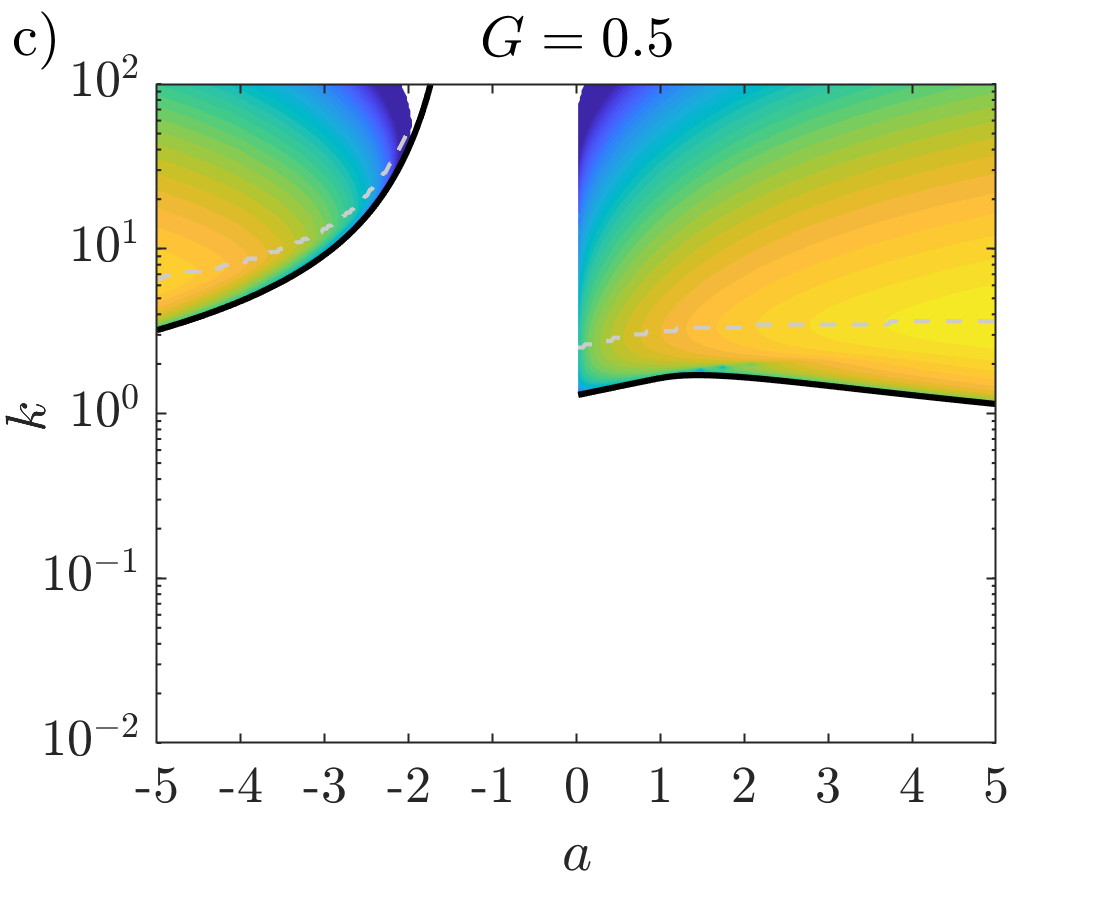}
    \end{minipage}
     \hspace{-0.4cm}
    \begin{minipage}[t]{0.57\textwidth}
      \centering
      \includegraphics[width=\textwidth]{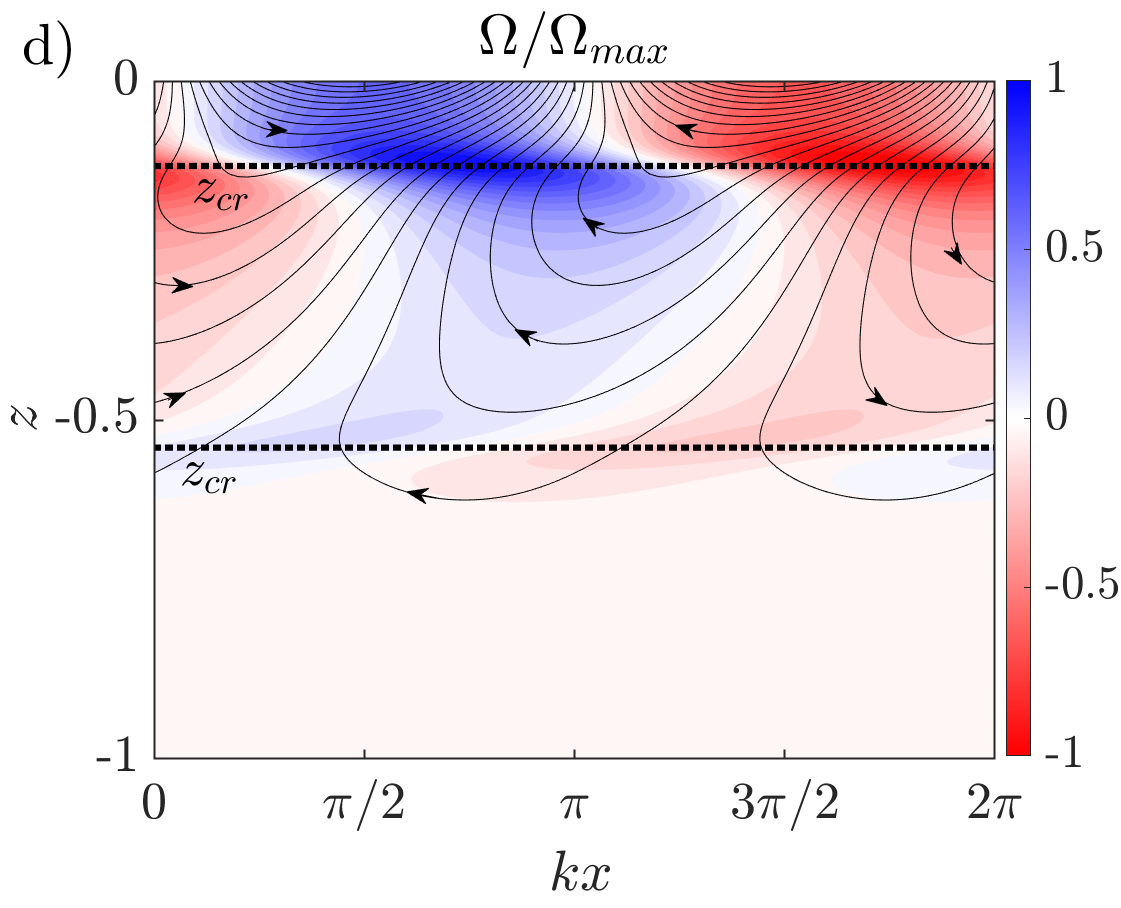}
    \end{minipage}
\caption{Growth-rate contours ($kc_i$) in the $(a,k)$ plane for $G=$ (a) 0, (b) 0.02, and (c) 0.5. Gray dashed lines in (a–c) trace the locus of maximum growth rates. Black curves in (b,c) mark the stability boundaries from \S~\ref{stabilityboundary}. Growth rates below $10^{-10}$ are omitted. Panel (d) shows streamlines overlaid on normalized vorticity contours ($\Omega/\Omega_{max}$) for $G=0$, $a=-3$, and $k=6.31$ (black star in panel a). Black dashed lines indicate critical-layer locations.}
\label{wi vs k}
\end{figure}

\subsection{Complete solution of the dispersion relation} \label{complete solution}

As outlined in \S~\ref{sec:analytical solution} and Appendix \ref{appA}, the dispersion relation \eqref{dispersionrelation} can be solved numerically across a wide parameter space using a root-finding routine in {\it Mathematica} \citep{Mathematica}, yielding the complex phase speed. Figures \ref{Burns integral cr vs a}b-c show $c_r$ and $c_i$ versus $k$ for $G=0$ and $a=-3$ (blue), $3$ (red), and $0.3$ (green). Corresponding results for $G=0.02$ and $a=\pm 3$ (blue), $\pm 5$ (red) are shown in figures \ref{Stability boundary contour}b-c. As discussed earlier, neutral modes meet the unstable branch at the long-wave cutoff $k_{min}$. At $k=k_{min}$, $c_r$ crosses the velocity extrema $U_{min/max}$ (gray dashed lines in figure \ref{Stability boundary contour}b). For $G=0$, $k_{min}=0$, so $c_r$ coincides with $U_{min/max}$ at onset. For $k>k_{min}$, modes are unstable, and both figure sets (\ref{Burns integral cr vs a}, \ref{Stability boundary contour}) show the development of a critical layer. Unlike the $G=0$ case (fig. \ref{Burns integral cr vs a}b), the critical layer does not approach the surface asymptotically when $G>0$. At large $k$, $c_r$ departs from the surface velocity, recrossing $U_{min/max}$ and recovering neutral stability. This short-wave cutoff is due to surface tension. It is not displayed in figure \ref{Stability boundary contour} since the cutoff occurs at very large $k$, obscuring the long-wave features of interest. In figure \ref{Stability boundary contour}b, the $U_{max}$ line (gray) moves downward with $a$, eventually coinciding with $c_r=1$ at $a=-1$. The result is a larger long-wave cutoff, smaller short-wave cutoff, and a shift in the most unstable mode.

Figures \ref{wi vs k}a-c show growth-rate contours ($kc_i$) in the $(a,k)$ plane for $G=0$, $0.02$, and $0.5$. For $a<0$, only the prograde branch is unstable, whereas for $a>0$ the retrograde branch is unstable. As anticipated from \eqref{stability boundary dispersion 3}, profiles with $-1<a<0$ are always stable. For $G=0$, instability exists for all $k$ whenever $a<-1$ or $a>0$. Any nonzero $G$ produces a finite region of stability, which grows with increasing $G$. Thus, quadratic velocity profiles with $a<-1$ or $a>0$ are always unstable in the inviscid limit, and gravity and surface tension act together to stabilize disturbances when $G\neq 0$. The black curves in figures \ref{wi vs k}a-c show the stability boundaries derived in \S~\ref{stabilityboundary}, in good agreement with the growth-rate contours. The gray dashed lines mark the locus of maximum growth rate in the $(a,k)$ plane. For retrograde modes ($a>0$), the maximum growth rate occurs at wavenumbers $O(1)$. For prograde modes ($a<-1$), the most unstable wavenumber increases as $a$ increases toward $-1$. This reflects the diminishing forward bulge of the velocity profile, which vanishes at $a=-1$. Away from the stability boundary in figures \ref{wi vs k}a-c, growth rates increase with $a$ (see also fig. \ref{stabilityboundary}c), consistent with the dependence of instability strength on the velocity curvature at the critical layer \citep{shrira1993surface, bonfils2023flow}.

Figure \ref{wi vs k}d illustrates the streamline pattern overlaid on normalized vorticity contours ($\Omega/\Omega_{max}$) for $G=0$, $a=-3$, and $k=6.31$ (black star in fig. \ref{wi vs k}a). Two critical layers are visible (black dashed lines), since $c_r=U(z)$ is satisfied at two distinct depths for forward-bulging profiles ($a<-1$). Around these layers, the streamfunction tilts and vorticity contours distort strongly, consistent with the negative Reynolds stress correlation $\overline{uw}<0$ that drives the instability \citep{young2014generation}.

To summarize, the inviscid stability analysis demonstrates that for $G=0$, $c_i\sim k^{4/3}$ when $|a|>1$ and $c_i\sim k^4$ when $|a|<1$, consistent with asymptotic results. These scalings were confirmed using solutions expressed in terms of confluent Heun functions. All quadratic profiles are unstable except those with $-1<a<0$. With increasing $G$, long-wave cutoffs emerge, stabilizing low-wavenumber disturbances. Instabilities propagate faster than the surface velocity when $a<-1$ and slower when $a>0$. In both cases, the peak growth occurs at wavenumbers $O(1)$.

The inviscid framework adopted here is the same as in earlier studies of wind-generated surface waves \citep{miles1957generation, stern1973capillary, young2014generation}, justified by the typically large Reynolds numbers of geophysical flows. In contrast, laboratory experiments on thin layers \citep{paquier2015surface, paquier2016viscosity} involve Reynolds numbers of $O(10^3)$, making viscous effects non-negligible. While viscous stability analyses are common in engineering, they have rarely addressed the rippling instability directly \citep[see Table 1 of ][for a summary]{young2014generation}. In the following section, we extend the analysis to finite Reynolds numbers, to ask: How does viscosity influence the instability? Does it merely damp the inviscid mode, or does it qualitatively alter the dynamics? And how does the inviscid rippling instability compare with other viscous modes?

\section{Viscous linear stability analysis}\label{vislsa}

The influence of viscosity on linear stability is well established in single-phase parallel shear flows, where it generally acts as a dissipative agent \citep{drazin2004hydrodynamic}. In air–water two-phase systems, however, its role is more subtle. As discussed in the introduction, \citet{zeisel2008viscous} and \citet{kadam2023wind} showed that viscosity can amplify Miles-type instability growth at short wavelengths, while \citet{kadam2023wind} also reported viscous damping of rippling instability under conditions corresponding to the laboratory experiments of \citet{paquier2015surface, paquier2016viscosity}. Thus, it is not obvious in which regions of parameter space viscosity suppresses, and in which it enhances, the rippling instability. The picture is further complicated by the fact that viscosity stratification alone, as in air over water, can destabilize the interface even at low Reynolds numbers \citep[see][]{yih1967instability}. Moreover, \citet{miles1960hydrodynamic} showed that a linearly sheared liquid film with a free surface is unstable to viscous perturbations, with growth rates that decay as $Re^{-1/2}$ and vanish in the inviscid limit. This is consistent with the stability of the linear profile in \S\ref{invlsa}, but emphasizes that viscosity is not always simply stabilizing: depending on the configuration, it can alter the character of existing instabilities or give rise to new ones.

In this section, we present the viscous linear stability results in the $(k,Re)$ parameter plane for different values of $a$ and $G$. The Orr–Sommerfeld equation \eqref{OSE} with the base velocity profiles \eqref{base-state profiles} does not admit closed-form solutions. Accordingly, asymptotic results for the long- and short-wave limits are first derived in \S\ref{asymptotics}. Numerical solutions of the eigenvalue problem \eqref{OSE}–\eqref{nor} are then presented in \S\ref{Numerical OSE}, followed by a synthesis of the results into a complete picture of growth-rate behavior across the parameter space in \S\ref{complete picture}.

\subsection{Asymptotic evaluation of growth rates}\label{asymptotics}
\subsubsection{Long wave asymptotics ($k \ll 1$)}\label{vis long wave}
In the long-wave limit, following \citet{yih1967instability}, the eigenfunction $\phi$ and eigenvalue $c$ may be expanded in a regular perturbation series in $k$,  
\begin{equation}\label{vis longwave series}
    \phi(z) = \sum_{n = 0}^{\infty} k^n \phi^{(n)}_{k \ll 1}(z),  
    \qquad 
    c = \sum_{n = 0}^{\infty} k^n c^{(n)}_{k \ll 1}.
\end{equation}
Substituting these expansions into \eqref{OSE}–\eqref{nor} and solving the resulting equations order by order in $k$, as outlined in Appendix \ref{AppC} \citep[see also][]{yih1967instability}, yields explicit expressions for the first three terms in the eigenvalue,  
\begin{equation}\label{vis longwave eig}
    c^{(0)}_{k \ll 1} = 1-a, 
    \qquad 
    c^{(1)}_{k \ll 1} = \frac{iRe}{15}\left(4a(a-1) - 5G\right), 
\end{equation}
\begin{equation}
    \nonumber
    c^{(2)}_{k \ll 1} = \frac{1}{1008} \left(179 a^3 Re^2 - 256 a^2 Re^2 - 222 a G Re^2 + 77 a Re^2 + 2016 a + 98 G Re^2\right).
\end{equation}

The leading-order and second-order contributions, $c^{(0)}_{k \ll 1}$ and $c^{(2)}_{k \ll 1}$, are purely real and thus do not affect the growth rate. The imaginary part arises solely from the $O(k)$ term, which controls the instability. In the limit of a strongly forced interface or in the absence of gravity ($G=0$), the imaginary part of the eigenvalue reduces to  
\begin{equation}\label{G=0 long wave asymptotic}
    c_i = \frac{4}{15}\,kRe\,a(a-1) + O(k^3). 
\end{equation}
Equation \eqref{G=0 long wave asymptotic} shows that long-wave (Yih-type) instability occurs when $a<0$ or $a>1$, whereas velocity profiles between the linear case ($a=0$) and the half-parabolic case ($a=1$) remain stable, in agreement with figure 15 of \citet{smith1982instability}. Moreover, $c_i$ scales linearly with $kRe$, the natural small parameter in this expansion \citep{yih1967instability}. Finally, from $c^{(0)}_{k \ll 1}$ in \eqref{vis longwave eig}, the leading order phase speed exceeds the maximum velocity in the domain for $a<0$ and falls below the minimum velocity for $a>1$, confirming the asymmetry of propagation depending on profile curvature.

\subsubsection{Short wave asymptotics ($k \gg 1$)}\label{vis short wave}

A parallel analysis to the long-wave limit can be carried out for short waves. This approach was first developed by \citet{hooper1983shear} for two-layer linear–linear profiles and later generalized by \citet{yiantsios1988linear} to channel flow. In this regime the small parameter is taken as $\epsilon = k^{-1}$, and the vertical coordinate is rescaled as $\tilde{z} = kz$. The eigenfunction and eigenvalue are then expanded in regular series of $\epsilon$,  
\begin{equation}\label{short wave expansion}
    \phi(\tilde{z}) = \sum_{n=0}^{\infty} \epsilon^n \phi_{k \gg 1}^{(n)}(\tilde{z}), 
    \qquad 
    c = \sum_{n=0}^{\infty} \epsilon^n c_{k \gg 1}^{(n)}.
\end{equation}
Substitution of these expansions into the system \eqref{OSE}–\eqref{nor} and collection of terms at successive orders in $\epsilon$ gives an iterative hierarchy, the details of which are outlined in Appendix \ref{AppD}. The first four contributions to the eigenvalue are,  
\begin{equation}\label{vis shortwave eig}
    \nonumber
    c_{k \gg 1}^{(0)} = 1 - \frac{i}{2Ca}, 
    \qquad 
    c_{k \gg 1}^{(1)} = - \frac{3 i Re}{16 Ca^2}, 
    \qquad  
    c_{k \gg 1}^{(2)} = \frac{1}{32 Ca}\left[16(a+1)Re - 16 i Bo - 5 i \frac{Re^2}{Ca^2}\right],
\end{equation}
\begin{align}
    \nonumber
    c_{k \gg 1}^{(3)} &= \frac{Re}{512 Ca}\left[-416 a + 348 (a+1)\frac{Re}{Ca} - 192 i G Re - 85 i \frac{Re^2}{Ca^3}\right], \\ \nonumber
    c_{k \gg 1}^{(4)} &= \frac{(a+1)Re^3}{Ca^3} + \frac{23 i (a+1)^2 Re^2}{32 Ca} + \frac{(a+1)}{2}GRe^2 + \frac{i Re a (a+1)}{2} - \frac{79 a Re^2}{64 Ca^2} 
    \\  &- \frac{407 i Re^4}{2048 Ca^5} - \frac{15 i G Re^3}{32 Ca^2}.
\end{align}
Here $Ca = Bo/(GRe)$ is the Capillary number.  

From \eqref{vis shortwave eig} it is evident that the imaginary parts up to $O(\epsilon^3)$ are negative, and therefore act to stabilize the interface. The first destabilizing contribution appears at $O(\epsilon^4)$. In the absence of gravity ($G=0$), the imaginary part of $c$ simplifies to  
\begin{equation}\label{G=0 short wave asymptotic}
    c_i = \tfrac{1}{2}\,a(a+1)Re\,\epsilon^4 + O(\epsilon^5).
\end{equation}
Thus all lower-order imaginary terms vanish at $G=0$, leaving only the destabilizing $O(\epsilon^4)$ contribution. This implies that short-wave instability is more widespread in parameter space when gravity is absent, and that nonzero $G$ suppresses it through the stabilizing terms in \eqref{vis shortwave eig}. According to \eqref{G=0 short wave asymptotic}, instability arises for $a>0$ and $a<-1$, while velocity profiles between the linear case ($a=0$) and the Nusselt profile ($a=-1$) remain stable to short-wave disturbances in the $G=0$ limit.

\begin{figure}
\centerline{\includegraphics[scale=0.4]{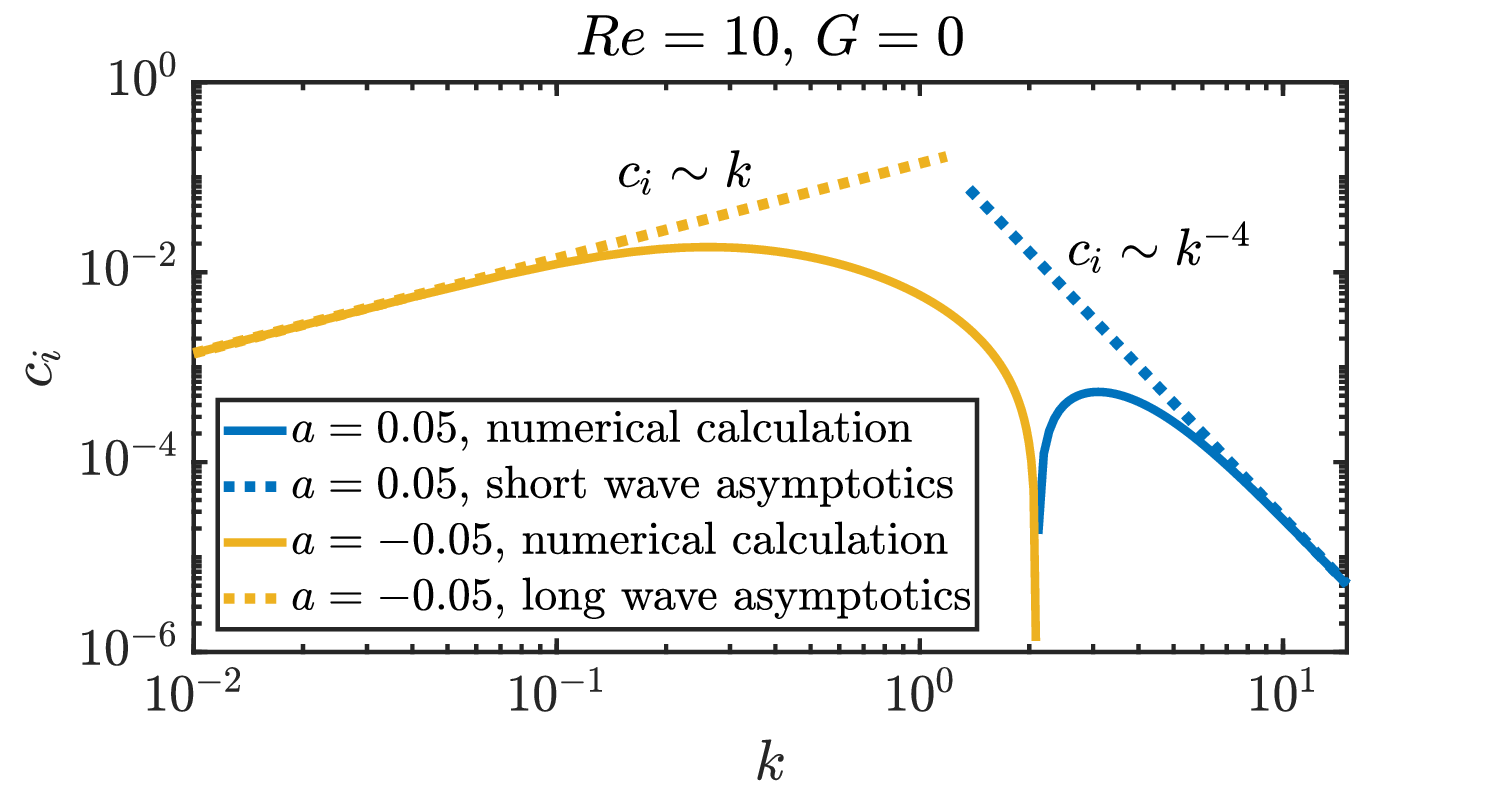}}% Images in 100% size
  \caption{The variation of $c_i$ with wavenumber $k$ shown for velocity profiles slightly perturbed from a linear profile: $a = 0.05$ corresponds to a mildly convex profile, and $a = -0.05$ corresponds to a mildly concave profile. Solid lines represent numerical results, and dashed lines indicate the asymptotic predictions from \S~\ref{vis long wave} and \S~\ref{vis short wave}.}
\label{Fig 5}
\end{figure}

\subsection{Numerical solution}\label{Numerical OSE}

Although the asymptotic solutions above reveal the existence of instabilities, their validity is confined to the limiting cases of small and large wavenumbers. To analyze the behavior at intermediate parameters, a numerical approach is required, since \eqref{OSE} admits no closed-form solution. We therefore employ a Chebyshev spectral collocation method following \citet{boomkamp1997chebyshev} and \citet{trefethen2000spectral} to solve the eigenvalue problem \eqref{OSE}--\eqref{nor}. 
%Spurious modes are eliminated by verifying that computed eigenfunctions satisfy the boundary conditions. Convergence tests show that 45 collocation points in $z$ are sufficient to obtain accurate and stable results for all cases reported below.  

\subsubsection{For small curvature parameters ($|a| \ll 1$)}\label{growth |a| < 1}

As established in \S\ref{vis long wave} and \S\ref{vis short wave}, long-wave instabilities are absent for $-1<a<0$ and short-wave instabilities are absent for $0<a<1$ in the $G=0$ limit. A direct implication is that a small perturbation of the curvature parameter $a$ away from zero (corresponding to the linear profile) results in an abrupt change in both the type of instability and the range of unstable parameters. Slightly convex profiles ($a>0$) develop short-wave instabilities, while slightly concave profiles ($a<0$) develop long-wave instabilities.  

This behavior is illustrated in figure \ref{Fig 5}, which shows $c_i$ as a function of $k$ for $a=-0.05$ and $a=0.05$ at $Re=10$. The solid lines denote the numerical solutions, while the dashed lines correspond to the asymptotic predictions \eqref{G=0 long wave asymptotic} and \eqref{G=0 short wave asymptotic}. The asymptotic scalings $c_i \sim k$ for the long-wave instability and $c_i \sim k^{-4}$ for the short-wave instability agree well with the numerical results. Notably, for small $|a|$, a sharp transition is observed from long-wave (small negative $a$) to short-wave (small positive $a$) instability at a critical wavenumber $k_{\text{cr}}$, which remains nearly constant across a range of Reynolds numbers. As $|a|$ increases, however, the transition becomes more gradual and a gap opens between the long- and short-wave cutoffs, which broadens progressively with curvature.  

 \begin{figure}
\centerline{\includegraphics[scale=0.4]{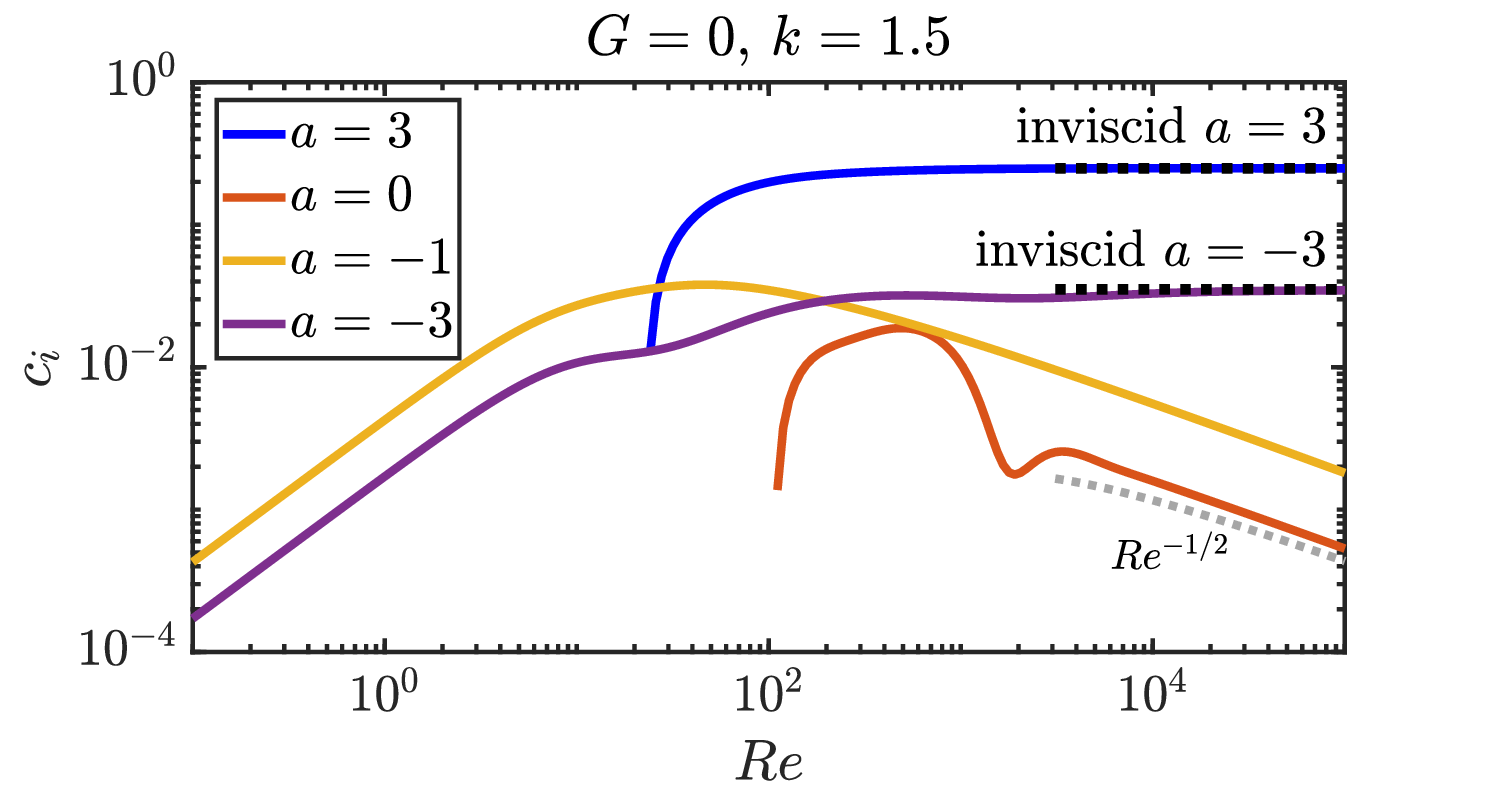}}
  \caption{The variation of $c_i$ as a function of $Re$ for four curvature parameters: $a = 3$ (blue curve), $a = 0$ (red curve), $a=-1$ (yellow curve) and $a = -3$ (purple curve). The black dashed lines overlapping the blue and purple solid lines represent the inviscid limits for $a = 3$ and $a = -3$, respectively. The gray dashed curve that scales as $Re^{-1/2}$ corresponds to the asymptotic solution for $a = 0$, as derived by \citet{miles1960hydrodynamic}.}
\label{Fig 6}
\end{figure}

\subsubsection{Transition from viscous to inviscid regimes}\label{Transition vis to inviscid}
 \begin{figure}
\centering
 \hspace{-1cm}
    \begin{minipage}[t]{0.52\textwidth}
      \centering
      \includegraphics[width=\textwidth]{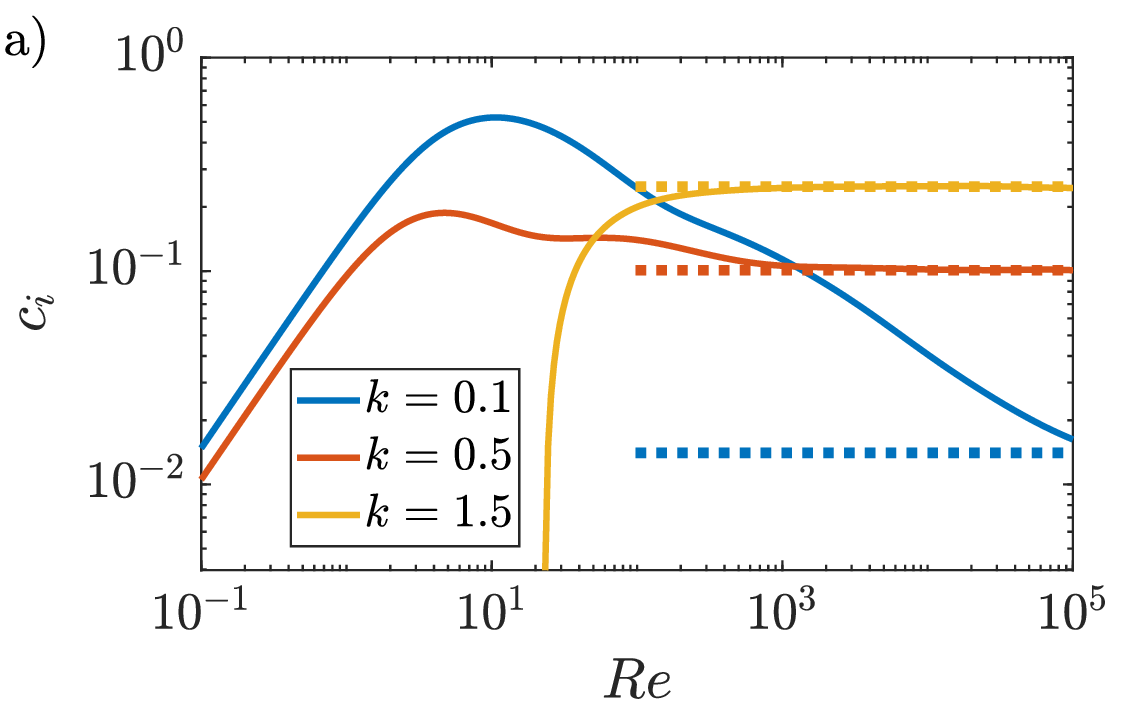}
    \end{minipage}
     \hspace{-0cm}
    \begin{minipage}[t]{0.52\textwidth}
      \centering
      \includegraphics[width=\textwidth]{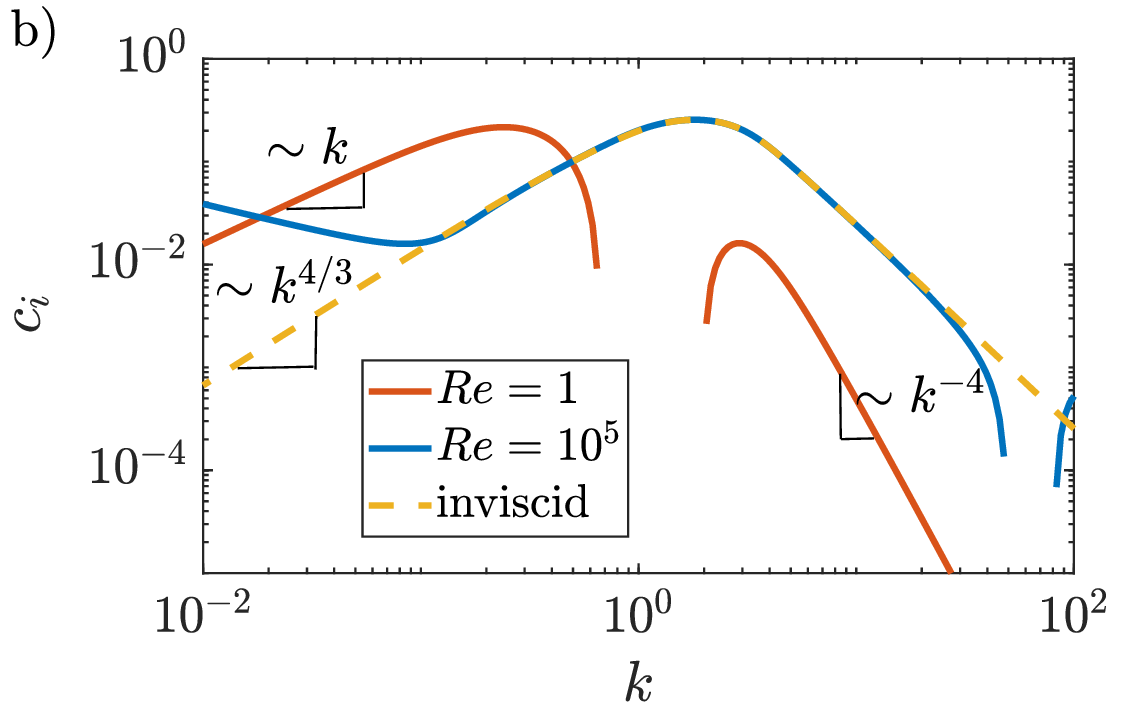}
    \end{minipage}

    \hspace{-0.5cm}
    \begin{minipage}{0.55\textwidth}
      \centering
      \includegraphics[width=\textwidth]{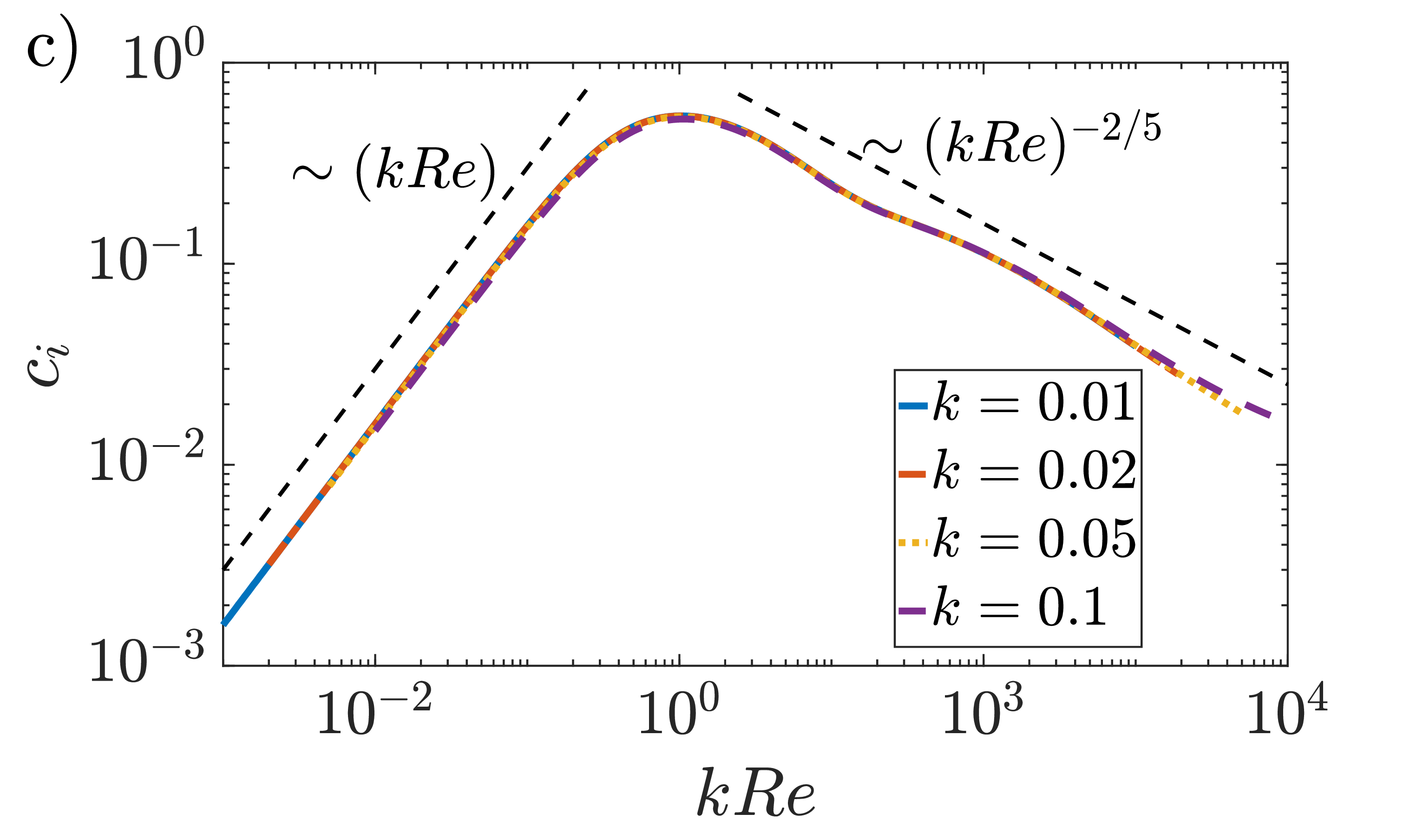}
    \end{minipage}

    \caption{For $G = 0$ and $a = 3$, the variation of $c_i$ as function of (a) Reynolds number $Re$, (b) wavenumber $k$, and (c) the scaled variable $kRe$. In (a), three cases are considered corresponding to wavenumbers $0.1$ (blue), $0.5$ (red) and $1.5$ (yellow), respectively. The continuous curves are from numerical calculation and dotted lines indicate the corresponding inviscid $c_i$ values. In (b), three Reynolds numbers are considered: $Re = 1$ (red continuous curve), $Re = 10^5$ (blue continuous curve), and  inviscid (yellow dashed curve). In (c), four wavenumbers are considered: $k = 0.01$ (blue continuous curve), $k = 0.02$ (red dashed curve), $k = 0.05$ (yellow dotted curve), and $k = 0.1$ (purple dot dashed curve). The thin black lines depict the asymptotes at small and large $kRe$.}
\label{Viscous a=3}
\end{figure}
To illustrate how viscous instabilities (\S\ref{asymptotics}) connect to the inviscid limit, figure \ref{Fig 6} shows the variation of $c_i$ with $Re$ for four representative velocity profiles: $a=3$, $0$, $-1$, and $-3$, at fixed wavenumber $k=1.5$ in the $G=0$ limit. Two distinct behaviors emerge. For $a=3$ (blue) and $a=-3$ (purple), $c_i$ increases with $Re$ and asymptotically approaches the value of $c_i$ in the inviscid limit (black dashed lines). By contrast, for $a=0$ (red) and $a=-1$ (yellow), $c_i$ decays with increasing $Re$, indicating viscous instabilities that vanish in the inviscid limit. As noted in \S\ref{invlsa} and \S\ref{asymptotics}, the linear profile ($a=0$) is stable to long-wave, short-wave, and rippling instabilities; the observed mode here corresponds to the viscous instability identified by \citet{miles1960hydrodynamic}, which decays as $Re^{-1/2}$, consistent with the numerical results. Similarly, the Nusselt profile ($a=-1$) is inviscidly stable, and the observed instability corresponds to the viscous continuation of the long-wave mode. In contrast, the profiles $a=\pm 3$ are unstable to all viscous mechanisms, though their small-$Re$ behavior differs. The matching slopes of the $a=-1$ and $a=-3$ curves at low $Re$ suggest a long-wave character for $a=-3$, followed by a transition to rippling instability as $Re$ increases.  

The case $a=3$ is particularly noteworthy, as it corresponds to velocity profiles observed in the experiments of \citet{paquier2015surface, paquier2016viscosity}. Figures \ref{Viscous a=3}a–c present its stability properties at $G=0$. Figure \ref{Viscous a=3}a shows $c_i$ versus $Re$ for $k=0.1,\,0.5,$ and $1.5$. While the $k=1.5$ case (yellow) is stable at low $Re$, both $k=0.1$ (blue) and $k=0.5$ (red) exhibit finite-$Re$ growth followed by a transition to rippling instability (dotted lines) at large $Re$. The non-monotonic variation of $c_i$ with $Re$ becomes more pronounced as the wavelength increases, demonstrating that viscosity can amplify rather than suppress disturbances—by more than an order of magnitude in some cases. Figure \ref{Viscous a=3}b shows $c_i$ as a function of $k$ for $Re=1$, $10^5$, and the inviscid limit. The $Re=1$ curve (red) displays asymptotic behavior at small and large $k$, consistent with the expressions of \S\ref{asymptotics}, and also reveals an interval at $O(1)$ wavenumbers where the instability is absent, in line with \S\ref{growth |a| < 1}. At $Re=10^5$ (blue), the spectrum overlaps with the inviscid prediction (yellow dashed) for $10^{-1}\lesssim k \lesssim 10$, while deviations at smaller $k$ reflect the delayed approach to the inviscid asymptote seen in figure \ref{Viscous a=3}a. This behavior is clarified in figure \ref{Viscous a=3}c, where $c_i$ is plotted against $kRe$ for several long-wave cases. The collapse of the curves onto a single master trend reveals two regimes: at small $kRe$, $c_i$ grows linearly with $kRe$, as predicted by the long-wave asymptotics in \S\ref{vis long wave}; at large $kRe$, $c_i$ decreases as $(kRe)^{-2/5}$ and approaches the inviscid value. The asymptote $-2/5$ is obtained through a numerical fit.

In summary, viscosity can either suppress instabilities (as for $a=0$ and $a=-1$) or enhance them (as for $a=\pm 3$), before smoothly connecting to the inviscid rippling mode. This dual role underlines that viscosity does not merely act as a dissipative effect but can also generate significant growth depending on the flow configuration and parameter regime.  

\subsection{Growth rate behavior: a complete picture}\label{complete picture}
\begin{figure}
    \hspace{-1cm}
    \begin{minipage}[t]{0.55\textwidth}
      \centering
      \includegraphics[width=\textwidth]{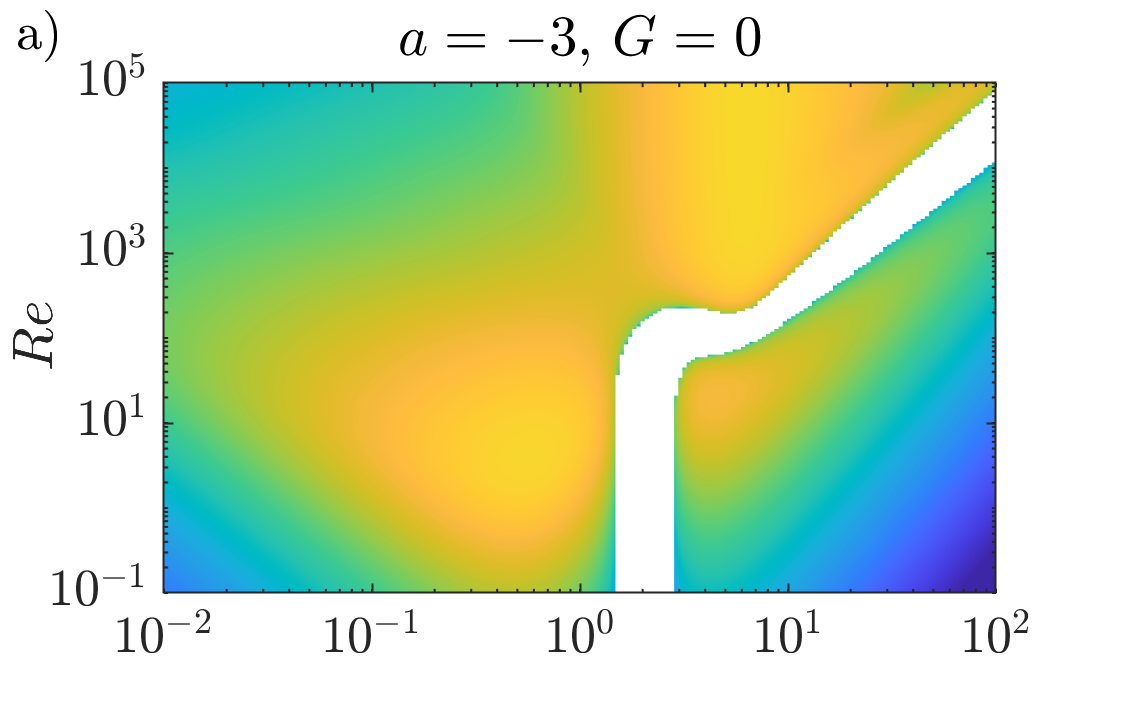}
    \end{minipage}
     \hspace{-0.4cm}
    \begin{minipage}[t]{0.55\textwidth}
      \centering
      \includegraphics[width=\textwidth]{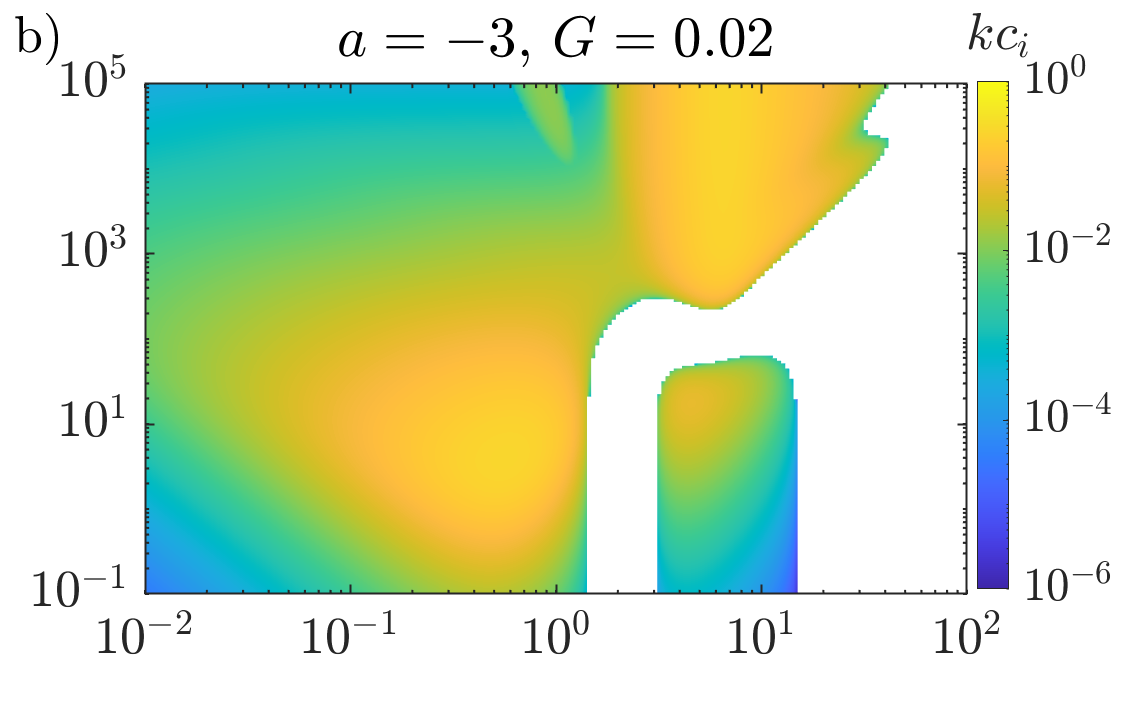}
    \end{minipage}

    \vspace{-0.1cm}
        \hspace{-1cm}
    \begin{minipage}[t]{0.55\textwidth}
      \centering
      \includegraphics[width=\textwidth]{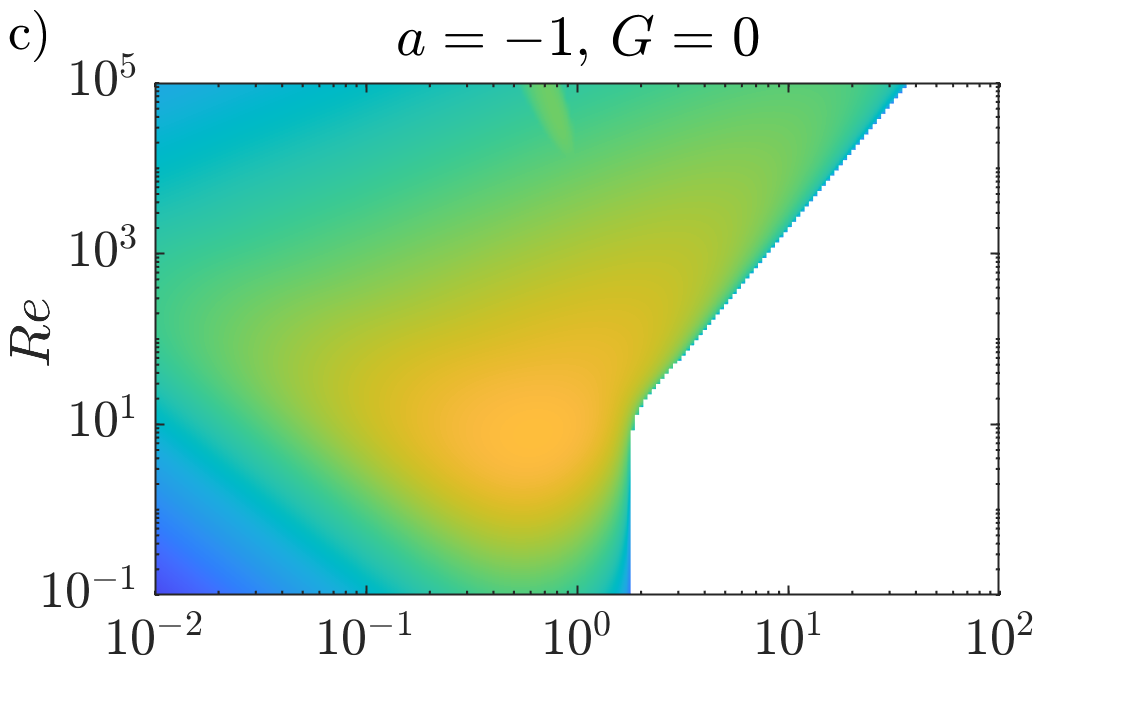}
    \end{minipage}
     \hspace{-0.4cm}
    \begin{minipage}[t]{0.55\textwidth}
      \centering
      \includegraphics[width=\textwidth]{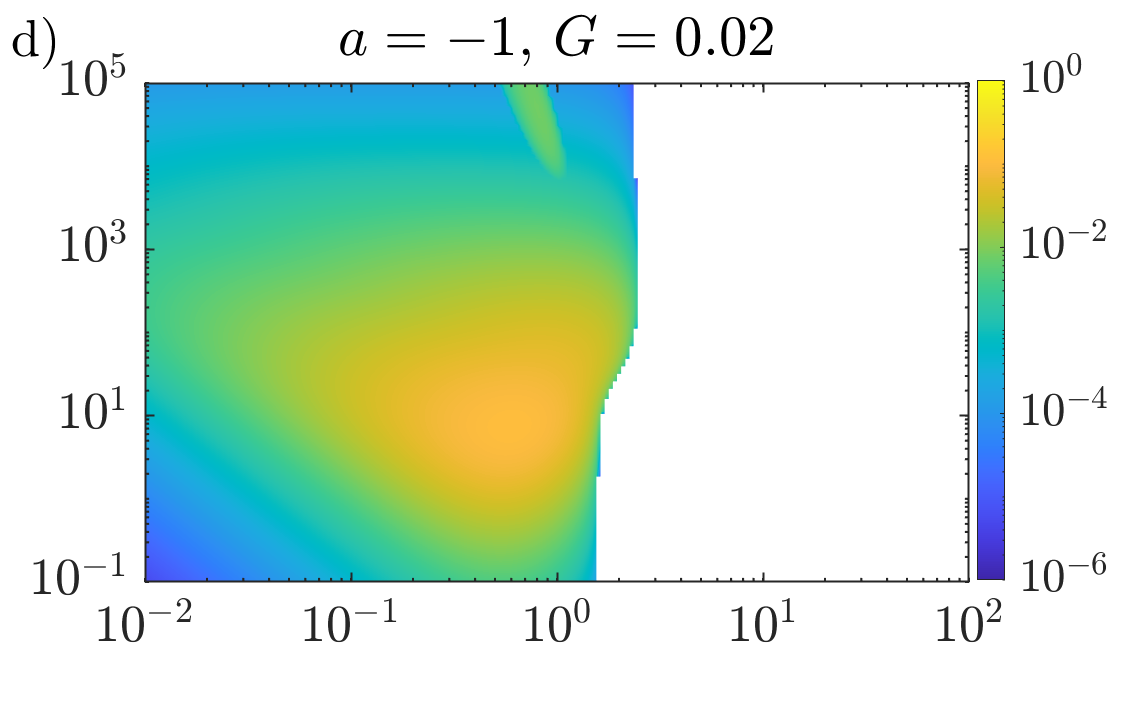}
    \end{minipage}

    \vspace{-0.1cm}
        \hspace{-1cm}
    \begin{minipage}[t]{0.55\textwidth}
      \centering
      \includegraphics[width=\textwidth]{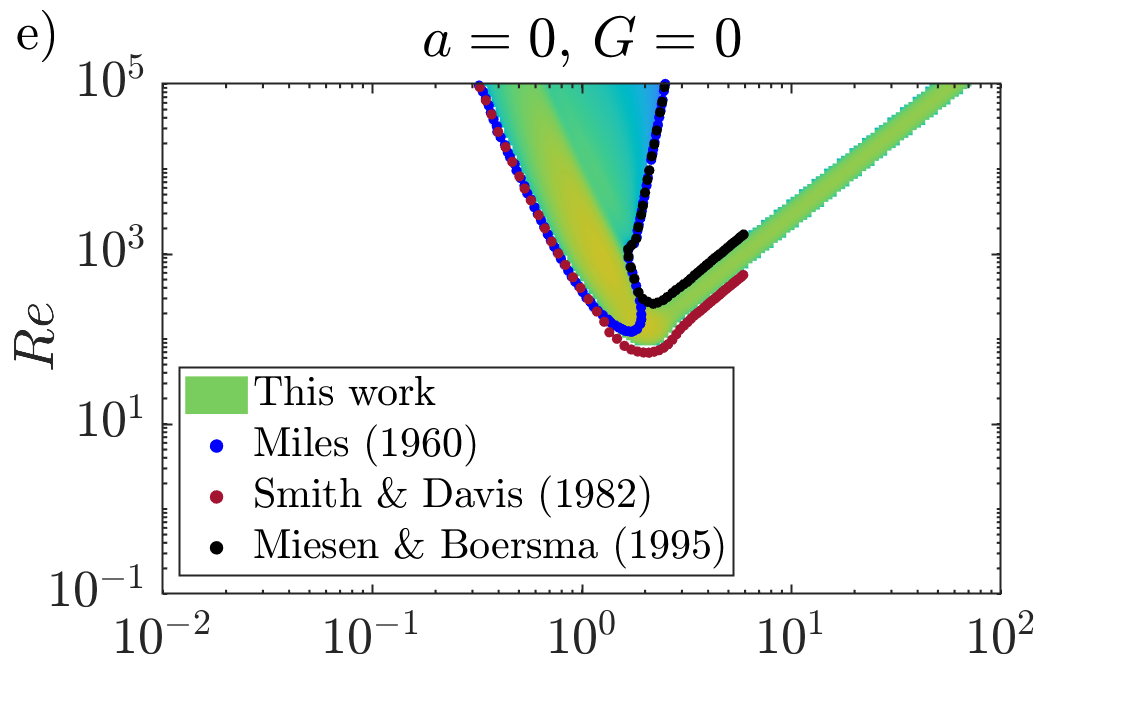}
    \end{minipage}
     \hspace{-0.4cm}
    \begin{minipage}[t]{0.55\textwidth}
      \centering
      \includegraphics[width=\textwidth]{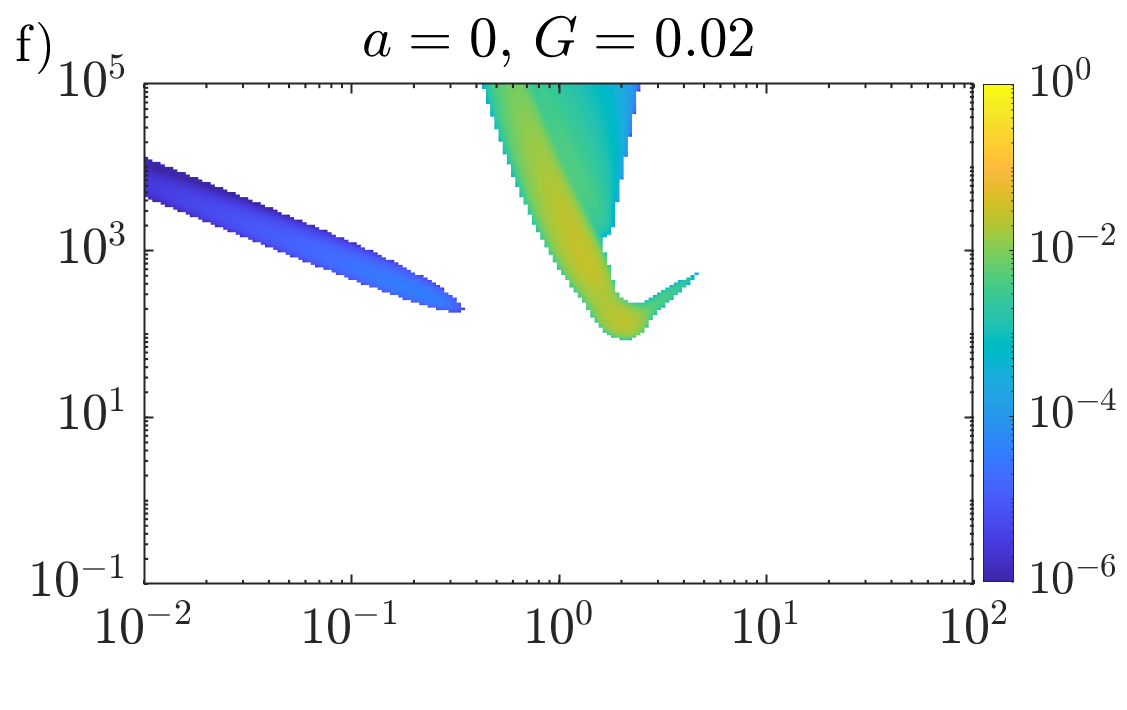}
    \end{minipage}
    
    \vspace{-0.1cm}
          \hspace{-1cm}
    \begin{minipage}[t]{0.55\textwidth}
      \centering
      \includegraphics[width=\textwidth]{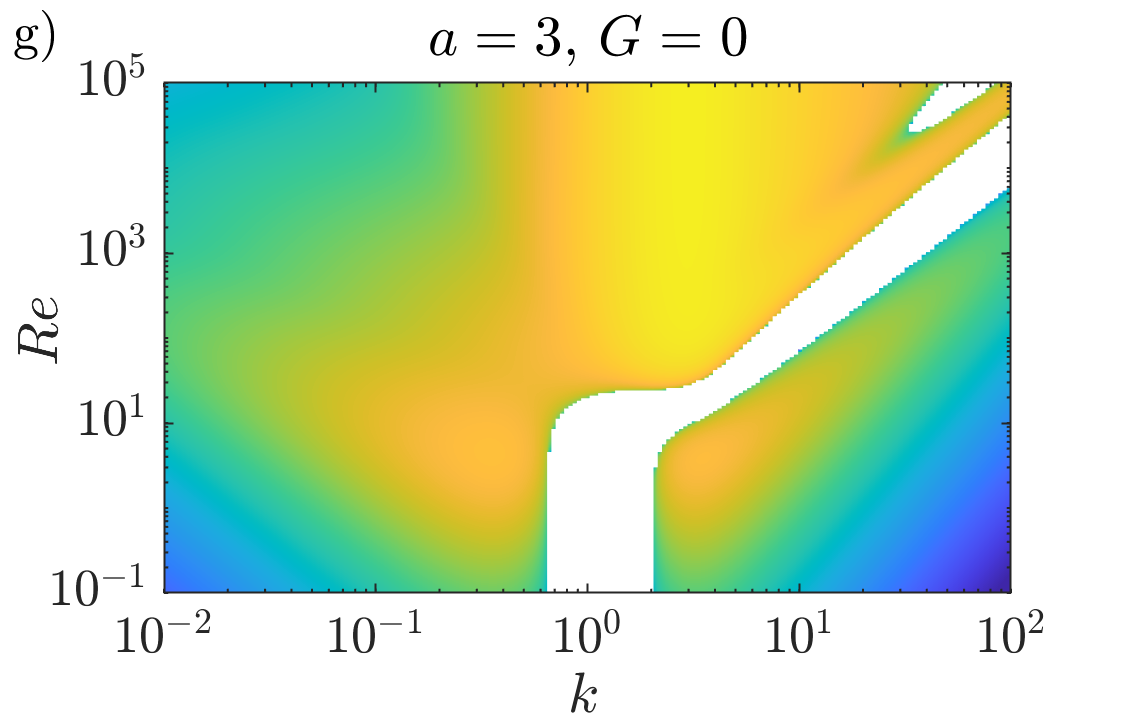}
    \end{minipage}
     \hspace{-0.4cm}
    \begin{minipage}[t]{0.55\textwidth}
      \centering
      \includegraphics[width=\textwidth]{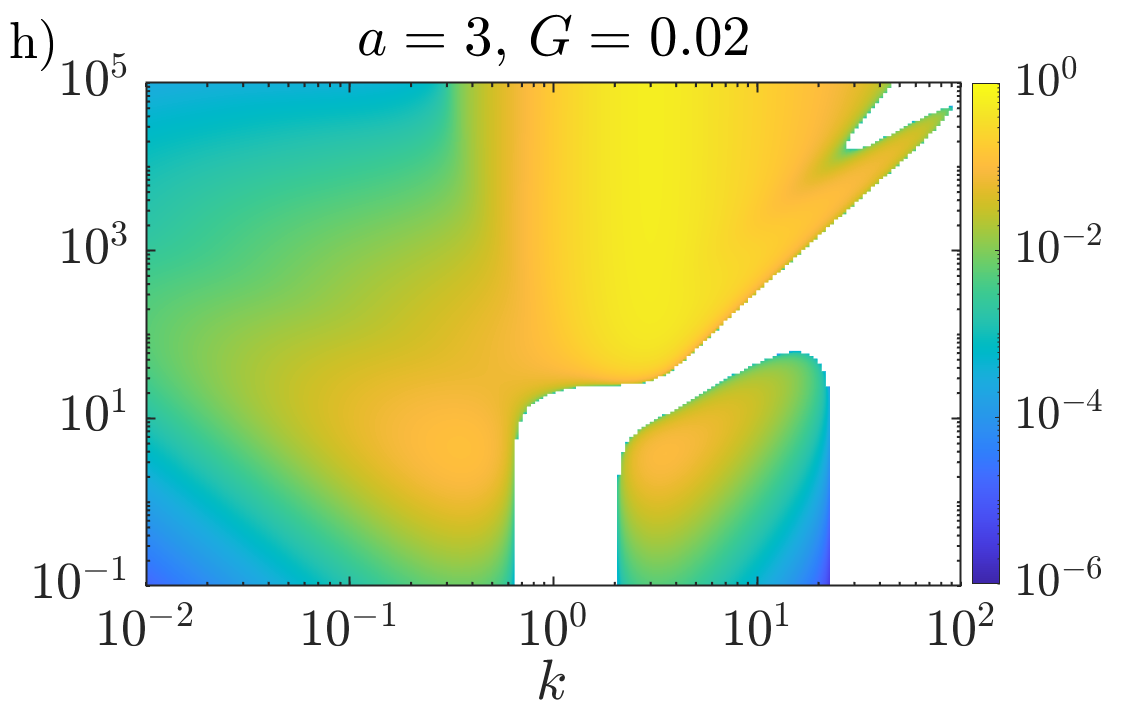}
    \end{minipage}  
\caption{Contours of the growth rate ($k c_i$) in the $k$–$Re$ plane for four representative velocity profile curvatures: (a,b) $a = -3$, (c,d) $a = -1$, (e,f) $a = 0$, and (g,h) $a = 3$. The left column (a, c, e, g) shows the results for $G = 0$ limit (i.e. no gravity or surface tension), while the right column (b, d, f, h) corresponds to $G = 0.02$. Panel (e) includes the neutral stability curves from \citet{miles1960hydrodynamic}, \citet{smith1982instability}, and \citet{miesen1995hydrodynamic} for comparison.
}
\label{Fig8}
\end{figure}

Given the variety of instabilities that emerge in different asymptotic regimes, as discussed in the previous subsections, it is useful to examine the contours of growth rate ($kc_i$) in the $(k,Re)$ plane to obtain a broader picture. These contours (together with neutral stability curves) reveal the dominant instability mechanisms, their parameter ranges, and their mutual transitions. Figures~\ref{Fig8}a–h present the growth rate contours of the most unstable mode for four representative velocity profiles—$a=-3,-1,0,$ and $3$—for $G=0$ (left panels) and $G=0.02$ (right panels).

For the forward-bulging profile ($a=-3$), two broad instability regions appear (figures~\ref{Fig8}a,b), separated by a narrow band of stability. Three distinct modes can be identified within them. At small $Re$ and $k$, the long-wave interfacial (Yih) mode dominates, with peak growth rates at $Re\sim O(10)$ and $k\sim O(1)$. At large $Re$ and moderate $k$, a bright region corresponds to the rippling instability, consistent with figure~\ref{wi vs k}a, where the wavenumber of maximum growth at $Re=10^5$ matches that of the inviscid case. A third zone, at higher $k$, corresponds to the short-wave interfacial mode, although its growth rates decay rapidly with increasing wavenumber. Including gravity and surface tension ($G=0.02$) strongly suppresses the short-wave mode and introduces a tongue-like region at very large $Re$, reminiscent of the shear mode described by \citet{floryan1987instabilities}. While present in the $G=0$ case as well, this mode is less visible there since it is not the most unstable. The overall growth rates, however, are diminished by the presence of gravity and surface tension.

The Nusselt profile ($a=-1$) lacks both rippling and short-wave instabilities, leaving only the long-wave interfacial mode and the shear mode at large $Re$ and $k\sim O(1)$ (figures~\ref{Fig8}c,d). Interestingly, at $G=0$, the long-wave mode extends into the short-wave regime at high $Re$, though this extension is curtailed when $G=0.02$, with gravity and surface tension acting to stabilize both long and short waves. The shear mode, however, remains relatively unaffected, reflecting its $O(1)$ wavenumber character.

For the linear profile ($a=0$), the instability landscape is quite different. Figures~\ref{Fig8}e,f show that the unstable region coincides precisely with the neutral curves of earlier studies: the outer boundary from \citet{miles1960hydrodynamic} (blue dots) and \citet{smith1982instability} (magenta dots), and the inner boundary from \citet{miesen1995hydrodynamic} (black dots). This agreement validates both the present formulation and numerical approach. Importantly, the instability observed for $a=0$ is not of the long-wave, short-wave, or rippling types; instead, it resembles a shear-type mode (often termed “internal mode” by \citet{boomkamp1996classification}), with properties more akin to those of the shear instabilities in figures~\ref{Fig8}b–d. The right-hand branch of the V-shaped contour, corresponding to short waves, disappears when $G=0.02$, while a new long-wave instability region emerges. Though weaker, this mode suggests that surface tension and gravity can, counterintuitively, play a destabilizing role, as also noted by \citet{yiantsios1988linear}. At higher $G$ (e.g. $0.5$), this mode vanishes, restoring the expected stabilizing effect.

The flow-reversal profile ($a=3$) displays contours (figures~\ref{Fig8}g,h) that qualitatively resemble those of $a=-3$: a long-wave interfacial instability, a short-wave interfacial instability, and a strong rippling mode at large $Re$. In this case, however, the rippling instability is more dominant, exhibiting growth rates that exceed those of the other modes, and persisting up to $Re\sim O(10^2)$. Interestingly, at large $Re$ and $k$, a narrow band of stability appears within the rippling region, consistent with the deviations from inviscid predictions shown earlier in figure~\ref{Viscous a=3}b. Unlike the other cases, no shear mode is observed for $a=3$.

In summary, the viscous stability analysis maps a diverse set of instabilities across the $(k,Re)$ parameter space. Long- and short-wave asymptotics identify the ranges of curvature $a$ that support interfacial instabilities, while the full growth rate contours clarify their coexistence and transitions. A key observation is that $a=0$ marks a sharp boundary in behavior, with small positive and negative curvatures favoring short- and long-wave instabilities, respectively. More generally, viscosity is not merely dissipative: for long waves it can enhance growth compared to inviscid predictions, while in other regimes it suppresses disturbances or generates new modes such as the shear-type instability.

These findings naturally raise several questions. What is the precise energy transfer mechanism that distinguishes the viscous long-wave mode from the inviscid rippling mode? How does the transition between interfacial and rippling instabilities proceed at finite Reynolds numbers, particularly for $a<-1$ and $a>0$? Under what conditions do gravity and surface tension play a destabilizing role rather than a stabilizing one? And finally, how robust are the shear-type modes in more realistic air–water systems with viscosity and density contrasts? These questions motivate a more detailed energy budget and eigenfunction analysis, which we turn to in the following section.

\section{A family of instabilities}\label{instabilityFamily}

\begin{figure}
    \hspace{-1cm}
    \begin{minipage}[t]{0.55\textwidth}
      \centering
      \includegraphics[width=\textwidth]{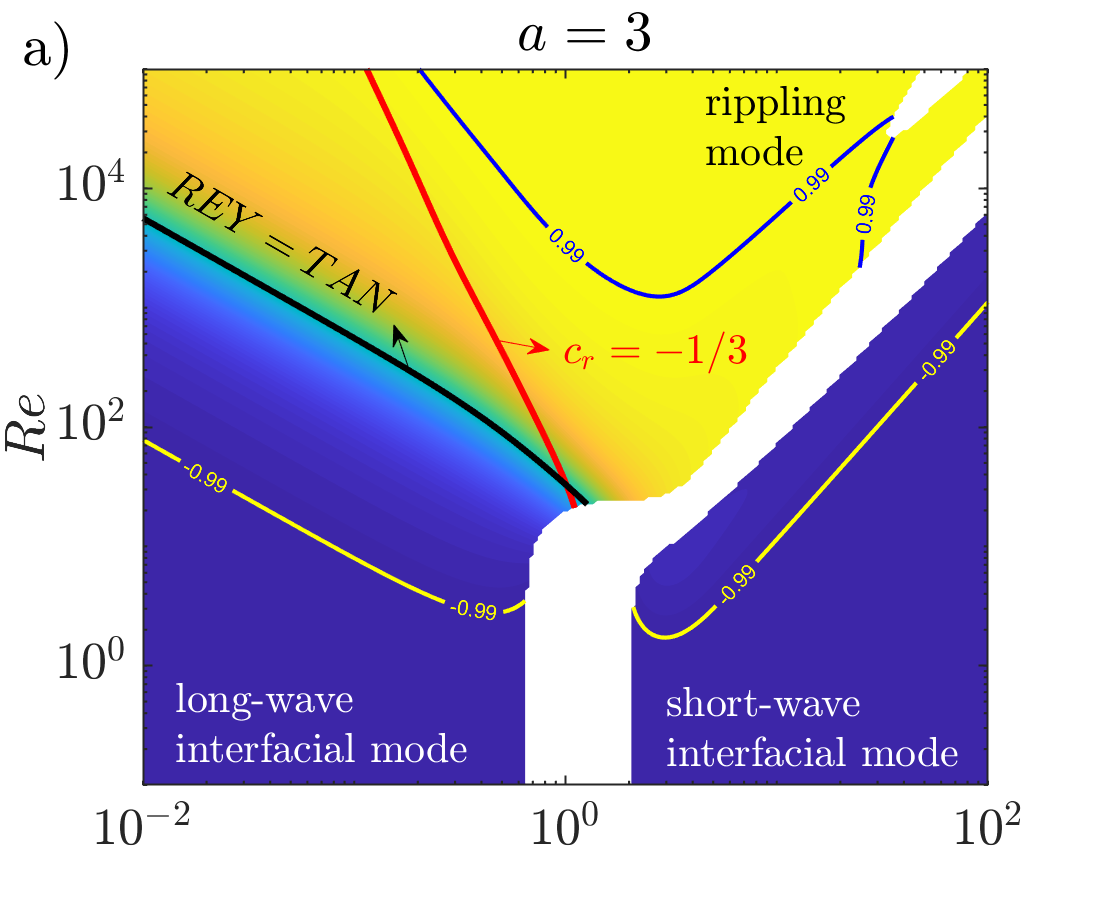}
    \end{minipage}
     \hspace{-0.4cm}
    \begin{minipage}[t]{0.55\textwidth}
      \centering
      \includegraphics[width=\textwidth]{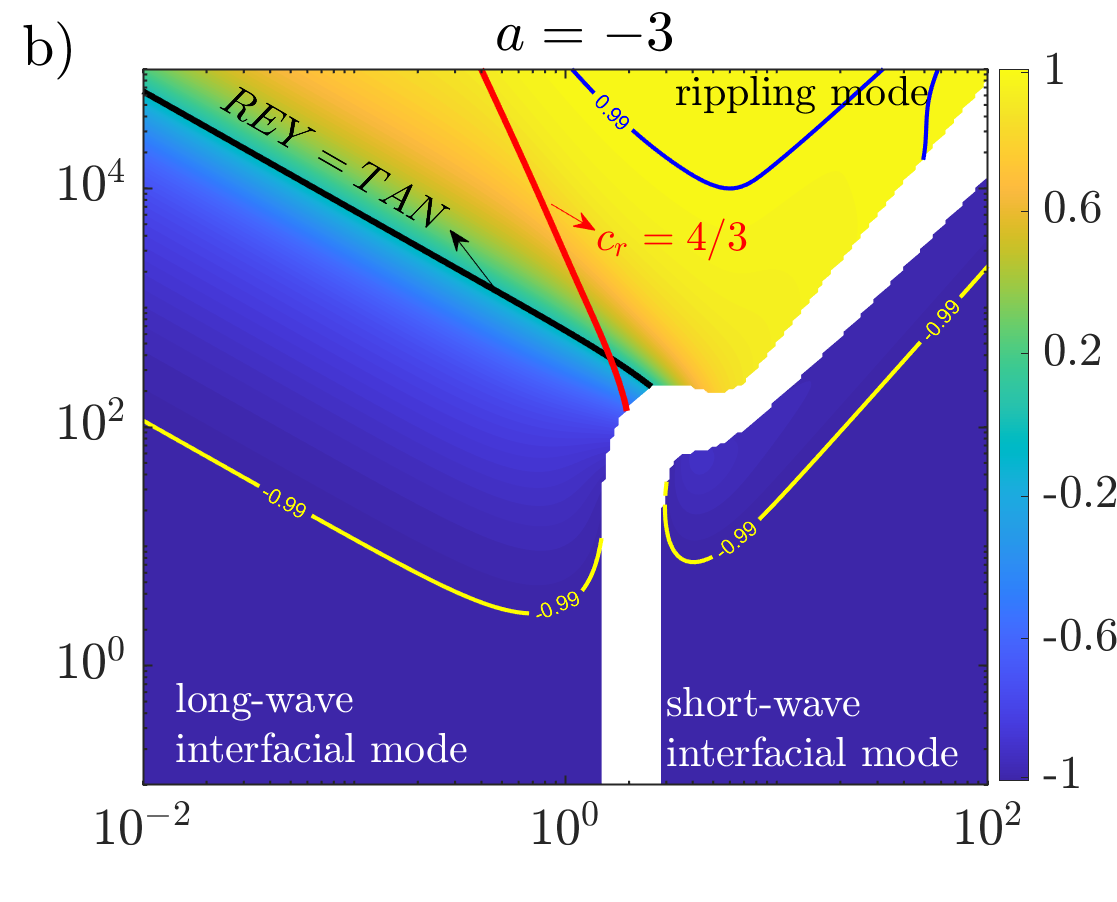}
    \end{minipage}

        \hspace{-1cm}
    \begin{minipage}[t]{0.55\textwidth}
      \centering
      \includegraphics[width=\textwidth]{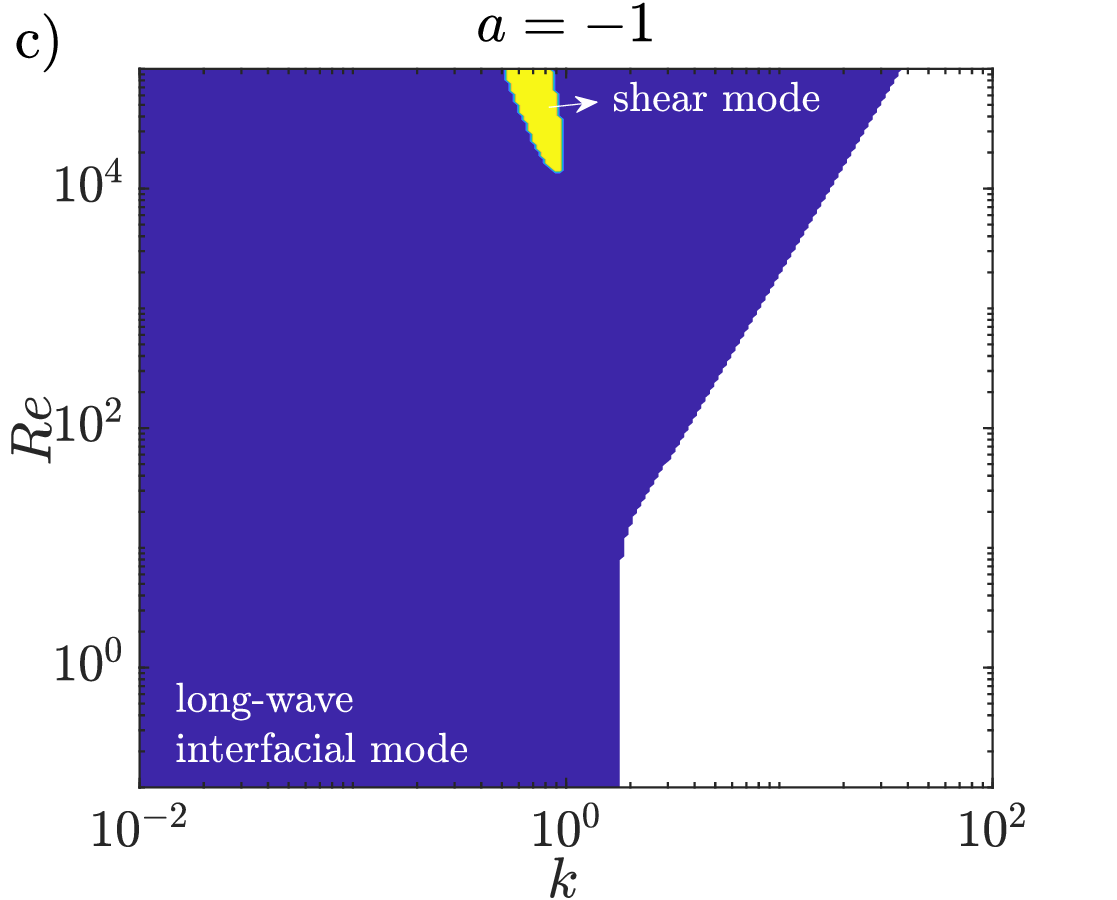}
    \end{minipage}
     \hspace{-0.4cm}
    \begin{minipage}[t]{0.55\textwidth}
      \centering
      \includegraphics[width=\textwidth]{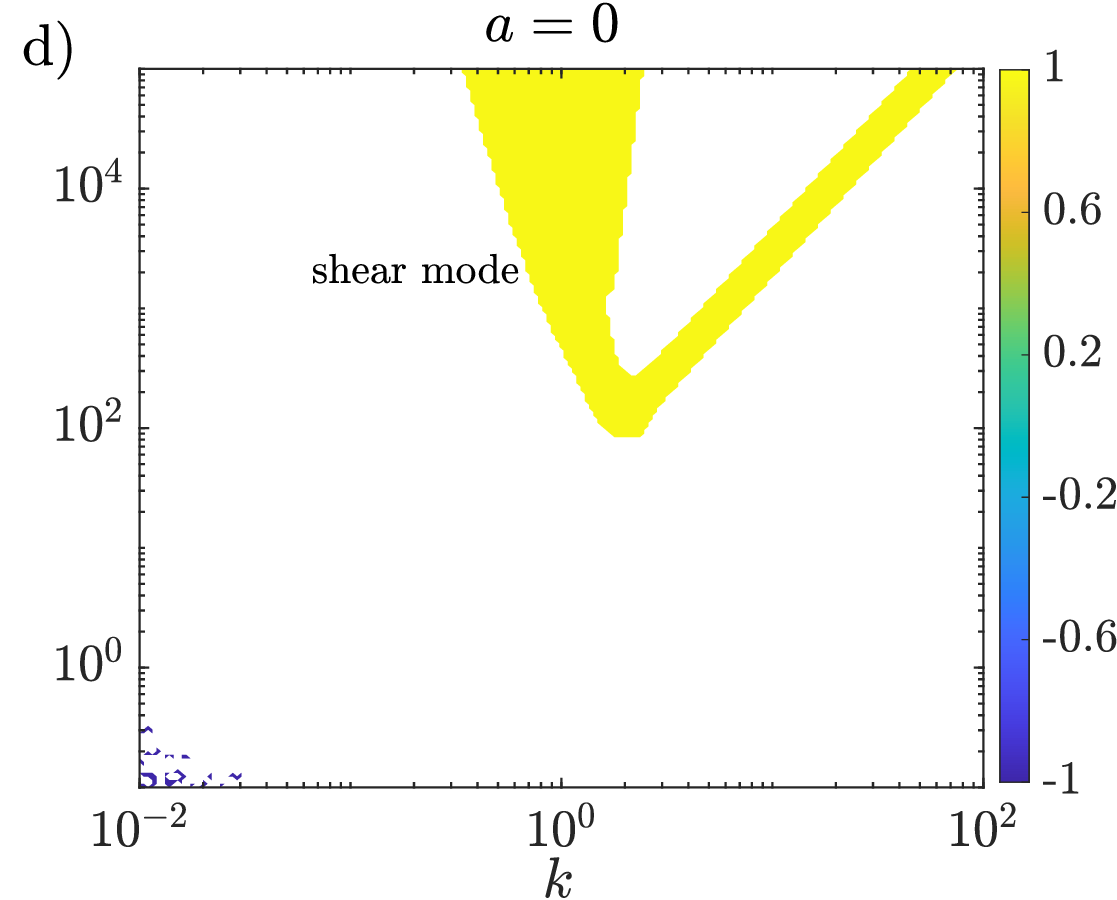}
    \end{minipage}

\caption{Contours of $\varepsilon = \left(\textrm{REY}-\textrm{TAN}\right)/\max{\left(\textrm{REY},\textrm{TAN}\right)}$ for the most-unstable mode in the $(k,Re)$ parameter space for $G=0$ and $a = $ (a) $3$, (b) $-3$, (c) $-1$, and (d) $0$. Black curves in both (a) \& (b) trace the locations where REY and TAN have the same magnitude. For $k$ and $Re$ to the right of the red line in (a) and (b), $c_r = U(z_c)$ is satisfied i.e., a critical layer exists at depth $z_c$. Yellow and blue curves indicate the contour lines where REY is 1\% of TAN, and vice versa. Note that if REY $<0$ (as for the blue region in $a=-1$), REY is substituted with $0$ in the expression above, resulting in $\varepsilon = -1$, which indicates that TAN alone is the energy source.}
\label{Energies}
\end{figure}
The characterization of instabilities in two-phase parallel flows is notoriously difficult. Even with the free-surface approximation (neglecting the top layer), the problem involves five parameters: $k, Re, G, Bo$, and $a$. A widely used diagnostic is the perturbation kinetic energy equation. \citet{boomkamp1996classification}, following \citet{hooper1983shear}, classified instabilities in two-phase flows by identifying the dominant contribution to perturbation kinetic energy production. For free-surface flows, \citet{kelly1989mechanism} derived the analogous energy balance in the context of falling films. In this framework, the perturbation kinetic energy (KE) is expressed as the sum of work done by tangential stresses (TAN), normal stresses (NOR), Reynolds stresses (REY), and viscous dissipation (DIS):  
\begin{equation}
    \textrm{KE} = \textrm{TAN} + \textrm{NOR} + \textrm{REY} + \textrm{DIS}.
\end{equation}
The individual terms are given by
\begin{equation}
    \textrm{KE} = \int_{-1}^0dz\dfrac{\partial}{\partial t}\left(\dfrac{\overline{u^2}+\overline{w^2}}{2}\right), \quad 
    \textrm{TAN} = \dfrac{1}{Re}\left(\overline{\dfrac{\partial u}{\partial z}u} + \overline{\dfrac{\partial w}{\partial x}u}\right)_{z=0}, \quad
    \textrm{NOR} = \left(-\overline{pw} + \dfrac{2}{Re}\overline{\dfrac{\partial w}{\partial z}w}\right)_{z=0}, \nonumber
\end{equation}
\begin{equation}
    \nonumber
    \textrm{REY} = \int_{-1}^0 dz\left(-\overline{uw}\dfrac{dU}{dz}\right), \qquad 
    \textrm{DIS} = -\dfrac{1}{Re}\int_{-1}^0dz\left[2\overline{\left(\dfrac{\partial u}{\partial x}\right)^2} + \overline{\left(\dfrac{\partial u}{\partial z} + \dfrac{\partial w}{\partial x}\right)^2} + 2\overline{\left(\dfrac{\partial w}{\partial z}\right)^2}\right],
\end{equation}
where overbars denote horizontal averages over a wavelength, and TAN and NOR are evaluated at the free surface ($z=0$). With the normal mode ansatz and the normal stress boundary condition, these expressions can be evaluated. For $G=0$, NOR vanishes identically. To compare the relative importance of TAN and REY, we define  
\[
\varepsilon = \frac{\textrm{REY}-\textrm{TAN}}{\max(\textrm{REY},\textrm{TAN})},
\]
so that $\varepsilon>0$ implies REY dominance and $\varepsilon<0$ implies TAN dominance.  

Figures~\ref{Energies}a-d show contour plots of $\varepsilon$ for the most unstable mode in the $(k,Re)$ plane, for $G=0$ and $a=3,-3,-1,$ and $0$. Yellow regions indicate REY dominance, blue regions TAN dominance. For $a=3$ and $a=-3$ (figures \ref{Energies}a,b), TAN dominates for long- and short-wave interfacial modes at low $Re$, consistent with \citet{boomkamp1996classification}. At higher $Re$, REY overtakes TAN, marking a change in mechanism \citep{charru2000phase}. The black curve corresponds to $\textrm{REY}=\textrm{TAN}$, while the red curve denotes the appearance of a critical layer ($c_r=U(z_c)$), and the arrow points towards the region where a critical layer exists. However, whether the mode in this regime is a rippling mode or a shear mode cannot be distinguished from these figures alone. The $a=-1$ (Nusselt flow) case, shown in figure \ref{Energies}c, illustrates a different behavior: TAN dominates across all $Re$, with no gradual transition to REY dominance. The small REY-dominated patch at large $Re$ and $O(1)$ wavenumbers corresponds to a shear mode, consistent with earlier discussions. For $a=0$ (figure \ref{Energies}d), the instability resembles the shear mode observed for $a=-1$ and is clearly REY-dominated. This conclusion differs from \citet{boomkamp1996classification}, who argued for TAN dominance with REY as a small but finite contribution, motivating their term “internal mode.” Our results suggest instead that the mode is primarily driven by REY, though a critical layer persists across the parameter space.  

These results highlight the limitations of energy analysis alone for classifying instabilities. In particular, the transition from long-wave interfacial to REY-dominated instability at high $Re$, and the precise distinction between rippling and shear instabilities, remain ambiguous. To refine this classification, \citet{charru2000phase}, building on \citet{hooper1987shear}, introduced two useful quantities: (i) the penetration depth of disturbances, and (ii) an effective Reynolds number measuring inertial influence via the ratio of the imaginary to real parts of the vorticity eigenfunction. They classified instabilities into three regimes: (a) a shallow-viscous regime for $k\ll 1$ (disturbances penetrate the depth), (b) a deep-viscous regime for $k\gg 1$ (disturbances confined to one wavelength), and (c) an inviscid regime where penetration depth is set by the viscous scale and the effective Reynolds number is $O(1)$. Transitions between these regimes are sharp and can be demonstrated by asymptotic calculations. Extending such a framework to free-surface flows lies beyond the scope of this study, but in the following subsections we adopt this terminology to describe representative eigenfunction structures.

\subsection{Long wave modes ($k \ll 1$)}\label{long wave modes char}
\begin{figure}
    \hspace{-1cm}
    \begin{minipage}[t]{0.57\textwidth}
      \centering
      \includegraphics[width=\textwidth]{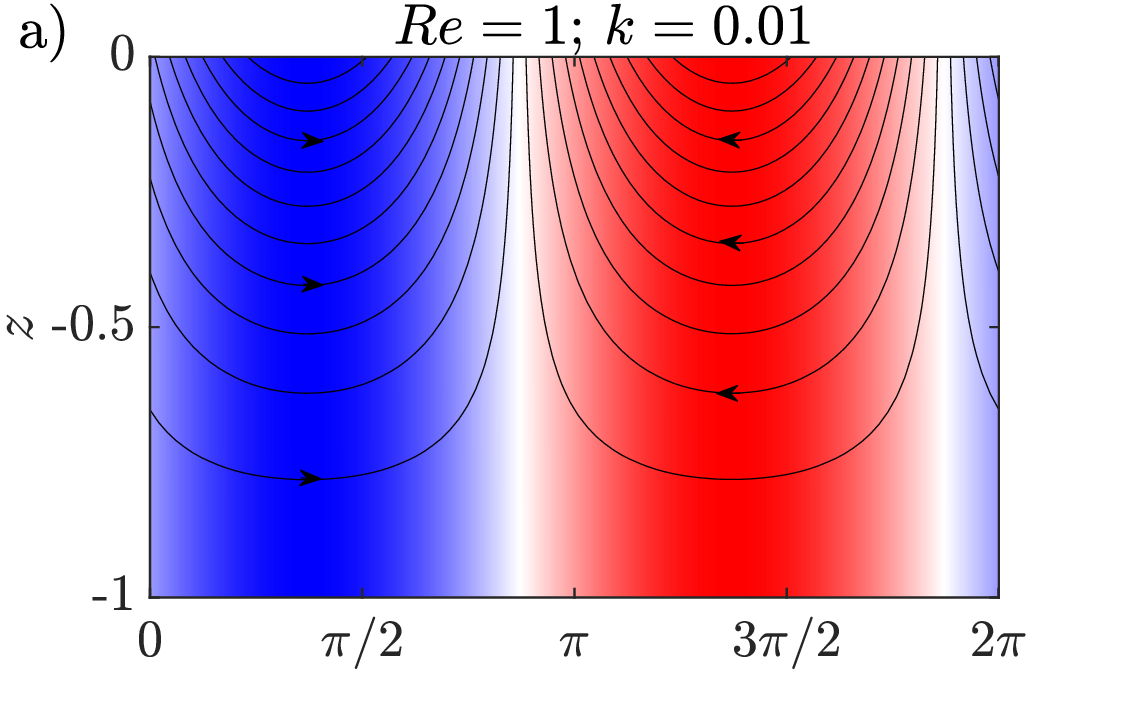}
    \end{minipage}
     \hspace{-0.4cm}
    \begin{minipage}[t]{0.57\textwidth}
      \centering
      \includegraphics[width=\textwidth]{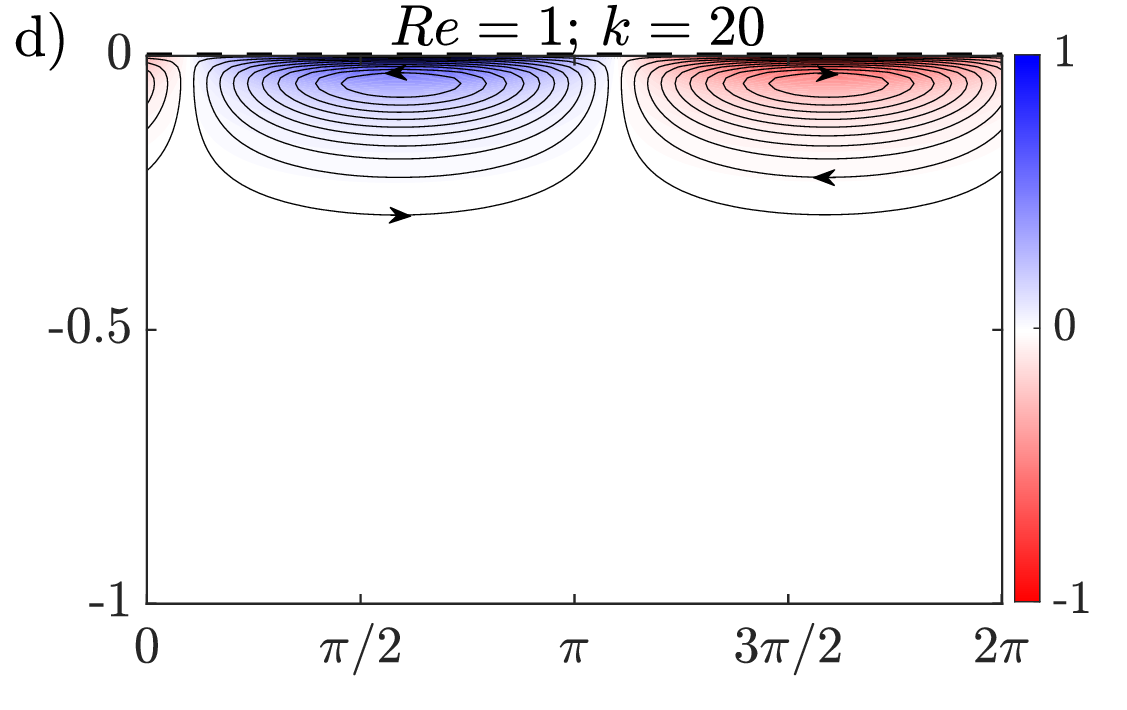}
    \end{minipage}

        \hspace{-1cm}
    \begin{minipage}[t]{0.57\textwidth}
      \centering
      \includegraphics[width=\textwidth]{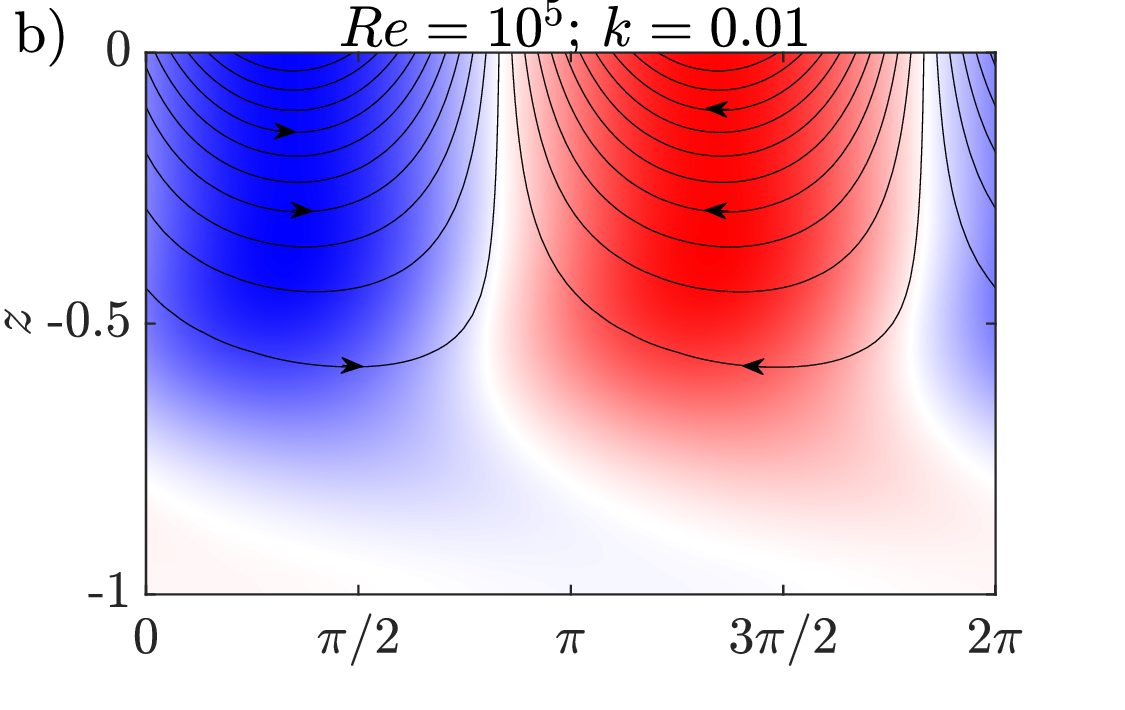}
    \end{minipage}
     \hspace{-0.4cm}
    \begin{minipage}[t]{0.57\textwidth}
      \centering
      \includegraphics[width=\textwidth]{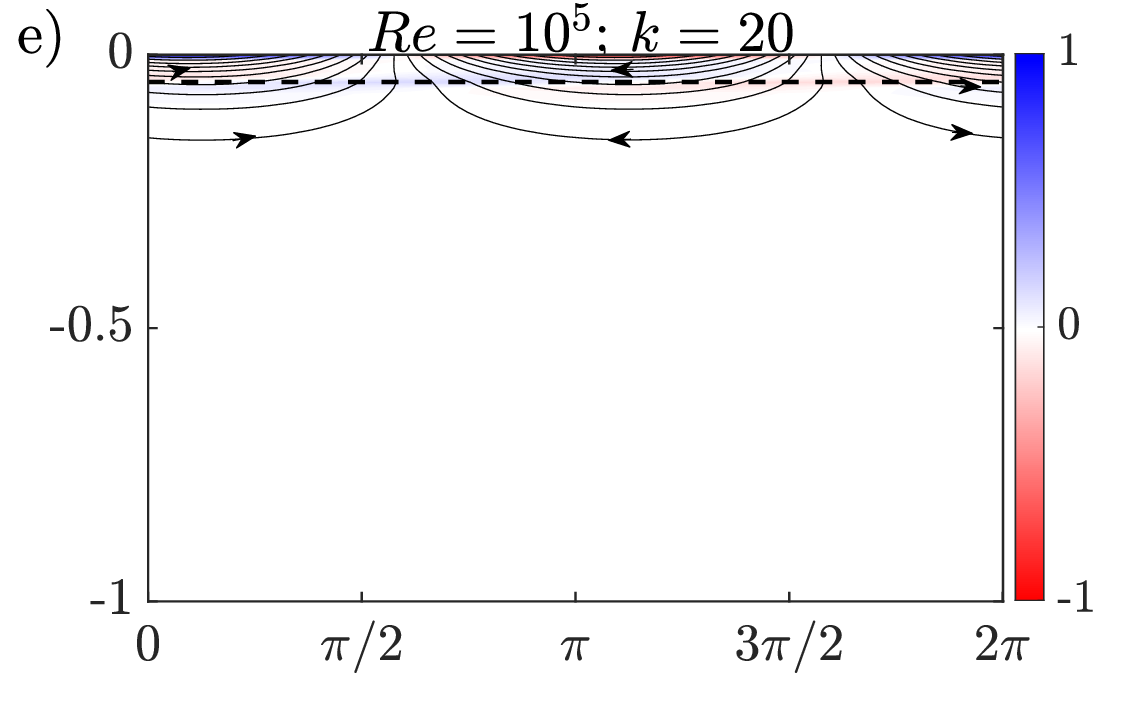}
    \end{minipage}

        \hspace{-1cm}
    \begin{minipage}[t]{0.57\textwidth}
      \centering
      \includegraphics[width=\textwidth]{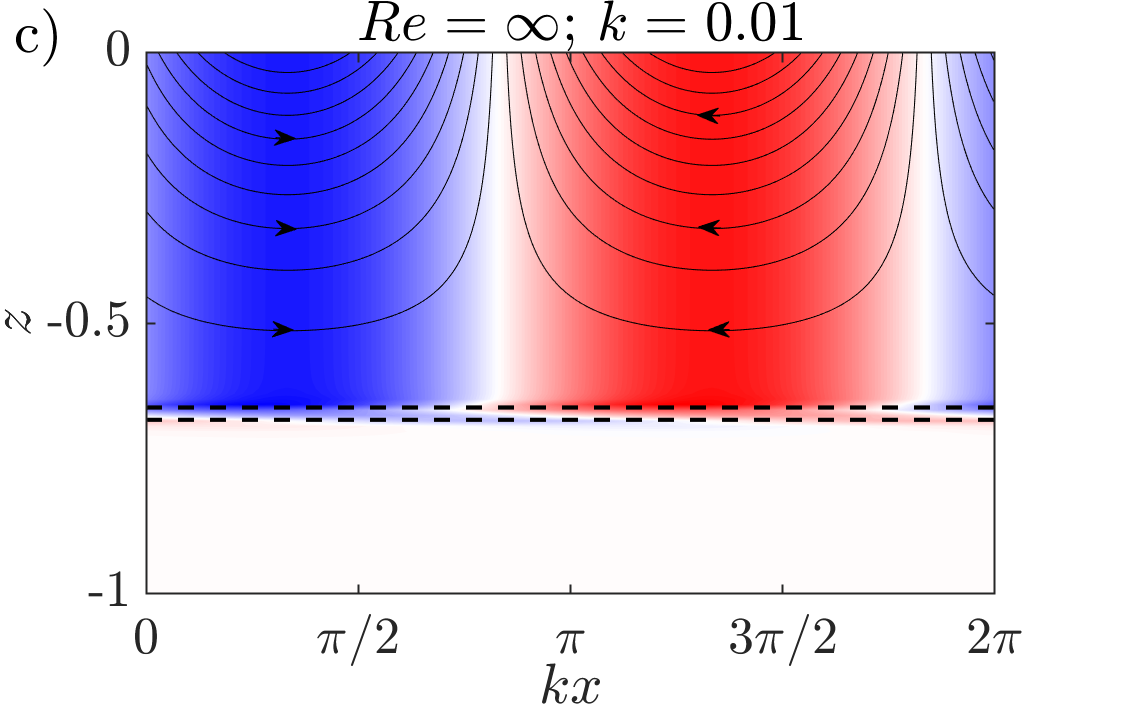}
    \end{minipage}
     \hspace{-0.4cm}
    \begin{minipage}[t]{0.57\textwidth}
      \centering
      \includegraphics[width=\textwidth]{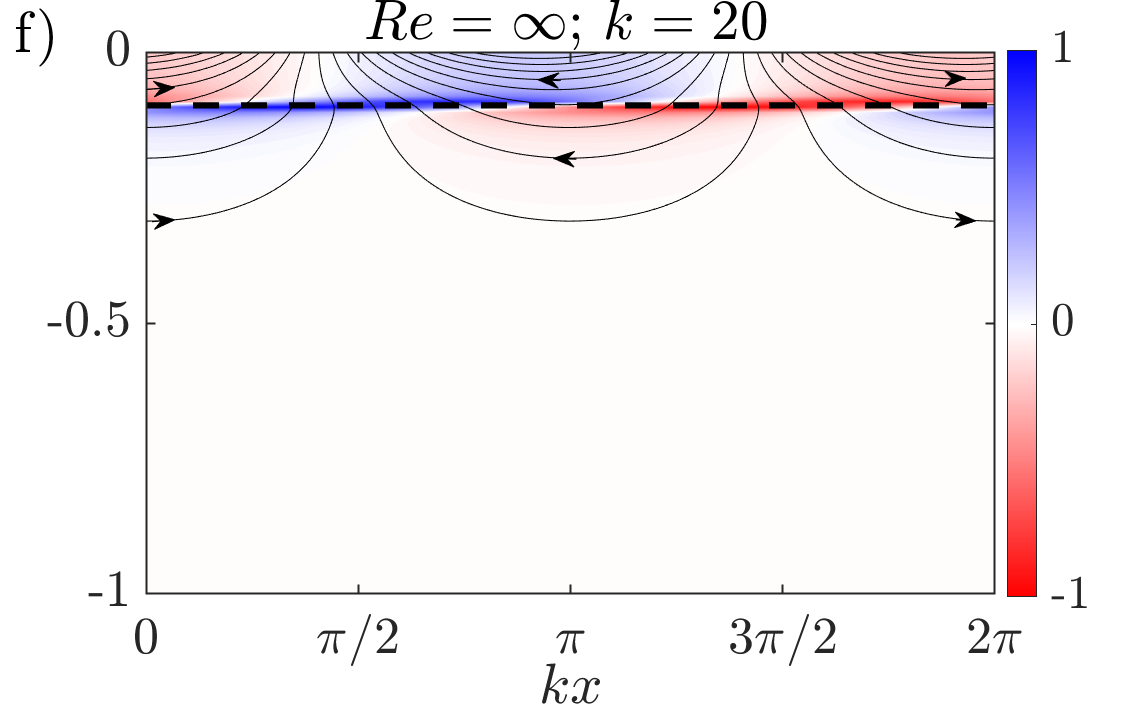}
    \end{minipage}
\caption{The streamlines of the disturbance flow field superimposed on the contours of the normalized disturbance vorticity ($\Omega/\Omega_{max}$) for \( G = 0 \) and \( a = 3 \). Panels a, b, and c correspond to $Re = 1$, $Re = 10^5$, and the inviscid limit, respectively, in the long wave limit ($k=0.01$). Panels d, e, and f correspond to $Re = 1$, $Re = 10^5$, and the inviscid limit ($Re = \infty$), respectively, in the short wave limit ($k=20$)}
\label{Viscous eigenstructures a=3}
\end{figure}

In \S\ref{Transition vis to inviscid}, it was shown that the long-wave growth rates for the flow-reversal profile ($a=3$) behave differently at small and large Reynolds numbers, especially when compared with the inviscid limit. This difference can be partly attributed to the transition of the energy source from tangential stress (TAN) to Reynolds stress (REY) for long waves (figure \ref{Energies}a). To explore this further, figures \ref{Viscous eigenstructures a=3}a–c present streamlines and normalized vorticity contours at $k=0.01$ for $Re=1$, $10^5$, and the inviscid limit, respectively.  

At $Re=1$ (figure \ref{Viscous eigenstructures a=3}a), the vorticity is nearly uniform across the depth, with no appreciable horizontal displacement, and the streamfunction is localized near the surface, in-phase with both vorticity and surface displacement $\eta$. This is characteristic of the shallow-viscous regime of \citet{charru2000phase} and agrees with the mechanism described by \citet{smith1990mechanism}: in the long-wave, low-$Re$ limit, the leading-order solution is a constant vorticity disturbance (see Appendix \ref{AppC}). Here the fluid depth is much smaller than both the wavelength and the viscous length scale, so vertical variations are $O(1/k)$ and horizontal shifts are $O(kRe)$. In contrast, the $Re=10^5$ case (figure \ref{Viscous eigenstructures a=3}b)—which lies in the parameter region where growth rates exceed the inviscid values (figure \ref{Viscous a=3}b) and where REY dominates over TAN without a critical layer (figure \ref{Energies}a)—shows markedly different eigenstructures. The vorticity no longer spans the entire depth but instead peaks in the interior, decays smoothly below this point, and exhibits a horizontal shift indicative of inertia. The streamlines remain in-phase with vorticity but tilt rightward and concentrate more strongly near the surface. The reduced penetration depth suggests correspondence with the “inviscid regime” of \citet{charru2000phase}, in which disturbances are confined to scales smaller than both the wavelength and the fluid depth. Finally, the inviscid solution (figure \ref{Viscous eigenstructures a=3}c) reveals yet another distinct structure: a critical layer is now present (black dashed lines), unlike at $Re=10^5$, where vorticity simply peaked in the bulk. Outside this layer, vorticity decays rapidly, while the streamfunction resembles the $Re=1$ case away from the critical layer but acquires a pronounced tilt in its vicinity, much like the behavior shown in figure \ref{wi vs k}d.  

In summary, the $Re=10^5$ eigenstructure is not merely a midpoint between the long-wave interfacial mode ($Re=1$) and the rippling mode (Inviscid), but rather represents a qualitatively distinct regime with its own dynamical character.

\subsection{Short wave modes ($k\gg 1$)}\label{Short wave modes char}

Unlike the long-wave case, short waves do not undergo a gradual transition from TAN- to REY-dominated regimes (figure \ref{Energies}a). Instead, a distinct region of stability separates the two, suggesting that the underlying mechanisms are fundamentally different instabilities rather than a continuous evolution. To clarify this, we examine the eigenstructures at short wavelengths. Figures \ref{Viscous eigenstructures a=3}d–f present streamlines and normalized vorticity contours for $k=20$ at $Re=1$, $10^5$, and the inviscid case, respectively.  

At $Re=1$ (figure \ref{Viscous eigenstructures a=3}d), the streamlines remain in phase with vorticity and are concentrated near the surface, with negligible horizontal shift due to weak inertia. However, unlike the long-wave case, the streamfunction exhibits a cellular pattern of nearly circular streamlines rather than surface-attached structures. This agrees with the physical mechanism of the short-wave interfacial instability described by \citet{hinch1984note}: the disturbance horizontal velocity is positive below crests and negative below troughs, a consequence of viscosity contrast across the interface. The vorticity penetration depth is of order one wavelength, consistent with the deep-viscous regime of \citet{charru2000phase}, where the viscous length scale exceeds the wavelength. The critical layer (black dashed line) lies close to the surface, in line with the eigenvalue expression \eqref{G=0 short wave asymptotic}.  

At $Re=10^5$ (figure \ref{Viscous eigenstructures a=3}e), the critical layer shifts slightly below the surface and exhibits a localized vorticity maximum, reminiscent of a rippling mode, but concentrated within a narrow interfacial region like a shear mode. Thus, the eigenstructure combines features of both instabilities. The streamlines, anchored at the surface, show sharp gradients across the critical layer and mark a transition to the inviscid regime of \citet{charru2000phase}, where penetration depths are smaller than a wavelength and comparable to the viscous length scale. In the inviscid case (figure \ref{Viscous eigenstructures a=3}f), the vorticity maximum resides exclusively at the critical layer, and not at the surface, identifying the instability as a purely rippling mode. Compared with the long-wave inviscid case (figure \ref{Viscous eigenstructures a=3}c), the vorticity and streamfunction structures are confined to a shallower region of the domain due to the large $k$ and the proximity of the critical layer to the interface.  

Taken together, these results show that short waves highlight a sharper distinction between viscous and inviscid mechanisms than long waves. At low $Re$, they correspond to the deep-viscous interfacial mode; at moderate to high $Re$, they acquire mixed features of shear and rippling modes; and in the inviscid limit, they reduce to the classical rippling instability.

\begin{figure}
\centering
 \hspace{-1cm}
    \begin{minipage}[t]{0.52\textwidth}
      \centering
      \includegraphics[width=\textwidth]{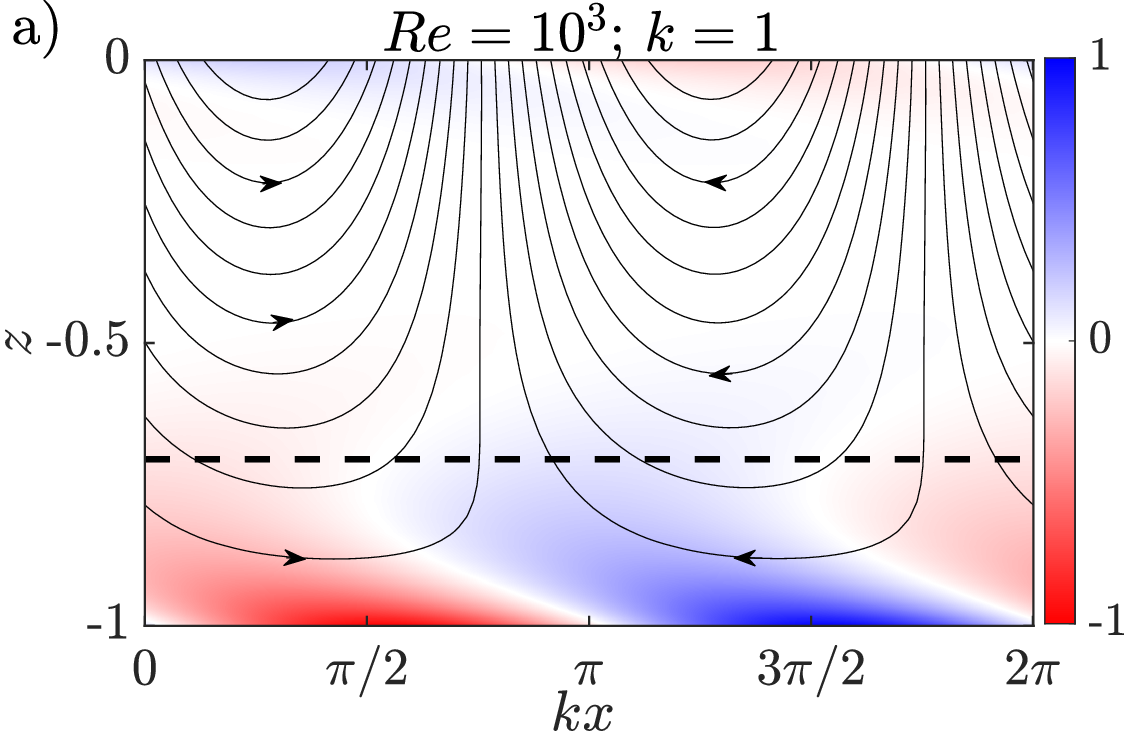}
    \end{minipage}
     \hspace{-0cm}
    \begin{minipage}[t]{0.52\textwidth}
      \centering
      \includegraphics[width=\textwidth]{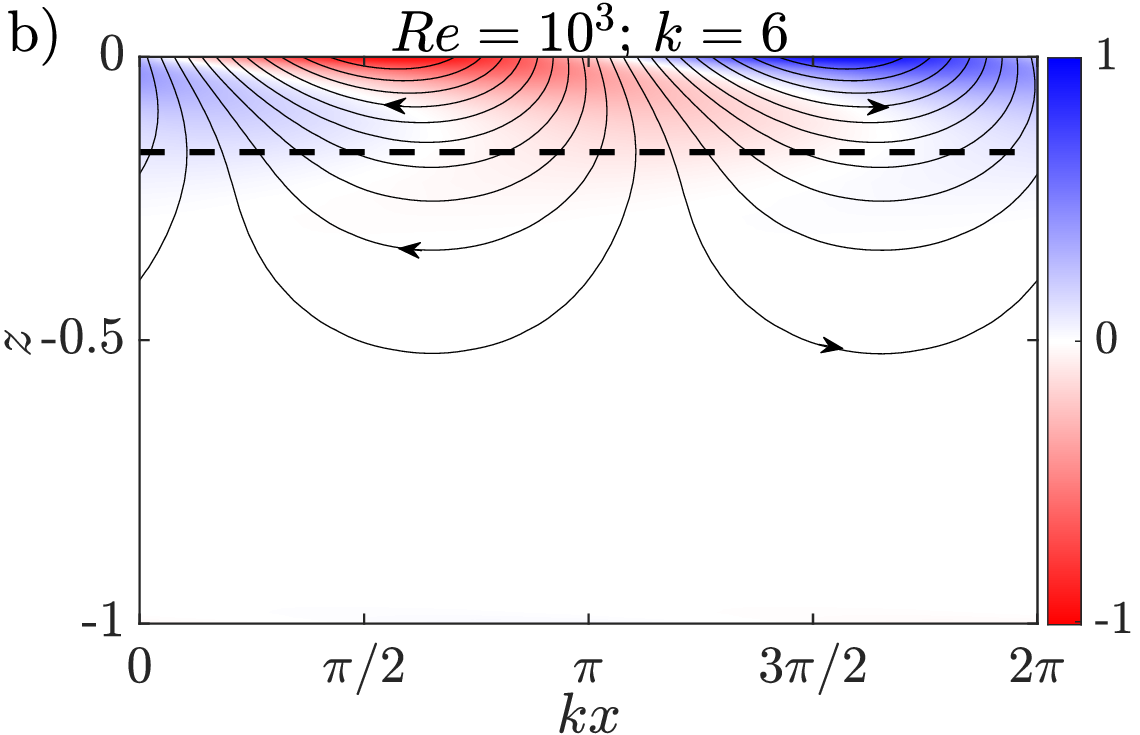}
    \end{minipage}

    \hspace{-0.5cm}
    \begin{minipage}{0.52\textwidth}
      \centering
      \includegraphics[width=\textwidth]{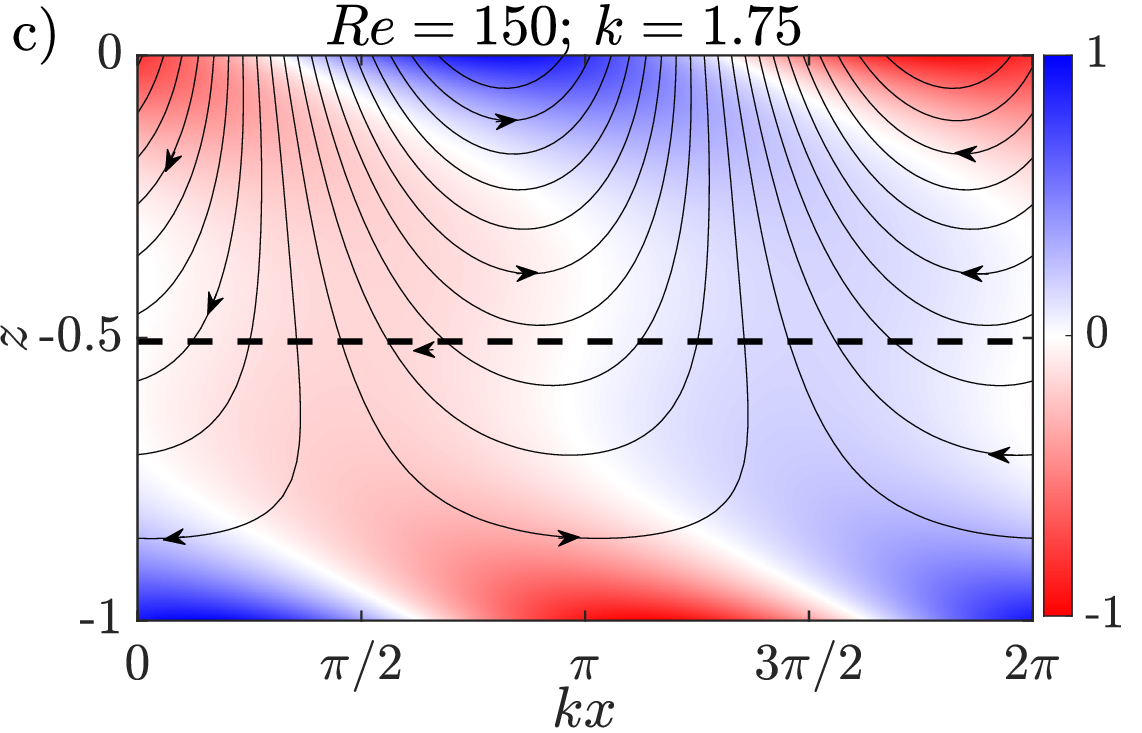}
    \end{minipage}
    \caption{The streamlines of the disturbance flow-field superimposed on contours of the normalized vorticity ($\Omega/\Omega_{max}$) for $G = 0, a = 0$, and for representative parameters given by $(k,Re) = $: (a) ($1,10^3$), (b) ($6, 10^3$), and (c) ($1.75, 150$). The black dashed lines indicate the critical layer.}
\label{Viscous eigenstructures a=0}
\end{figure}
\subsection{Shear mode at $a = 0$}\label{Miles_1960}
\begin{figure}
     \hspace{-1cm}
    \begin{minipage}[t]{0.57\textwidth}
      \centering
      \includegraphics[width=\textwidth]{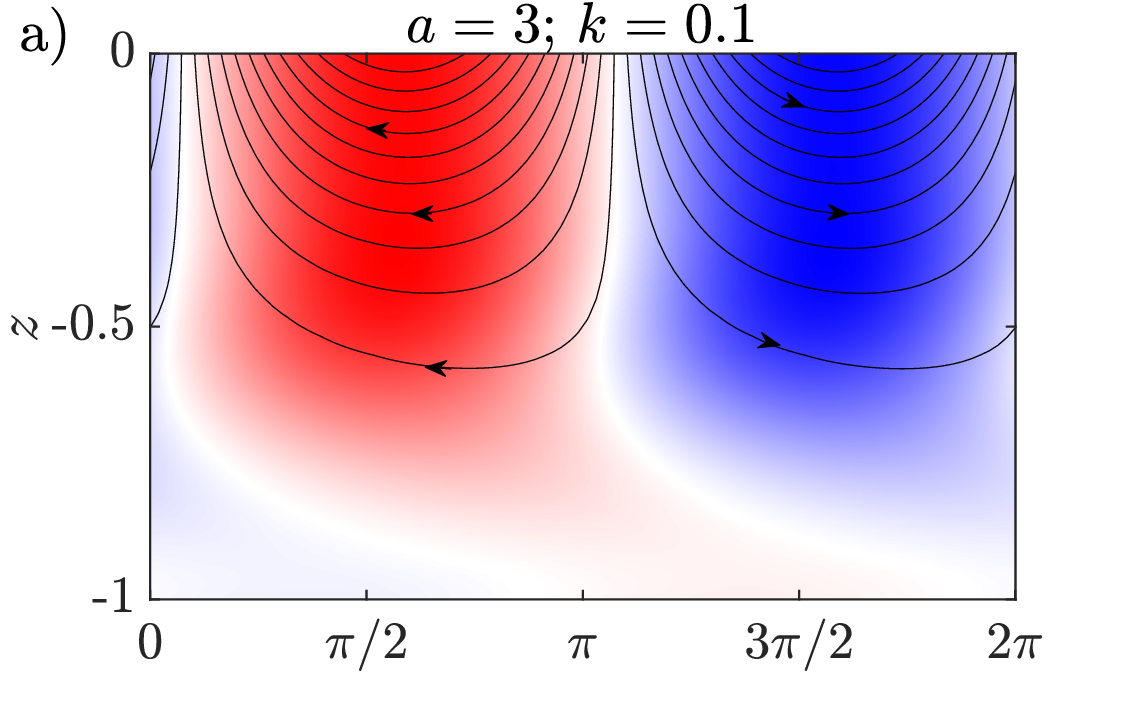}
    \end{minipage}
     \hspace{-0.4cm}
    \begin{minipage}[t]{0.57\textwidth}
      \centering
      \includegraphics[width=\textwidth]{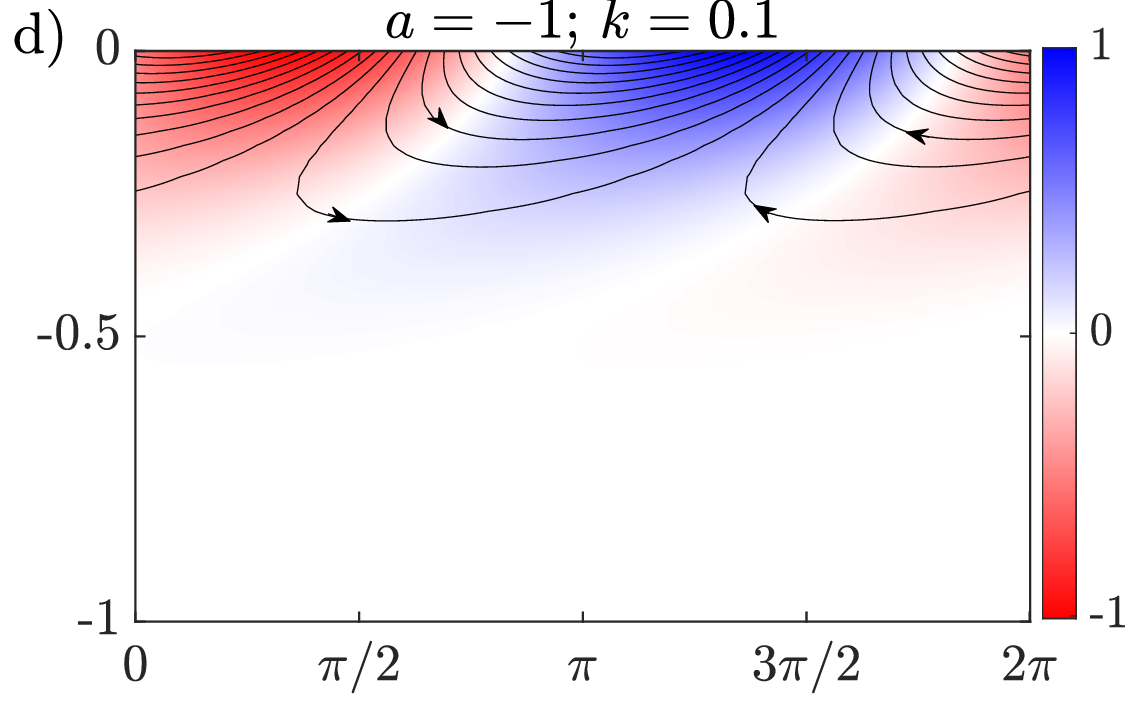}
    \end{minipage}

        \hspace{-1cm}
    \begin{minipage}[t]{0.57\textwidth}
      \centering
      \includegraphics[width=\textwidth]{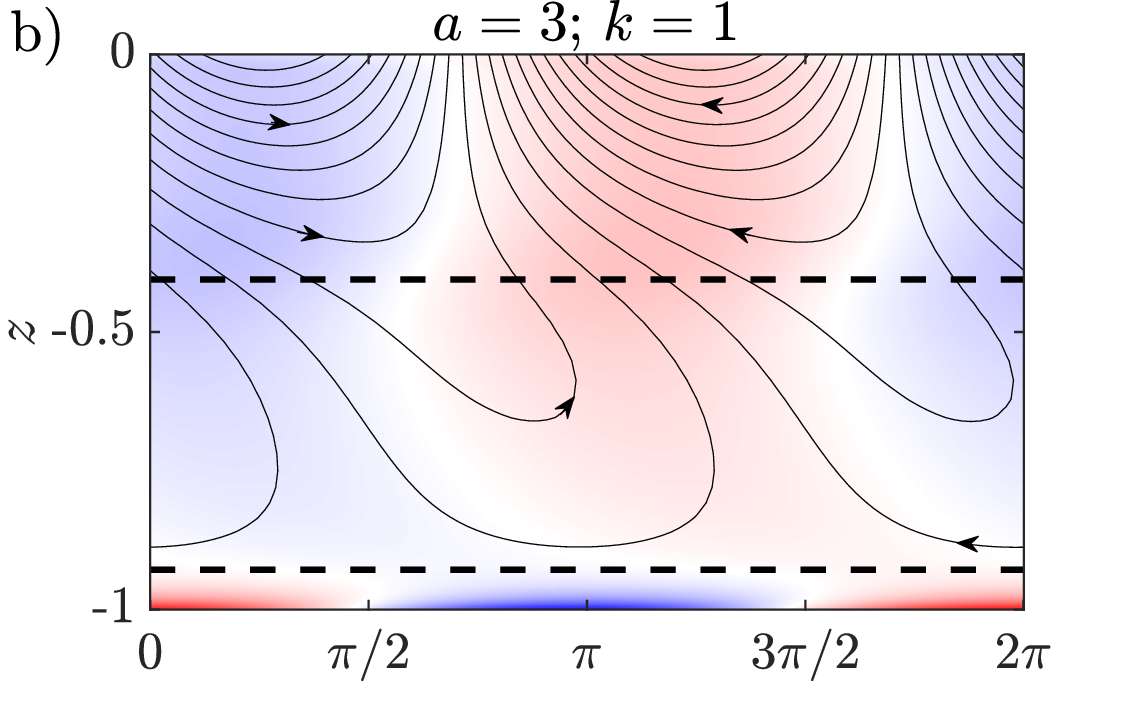}
    \end{minipage}
     \hspace{-0.4cm}
    \begin{minipage}[t]{0.57\textwidth}
      \centering
      \includegraphics[width=\textwidth]{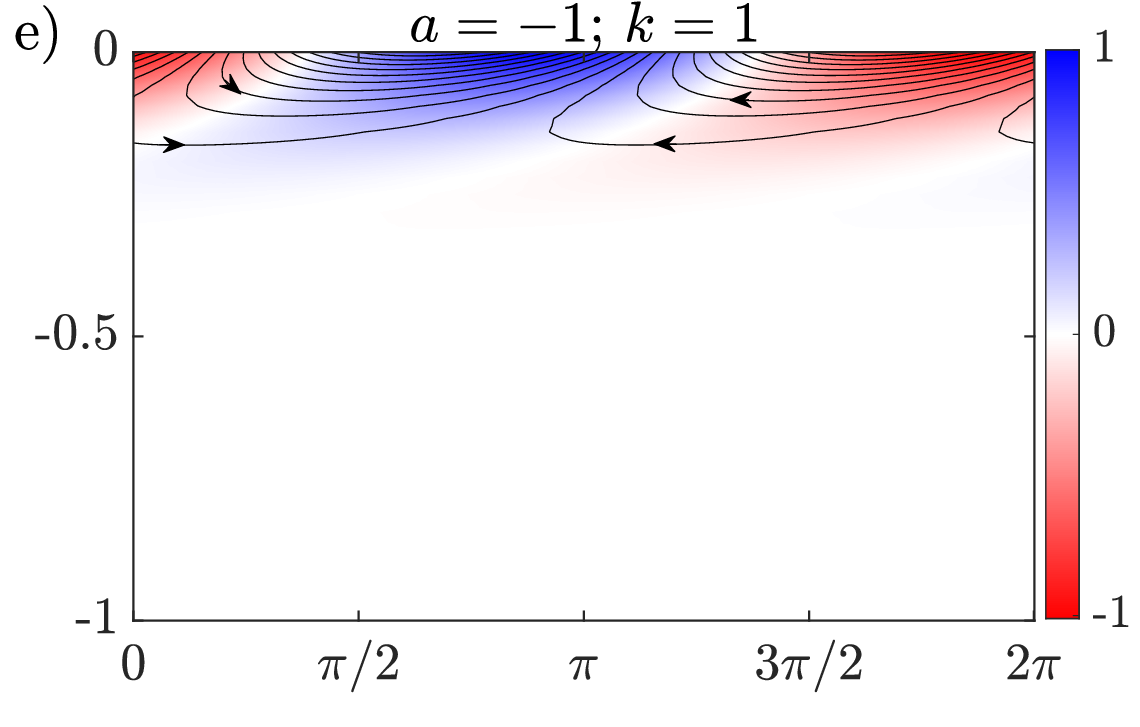}
    \end{minipage}

        \hspace{-1cm}
    \begin{minipage}[t]{0.57\textwidth}
      \centering
      \includegraphics[width=\textwidth]{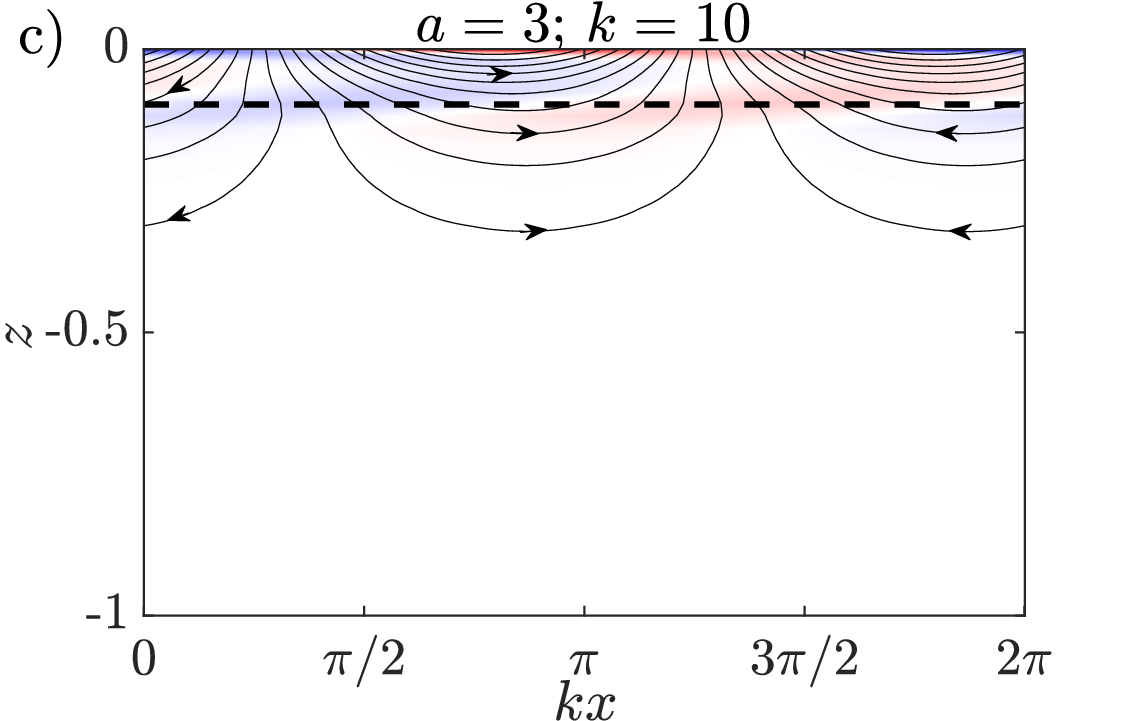}
    \end{minipage}
     \hspace{-0.4cm}
    \begin{minipage}[t]{0.57\textwidth}
      \centering
      \includegraphics[width=\textwidth]{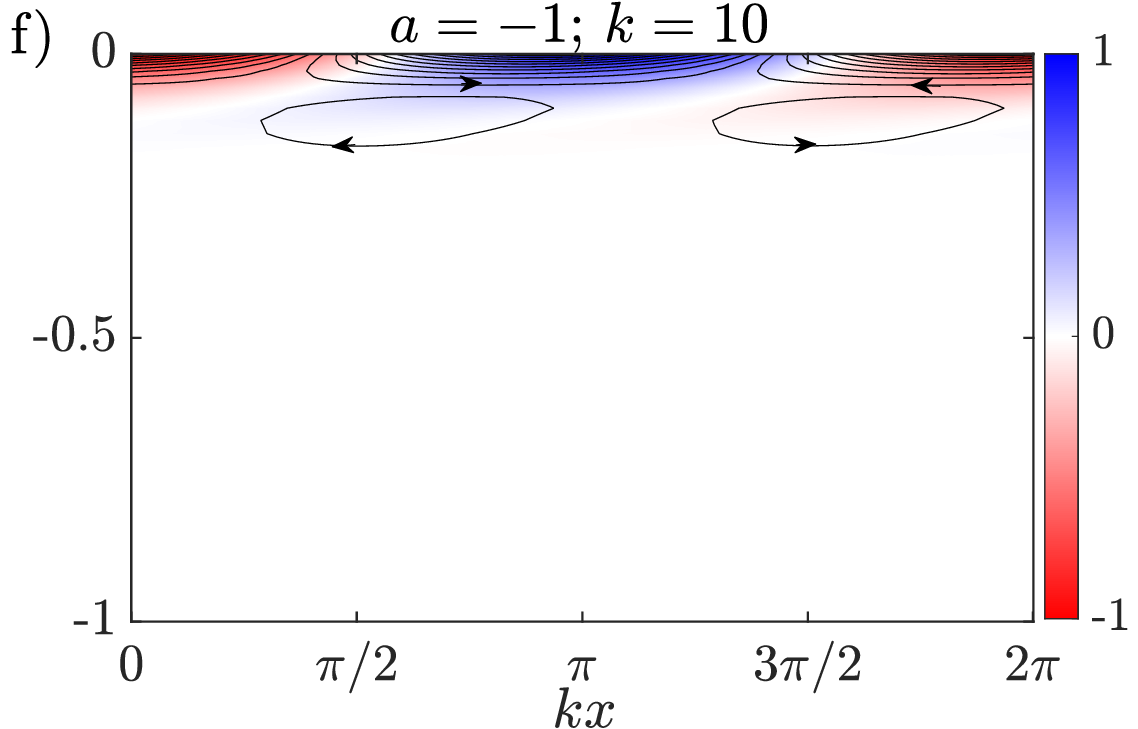}
    \end{minipage}
\caption{The streamlines of the disturbance flow field superimposed on the contours of the normalized disturbance vorticity ($\Omega/\Omega_{max}$) for \( G = 0 \) and \( Re = 10^{4} \). Panels a, b, and c correspond to $k = 0.1$, $k=1$, and $k=10$, respectively, with $a=3$. Panels d, e, and f correspond to $k = 0.1$, $k=1$, and $k=10$, respectively, with $a=-1$.}
\label{Viscous eigenstructures a=3m1}
\end{figure}
Before examining eigenfunctions in the REY-dominant region for $a=3$, it is useful to first consider the shear mode for the linear profile ($a=0$). Figures \ref{Viscous eigenstructures a=0}a–c show streamlines and normalized vorticity contours for three representative points: $(k,Re)=(1,10^3)$, $(6,10^3)$, and $(1.75,150)$. The first two correspond to the left and right branches of the `V’-shaped instability region in figure \ref{Fig8}e, while the third lies at their intersection. The case $(k,Re)=(1,10^3)$ (figure \ref{Viscous eigenstructures a=0}a) exhibits the hallmark features of a shear mode: vorticity concentrated near the bottom wall with a horizontal shift \citep{hooper1987shear}, the presence of a critical layer \citep{miles1960hydrodynamic}, and streamfunction contours filling most of the domain, consistent with the `A’-family of modes \citep{kaffel2015eigenspectra}. The Reynolds stress $(-\overline{uw})$ shows a local extremum near the critical layer \citep{hooper1989stability}, and the perturbation kinetic energy is supplied entirely by REY, confirming that this instability is a shear mode and not the “internal mode” suggested by \citet{boomkamp1996classification}.  

For $(k,Re)=(6,10^3)$ (figure \ref{Viscous eigenstructures a=0}b), the eigenstructure shifts to the surface: vorticity is concentrated near the interface, the critical layer sits close to the surface, and the Reynolds stress maximum is also located there. In this regime, where the viscous length scale is smaller than both the wavelength and the fluid depth, \citet{hooper1987shear} identified three possible sources of growth: boundary layers at the wall, the interface, and the critical layer. For the linear profile, the critical layer plays no role, while the interface boundary layer was expected to be stabilizing, yet here the right-branch instability clearly originates at the surface. The intersection case $(k,Re)=(1.75,150)$ (figure \ref{Viscous eigenstructures a=0}c) combines wall- and surface-driven mechanisms: the critical layer is at mid-depth, and vorticity is localized at both boundaries but with a $180^\circ$ phase difference. Similar eigenstructures appear in the yellow region of figure \ref{Energies}c, confirming that it too corresponds to a shear mode, albeit without the right branch. In general, both left and right branches occur for curvature values close to $a=0$, but their extent diminishes differently as $a$ moves away from zero, leaving only the left branch at larger $a$ (as in figure \ref{Energies}c).

\subsection{A composite mode at large $Re$}\label{CompositeMode}
A deviation between the growth rates at $Re = 10^5$ and the inviscid case was attributed in figure \ref{Viscous a=3}b to the large-$Re$ asymptote at long waves. Hence, the large-$Re$, small-$k$ region does not correspond to a pure rippling instability. Figure \ref{Energies}a further shows that REY dominates the perturbation KE in this regime, which rules out a purely long-wave interfacial mechanism, while the absence of a critical layer and the lack of vorticity concentration at the wall or interface exclude a shear mode. To probe this mixed behavior, figures \ref{Viscous eigenstructures a=3m1}a–c present eigenfunctions at $Re=10^4$ for $k=0.1,1,$ and $10$. At $k=0.1$ (figure \ref{Viscous eigenstructures a=3m1}a), the vorticity and streamfunction resemble a long-wave interfacial mode but with a horizontal tilt at depth, similar to figure \ref{Viscous eigenstructures a=3}b despite the order-of-magnitude difference in parameters. Increasing the wavenumber to $k=1$ (figure \ref{Viscous eigenstructures a=3m1}b) introduces two critical layers, one near the wall and another in mid-depth. Here, the vorticity shows a wall maximum like a shear mode, while the streamlines extend across the domain like a long-wave mode but tilt as in a rippling instability. At $k=10$ (figure \ref{Viscous eigenstructures a=3m1}c), the eigenstructure closely resembles figure \ref{Viscous eigenstructures a=3}e: vorticity peaks at the surface and also has a local maximum near a surface-critical layer, combining surface-driven shear behavior (cf. figure \ref{Viscous eigenstructures a=0}b) with rippling features (cf. figure \ref{Viscous eigenstructures a=3}f).  

A comparison with the $a=-1$ case (figures \ref{Viscous eigenstructures a=3m1}d–f), which lacks a rippling instability, highlights the difference. Although the eigenfunctions deviate from the classical long-wave structure of figure \ref{Viscous eigenstructures a=3}a, the variations in tilt and penetration depth are monotonic with $k$ and $Re$, consistent with inertia-modified long-wave interfacial instability \citep{charru2000phase}. By contrast, the $a=3$ case exhibits a composite mode: behaving like a shear–interfacial hybrid at small $k$, showing properties of all three modes at moderate $k$, and resembling a shear mode modified by rippling at large $k$. Examination of the eigenspectrum confirms that this is not due to coalescence of separate modes \citep{ozgen2008coalescence}, but rather a single mode with varying character across parameter space. Similar dual behavior was noted by \citet{albert2000small} in two-phase Couette flows, where an interfacial mode at small $Re$ evolved into a shear mode at large $Re$. The richer composite nature described here, however, does not appear to have been reported previously.

\section{Conclusion}\label{conclusion}

This work has presented a linear stability analysis of a class of Couette–Poiseuille (quadratic) shear flows in a liquid layer with a free surface. The base-state profiles are characterized by a curvature parameter $a$, which controls the degree of profile convexity or concavity. A systematic study was carried out in the $(a,k,Re)$ parameter space, with fixed inverse squared Froude number $G$ and Bond number $Bo$. Both inviscid and viscous mechanisms were analyzed, with emphasis on how viscous modes transition into their inviscid counterparts as $Re$ increases. Growth rate contours were mapped in $(k,Re)$ space, and eigenfunction structures were examined to classify the instabilities into five types: long-wave interfacial, short-wave interfacial, shear, rippling, and composite modes. The composite mode was found to exhibit hybrid characteristics of long-wave, shear, and rippling instabilities, bridging transitional regions between them.  

\begin{figure}
    \hspace{-1.5cm}
    \begin{minipage}[t]{0.6\textwidth}
      \centering
      \includegraphics[width=\textwidth]{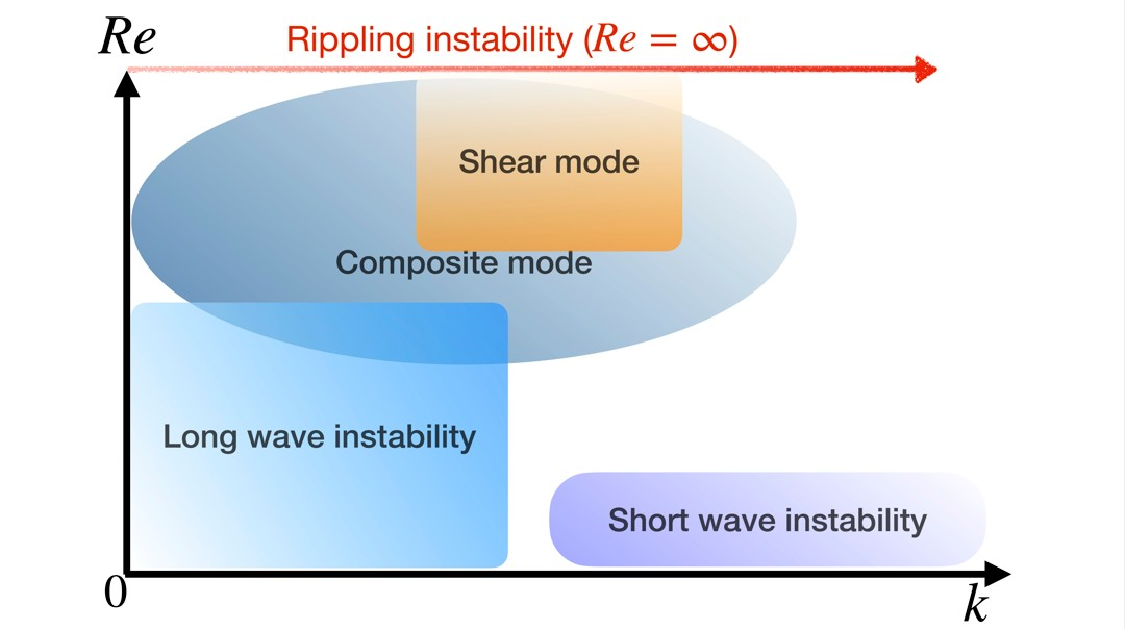}
      \put(-220,130){a)} 
    \end{minipage}
     \hspace{-0.5cm}
    \begin{minipage}[t]{0.6\textwidth}
      \centering
      \includegraphics[width=\textwidth]{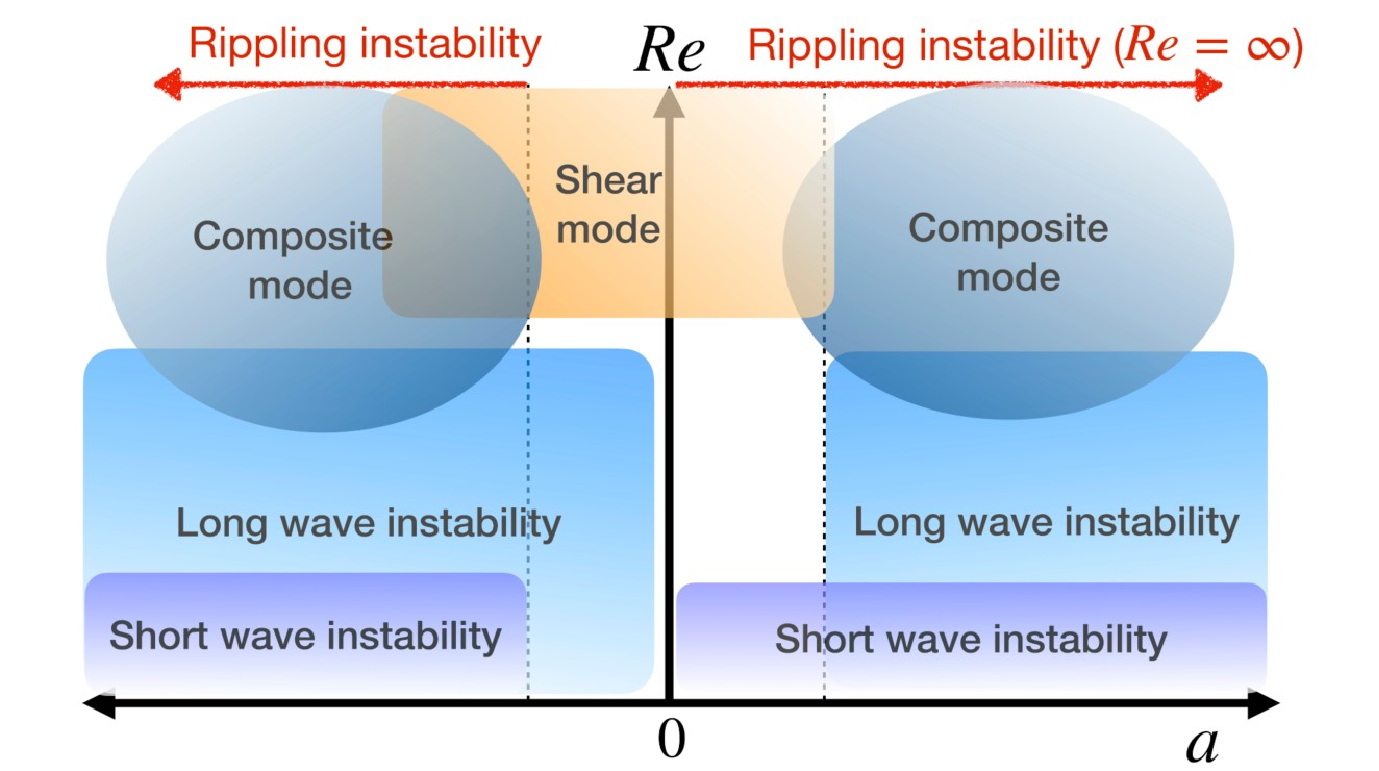}
      \put(-220,130){b)}
    \end{minipage}
\caption{A schematic illustrating the regions in the (a) $(k,Re)$ and (b) $(a,Re)$ parameter space where the five identified instability modes dominate. The rippling mode forms the upper boundary of the unstable region, while the composite mode connects the long-wave, shear, and rippling instabilities across transitional zones.}
\label{Instability cartoon}
\end{figure}

In the inviscid limit, closed-form solutions of the Rayleigh equation were obtained in terms of confluent Heun functions (§\ref{sec:analytical solution}). Long-wave asymptotics for $G=0$ showed $c_i \sim k^{4/3}$ for $|a|>1$ and $c_i \sim k^4$ for $0<a<1$, while $a \in [-1,0]$ were stable. Both asymptotic and small-$a$ analyses highlighted the role of the critical layer in driving rippling instability. Modes were found to be prograde for $a<0$ and retrograde for $a>0$. Numerical results showed that short waves satisfy $c_i \sim k^{-2}$ for all $a \notin [-1,0]$. For finite $G$, both long- and short-wave cutoffs appear, and analytical expressions for the neutral stability boundary were derived (§\ref{stabilityboundary}). Growth rate maps in $(a,k)$ revealed that instability is strongest for large $|a|$ and moderate $k$.

Viscous effects were explored using asymptotic analyses for long ($k \ll 1$) and short ($k \gg 1$) waves (§\ref{asymptotics}). For $G=0$, velocity profiles near $a=0$ exhibited a sharp change: slightly concave profiles ($a<0$) supported long-wave instabilities, while slightly convex profiles ($a>0$) supported short-wave instabilities (figure~\ref{Fig 5}). Asymptotic results further showed that $a \in [0,1]$ are stable to long waves, while $a \in [-1,0]$ are stable to short waves. Numerical solutions of the Orr–Sommerfeld problem confirmed these findings. For long waves, $c_i$ grew linearly with $Re$ at small values, peaked at moderate $Re$, and decayed towards the inviscid asymptote at large $Re$; viscosity thus enhanced long-wave growth by nearly an order of magnitude (figure~\ref{Viscous a=3}). In contrast, for moderate and short waves, viscous growth rates increased monotonically and then converged to inviscid values. Complete stability diagrams in $(k,Re)$ (§\ref{complete picture}, figure~\ref{Fig8}) revealed four main regimes: long-wave interfacial, short-wave interfacial, shear, and rippling instabilities. The rippling and long-wave regions overlapped smoothly, while shear and short-wave instabilities remained sharply distinct.

To clarify the nature of these instabilities, we utilize the energy-budget analysis \citep{boomkamp1996classification}, the presence of a critical layer, and eigenfunction structure \citep{charru2000phase}. For $a=3$, the high-$Re$, low-$k$ regime with unusually high growth rates was shown to be driven mainly by Reynolds stress, distinguishing it from both rippling and long-wave interfacial modes. Eigenfunction analysis confirmed the physical mechanisms of long- and short-wave instabilities described by \citet{smith1990mechanism} and \citet{hinch1984note} respectively. For $a=0$, the unstable mode was identified as a shear mode powered solely by Reynolds stress, in contrast to the “internal mode” classification of \citet{boomkamp1996classification}. For $a=3$, the high-$Re$ unstable branch displayed a mixture of long-wave, shear, and rippling characteristics, leading us to define a new composite mode (§\ref{CompositeMode}). 

Figure~\ref{Instability cartoon}a,b present a schematic summary of the identified instability regimes in the parameter spaces \((Re,k)\) and \((Re,a)\). The colour gradients indicate the relative strength of the instabilities. In both cases, the rippling instability, which arises in the inviscid limit, forms the upper bound of the instability map. In Figure \ref{Instability cartoon}a (\(Re\)–\(k\) plane), the long- and short-wave instabilities appear at small to moderate \(Re\), with the long-wave mode persisting to slightly higher \(Re\). The shear mode dominates at large \(Re\) and \(O(1)\) wavenumbers, while the composite mode serves as a transition between the long-wave instability, the shear mode, and the rippling limit. In Figure \ref{Instability cartoon}b (\(Re\)–\(a\) plane), the rippling instability is prominent for \(a > 0\) and \(a < -1\). The long-wave mode is relevant in the ranges \(a > 1\) and \(a < 0\), while the short-wave mode is active for \(a > 0\) and \(a < -1\). The shear mode is restricted to small values of \(a\), with a slight asymmetry of greater importance for negative values of $a$ with its influence diminishing as \(|a|\) increases.

The present study has focused on linear instabilities in a single-layer system. While many of the same mechanisms appear in two-layer flows \citep{charru2000phase}, the addition of a top layer introduces further modes such as Miles’ instability \citep{miles1957generation} and upper-wall shear instabilities \citep{shapiro2005patterns}, and can alter the stability boundary and critical Reynolds number of the bottom layer modes \citep{miesen1995hydrodynamic}. Open questions remain: How does a top layer modify bottom-layer instabilities? What new composite modes arise from coupling between layers? Beyond this, the current analysis was restricted to two-dimensional disturbances, yet in regimes where $c_i$ grows with $Re$, three-dimensional or oblique instabilities may dominate \citep{yiantsios1988linear, mohammadi2017linear, guha2010stability, cheng2024linear}. Non-modal growth \citep{abdullah2024linear, deng2025mechanisms} and the competition between linear and nonlinear effects \citep{francius2017two} should also be explored further. An initial value problem approach \citep{patibandla2023surface} could reveal the role of different unstable modes in amplifying an initial disturbance and its later time evolution. While the present work is restricted to the linear regime, it serves as a precursor by identifying parameter ranges and destabilization mechanisms that can be further explored through full direct numerical simulations (DNS) to capture their nonlinear evolution. 
%These questions point toward future work involving three-dimensional stability analysis, weakly nonlinear theory, and DNS to capture fully the interplay between viscous, inviscid, and nonlinear mechanisms.

\backsection[Supplementary data]{\label{SupMat}}

\backsection[Acknowledgements]{}

\backsection[Funding]{A.R. acknowledges financial support of the Science and Engineering Research Board, Government of India, under project nos. SPR/2021/000536 and MTR/2021/000706.}

\backsection[Declaration of interests]{The authors report no conflict of interest.}

\backsection[Data availability statement]{}

\backsection[Author ORCIDs]{}

\backsection[Author contributions]{H.M. and R.P. share equal contributions to the work}

\appendix
\section{Various confluent Heun function solutions of the Rayleigh equation with a quadratic background velocity profile}\label{appA}
As mentioned earlier, the Rayleigh equation \eqref{Rayleigh eqn} with a quadratic velocity profile can be simplified to a confluent Heun differential equation \citep{kadam2023wind}
\begin{equation}\label{eq:heun}
    x (x -1) f''(x) + (2(x -1) + 2 x + 2\beta x (x -1)) f'(x) + 2\beta (2x -1) f(x) = 0,
\end{equation}
where, $\phi(z) = x(x-1)e^{\beta x}f(x)$, $\beta = k\zeta_c/a$, $x = (\zeta(z)+\zeta_c)/2\zeta_c$, $\zeta_c = \zeta(z_c)$, $\zeta(z)$ is the base-state vorticity and $z_c$ is defined as the location that satisfies $U(z_c)=c$ (here $c$ is complex). The auxiliary parameters in the confluent Heun equation above are $4\beta, 2\beta, 2, 2$ and $2\beta$, respectively. Equation \eqref{eq:heun} possesses two regular singularities at $0$ and $1$ and an irregular singularity at $\infty$. One can write a Frobenius solution to \eqref{eq:heun} at $x=0$, with indicial exponents $0$ and $-1$. The first linearly independent series solution can be extended for $|x|>1$ using analytic continuation \citep{motygin2018evaluation} and is defined as the confluent Heun function (`HeunC' in {\it Mathematica}). Note that the second series solution, with negative indicial exponent, indicates logarithmic terms in the series and is not available in popular software applications. One can instead use the series solution at the regular singularity $x=1$ as the second linearly independent solution of the differential equation \eqref{eq:heun}. A Mobius transformation $x' = x-1$, provides a confluent Heun function that is centered at $x=1$. Note that, due to the nature of the analytic continuation algorithm, evaluating the confluent Heun function at $x=1$ results in an indeterminate expression. This can be circumvented by using a confluent Heun function centered at $x=1$. Here, we present various solution forms of the Rayleigh equation \eqref{Rayleigh eqn}, with the quadratic velocity profile, that can be evaluated in {\it Mathematica}. 
\begin{equation}\label{eq:solused1}
    x(x-1)e^{\beta x}\textrm{H}_c[2 \beta ,4 \beta ,2,2,2 \beta ,x], \hspace{0.5cm} x(x-1)e^{\beta x}\textrm{H}_c[-2 \beta ,-4 \beta ,2,2,-2 \beta ,1-x],
\end{equation}
\begin{equation}\label{eq:solused2}
    xe^{\beta x}\textrm{H}_c[2+2 \beta ,2 \beta ,2,0,2 \beta ,x], \hspace{0.5cm} (x-1)e^{\beta x}\textrm{H}_c[2-2 \beta ,-2 \beta ,2,0,-2 \beta ,1-x]
\end{equation}
\begin{equation}\label{eq:solused3}
    \textrm{H}_c\left[\dfrac{1}{2}-\dfrac{\beta^2}{16} ,-\dfrac{\beta^2}{16} ,\dfrac{1}{2},0,0 ,(1-2x)^2\right], \hspace{0.1cm}(1-2x)\textrm{H}_c\left[\dfrac{1}{2}-\dfrac{\beta^2}{16} ,-\dfrac{\beta^2}{16} ,\dfrac{3}{2},0,0 ,(1-2x)^2\right]
\end{equation}
\begin{equation}\label{eq:solused4}
    x(1-x)\textrm{H}_c\left[0 ,\dfrac{\beta^2}{16} ,2,\dfrac{1}{2},0 ,4x(1-x)\right].
\end{equation}
The two terms in each of the expression sets above are linearly independent with each other. Solution forms \eqref{eq:solused2} are presented in \S~\ref{sec:analytical solution}. Expressions \eqref{eq:solused3}-\eqref{eq:solused4} are obtained following \citet{ishkhanyan2024quadratic}. One can write any two forms from the seven expressions above as a solution to \eqref{Rayleigh eqn} although their linear independence should be checked. One can also change the sign of $\beta$ in the expressions above, to obtain additional solutions. There are indeed many solution forms \citep{slavianov2000special} and listing them here is not the intention of the current work. Rather, it is to point out that different sets of solution forms can be chosen to ease the numerical calculation of the confluent Heun function at different regions in the parameter space. In the current work the seven solution forms above have been used.
\section{Small-curvature inviscid asymptotics}\label{appB}
\begin{figure}
\centering
    \begin{minipage}[t]{0.6\textwidth}
      \centering
      \includegraphics[width=\textwidth]{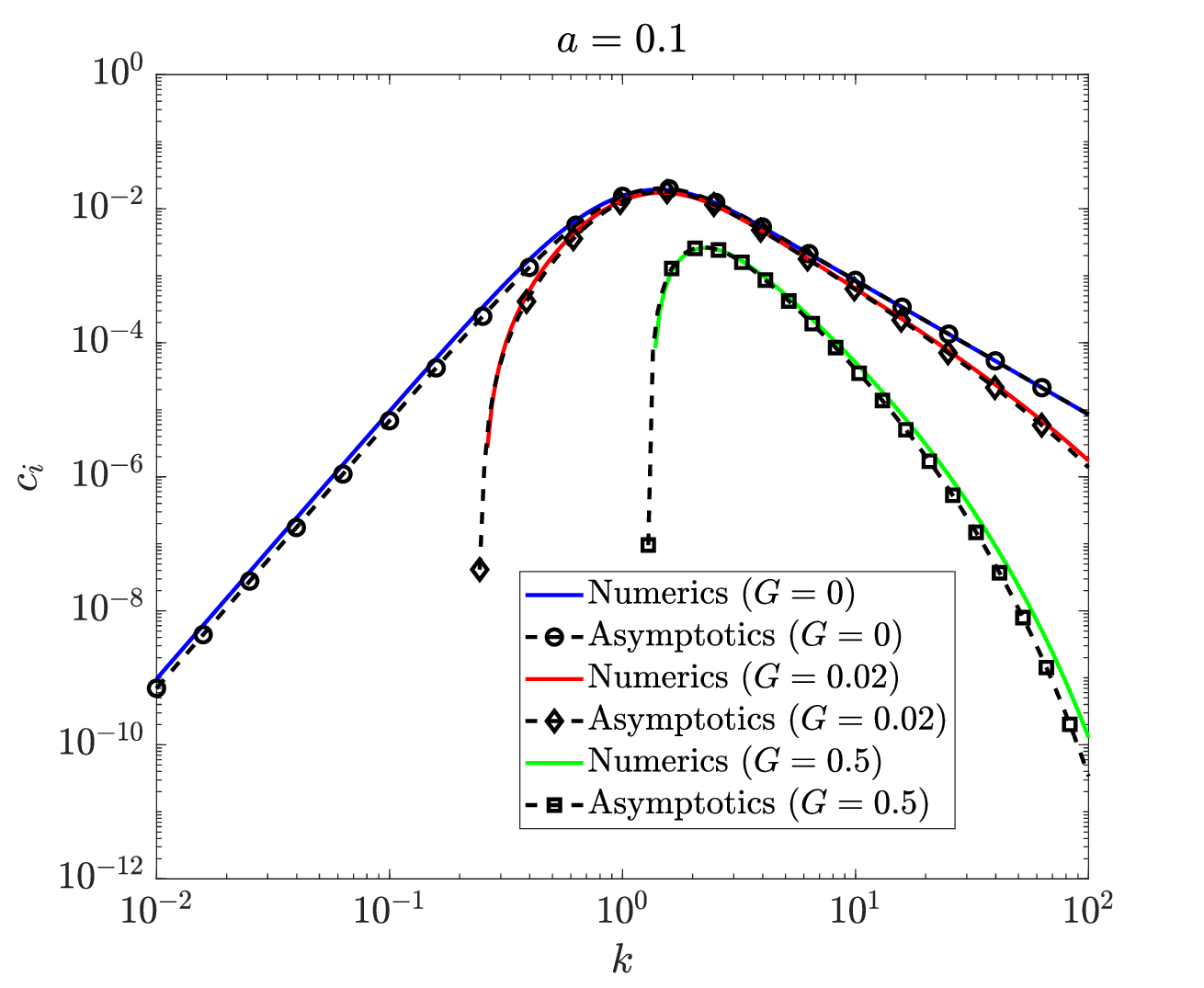}
    \end{minipage}
\caption{The imaginary part of the phase speed ($ci$) plotted as a function of wavenumber ($k$) for $a=0.1$ and $G=0, 0.02$ and $0.5$.. The black dashed curves with markers indicate $c_i$ calculated from \eqref{eq:asymptoticGrowth} and the continuous curves indicate the complete numerical solution}
\label{Asymptotic growth rates comparison small a}
\end{figure}

We assume $a$ to be a small parameter in the Rayleigh equation \eqref{Rayleigh eqn} and the boundary condition \eqref{inv dynamic bc}. A regular perturbation expansion in terms of $a$, of the eigenfunction $\phi$ and eigenvalue $c$ are given by
\begin{equation}
    {\phi} = {\phi}_{a\ll 1}^{(0)} + a {\phi}_{a\ll 1}^{(1)} + a^2 {\phi}_{a\ll 1}^{(2)} + ..., \hspace{1cm}{c} = {c}_{a\ll 1}^{(0)} + a {c}_{a\ll 1}^{(1)} + a^2 {c}_{a\ll 1}^{(2)} + ....
\end{equation}
The subscript $a \ll 1$ will be dropped henceforth for convenience. The leading order eigenfunction and eigenvalue correspond to the linear base-state flow and are given by
\begin{equation}
    {\phi}^{(0)}(z) = \dfrac{\sinh{{k}({z}+1)}}{\sinh{{k}}}, \hspace{1cm} {c}^{(0)} = 1 -\dfrac{\left(1+\sqrt{1+4G\left(1+Bo^{-1}{k}^2\right){k}\coth{{k}}}\right)}{2{k}\coth{{k}}}.
\end{equation}
To the next order, the Rayleigh equation can be simplified as
\begin{equation}
    \left(\frac{d^2}{d{z}^2} - {k}^2\right){\phi}^{(1)}(z)  = \dfrac{2\phi^{(0)}(z)}{(1+{z}-{c}^{(0)})}.
\end{equation}
Using the variation of parameters, a particular solution that satisfies the bottom boundary condition can be written as
\begin{equation}\label{eq:small_a_eigf}
    {\phi}^{(1)}(z) = \dfrac{-1}{{k}\sinh{{k}}}\left(-e^{{k}{z}}g_1({z}) + e^{-{k}{z}}g_2({z})\right),
\end{equation}
where,
\begin{equation}\label{eq:small_a_int}
    g_1({z}) = \int_{-1}^{{z}} d{z}'\dfrac{\sinh{k({z}'+1)}e^{-k{z}'}}{(1+{z}'-{c}^{(0)})},\hspace{1cm} g_2({z}) = \int_{-1}^{{z}} d{z}'\dfrac{\sinh{k({z}'+1)}e^{k{z}'}}{(1+{z}'-{c}^{(0)})}.
\end{equation}
The boundary condition at $O(a)$ is,
\begin{equation}
    \left(\dfrac{{\phi}'^{(1)}(0)}{{\phi}^{(0)}(0)} - \dfrac{{\phi}^{(1)}(0){\phi}'^{(0)}(0)}{{\phi}^{(0)2}(0)}\right) ({c}^{(0)} - 1)^2 + 2({c}^{(0)}-1){c}^{(1)} \dfrac{{\phi}'^{(0)}(0)}{{\phi}^{(0)}(0)} + {c}^{(1)} + {c}^{(0)}-1 = 0,
\end{equation}
where the apastrophe indicating a derivative with $z$. Simplifying the expression above yields
\begin{equation}\label{eq:small_a_eigv}
    {c}^{(1)} = \left(\dfrac{{\phi}'^{(1)}(0) - {k}\coth{{k}}~ {\phi}^{(1)}(0)}{2\left(1-{c}^{(0)}\right){k}\coth{{k}}-1}\right)\left({c}^{(0)}-1\right)^2 + \dfrac{\left({c}^{(0)}-1\right)}{2\left(1-{c}^{(0)}\right){k}\coth{{k}}-1}.
\end{equation}
Please note here that, $\phi^{(0)}(0) = 1$, $\phi'^{(0)}(0) = k\coth{k}$. For the existence of a growth rate, the RHS of \eqref{eq:small_a_eigv} should be complex. This, in-turn, requires complex $g_1(0)$ and $g_2(0)$. Inspecting the integrals in \eqref{eq:small_a_int}, one finds that if the value of $c_0$ doesn't match the flow velocity (somewhere in the domain) then equation \eqref{eq:small_a_eigf} and hence \eqref{eq:small_a_eigv} will be real. One can evaluate the imaginary parts of $g_1(0)$ and $g_2(0)$ using Plemelj theorem in the current case, as outlined by \citet{shrira1993surface}, as
\begin{equation}
    \Im(g_1(0)) = \dfrac{\pi e^{{k}}}{2}\left(1-e^{-2{k}{c}^{(0)}}\right), \hspace{0.5cm}\Im(g_2(0)) = \dfrac{\pi e^{-{k}}}{2}\left(e^{2{k}{c}^{(0)}}-1\right).
\end{equation}
Therefore, one obtains,
\begin{equation}
    \Im({\phi}'_1(0)) = \dfrac{1}{\sinh{{k}}}\left(\Im(g_1(0))+\Im(g_2(0))\right) =  \dfrac{2\pi}{\sinh{{k}}}\left(\sinh{{k}c^{(0)}}\cosh{{k}(1-{c}^{(0)})}\right),
\end{equation}
\begin{equation}
    \Im({\phi}_1(0)) = \dfrac{-1}{{k}\sinh{{k}}}\left(-\Im(g_1(0))+\Im(g_2(0))\right) = \dfrac{2\pi}{{k}\sinh{{k}}}\left(\sinh{{k}c^{(0)}}\sinh{{k}(1-{c}^{(0)})}\right).
\end{equation}
Finally, substituting the results above into \eqref{eq:small_a_eigv}, an expression for growth rate can be written as
\begin{equation}\label{eq:asymptoticGrowth}
    k{c}_i = \dfrac{2\pi k({c}^{(0)}-1)^2a}{\left((1-{c}^{(0)})2{k}\coth{{k}}-1\right)}\dfrac{\sinh^2{{k}{c^{(0)}}}}{\sinh^2{{k}}}.
\end{equation}
This expression is valid for all $k$ and $G$, but at small $a$. It depends on gravity and surface tension implicitly through $c^{(0)}$. Interestingly, one can find that $kc_i$ is positive for positive $a$ and vice versa. In otherwords, velocity profiles with $U''>0$ are unstable, however small the curvature may be \citep{shrira1993surface, bonfils2023flow}. Even a small negative curvature renders the mode stable. This explains the stark contrast across $a=0$ mentioned in \S~\ref{longwave}, and \S~\ref{complete solution}. Finally, as shown in figure \eqref{Asymptotic growth rates comparison small a}, $c_i$ calculated from the asymptotic expression \eqref{eq:asymptoticGrowth} matches well with the complete numerical solution for small $a$ (= 0.1).

\section{Viscous long wave asymptotics}\label{AppC}

Substituting \eqref{vis longwave series} into the Orr-Sommerfeld equation \eqref{OSE} and the corresponding viscous boundary conditions \eqref{no slip and no pen}-\eqref{nor}, one obtains the following set of equation at the leading order,
\begin{equation}\label{longwave O(1)}
    \phi''''^{(0)}_{k \ll 1}(z) =0,
\end{equation}
\begin{equation}
    \phi_{k \ll 1}^{(0)}(-1) =0, \quad \quad \phi'^{(0)}_{k \ll 1}(-1) =0
    \tag{\theequation{a,b}}
\end{equation}
\begin{equation}
    \phi''^{(0)}_{k \ll 1}(0)  +\dfrac{\phi_{k \ll 1}^{(0)}(0) U''(0)}{c_{k \ll 1}^{(0)} - U(0)} = 0, \quad \quad \phi'''^{(0)}_{k \ll 1}(0) = 0.
    \tag{\theequation{c,d}}
\end{equation}
The subscript $k \ll 1$ will be dropped henceforth for convenience. Solving equation \eqref{longwave O(1)} along with its boundary conditions yields the leading-order solution. An undetermined constant, which appears due to the homogeneity of the equation, is set to 1, resulting in:
\begin{equation}\label{Vis leading order sol}
    \phi^{(0)}(z) = (1+z)^2, \quad \text{and} \quad c^{(0)} = 1-a.
\end{equation}

The order $k$ governing equations are given by,
\begin{equation}\label{longwave O(k)}
     \phi''''^{(1)}(z) - i Re \Big( (U(z) - c^{(0)}) \phi''^{(0)}(z) - \phi^{(0)}(z) U''(z) \Big) =0,
\end{equation}
\begin{equation}
    \phi^{(1)}(-1) =0, \quad \quad \phi'^{(1)}(-1) =0,
    \tag{\theequation{a,b}}
\end{equation}
\begin{equation}
     \phi''^{(1)}(0) - \dfrac{U''(0) \left[ \phi^{(1)}(0) (U(0) - c^{(0)}) + c^{(1)} \phi^{(0)}(0) \right]}{(c^{(0)} - U(0))^2} = 0,
    \tag{\theequation{c}}
\end{equation}
\begin{equation}
    \phi'''^{(1)}(z) 
    - i Re \Big[ \phi^{(0)}(z) \Big( \frac{G}{c^{(0)} - U(z)} - U'(z) \Big) + (U(z) - c^{(0)}) \phi'^{(0)}(z) \Big] = 0.
    \tag{\theequation{d}}
\end{equation}

The zeroth order solution will appear in all subsequent orders. To eliminate repetition, the coefficient multiplying the zeroth order solution can be set to zero at each order, without loss of generality. The solution at $O(k)$ can then be written as
\begin{equation}
    \phi^{(1)}(z)=\frac{(z+1)^2 (-i Re z ((a-1) a ((z-2) z-7)+10 G))}{60 a}, \quad \text{with} \quad c^{(1)} = \frac{1}{15} i Re (4 (a-1) a- 5 G).
\end{equation}

The $O(k^2)$ governing equation is given by,
\begin{align}
    i Re \Big( (U(z) - c^{(0)}) \phi''^{(1)}(z) - c^{(1)} \phi''^{(0)}(z) - \phi^{(1)}(z) U''(z) \Big) + 2\phi''^{(0)}(z) + \phi''''^{(2)}(z)  = 0,
\end{align}
\begin{equation}
    \phi^{(2)}(-1) =0, \quad \quad \phi'^{(2)}(-1) =0,
    \tag{\theequation{a,b}}
\end{equation} 
\\
\begin{equation}
     \begin{split}
        &U''(0) \Big[\frac{\phi^{(0)}(0) \big( -c^{(0)} c^{(2)} + (c^{(1)})^2 + c^{(2)} U(0) \big)}{(c^{(0)} - U(0))^3} 
        \\ &+ \frac{(c^{(0)} - U(0)) \big( \phi^{(2)}(0) (c^{(0)} - U(0)) - c^{(1)} \phi_1(0) \big)}{(c^{(0)} - U(0))^3} \Big] 
        + \phi^{(0)}(0) + \phi''^{(2)}(0) = 0,
    \end{split}
    \tag{\theequation{c}}
\end{equation}
\begin{equation}
     \begin{split}
        &\frac{i G Re (\phi^{(1)}(0) (U(y)- c^{(0)})+ c^{(1)} \phi^{(0)}(0))}{(c^{(0)}-U(0))^2} 
        \\ &+ i Re \left((c^{(0)}-U(0)) \phi'^{(1)}(0) + c^{(1)} \phi'^{(0)}(0) + \phi^{(1)}(0) U'(0)\right) 
        - 3 \phi'^{(0)}(0) + \phi'''^{(2)}(0) = 0.
    \end{split}
    \tag{\theequation{d}}
\end{equation}
The calculation is terminated at $O(k^2)$ because an unstable mode is already identified at $O(k)$. This provides the solution for the eigenvalue $c$ up to $O(k^2)$. The process can, however, be extended further to determine higher-order solutions if needed. The leading order ($k=0$) solution for vorticity ($\Omega$) can be found by differentiating $\phi^{(0)}(z)$ from \eqref{Vis leading order sol} twice with respect to $z$, which shows that the vorticity ($\Omega$) is constant.

\section{Viscous short wave asymptotics}\label{AppD}

As outlined in \S \ref{vis short wave}, the governing equations for the short wave asymptotics are first scaled by $k$, with a small parameter ($\epsilon = k^{-1}$). The solution is then expressed as a series expansion in powers of $\epsilon$. By substituting the expansion \eqref{short wave expansion} into the scaled equations and the corresponding boundary conditions—where, in the scaled coordinate system, the domain extends from $0$ at the interface to $-\infty$ at the wall—the leading-order governing equation and the boundary conditions can be written as:

\begin{equation}\label{short wave O(1)}
    \phi''''^{(0)}_{k \gg 1}(z) - 2\phi''^{(0)}_{k \gg 1}(z) + \phi^{(0)}_{k \gg 1}(z) = 0,
\end{equation}
\begin{equation}
    \phi^{(0)}_{k \gg 1}(-\infty) =0 , \quad \quad \phi'^{(0)}_{k \gg 1}(-\infty) =0,
    \tag{\theequation{a,b}}
\end{equation}
\begin{equation}
    \phi''^{(0)}_{k \gg 1}(0)+ \phi^{(0)}_{k \gg 1}(0) = 0 , \quad \quad \phi'''^{(0)}_{k \gg 1}(0) - 3\phi'^{(0)}_{k \gg 1}(0) + \dfrac{i G Re }{Bo (1- c^{(0)}_{k \gg 1})}\phi^{(0)}_{k \gg 1}(0)= 0.
    \tag{\theequation{c,d}}
\end{equation}
Solving the set of equations \eqref{short wave O(1)}, one obtains the leading order solution as,
\begin{equation}
    \phi^{(0)}_{k \gg 1}(z) = (1-z) \exp{z}, \quad \quad c^{(0)}_{k \gg 1} = 1 - \dfrac{i G Re}{2 Bo}.
\end{equation}
Henceforth, the subscript $k \gg 1$ will be dropped for convenience. For the $O(\epsilon)$ terms, the governing equation and the boundary conditions become,
\begin{equation}
     \phi''''^{(1)}(y) - 2 \phi''^{(1)}(y) + \phi^{(1)}(y) -i Re(c^{(0)} - 1) \left( \phi^{(0)}(y) - \phi''^{(0)}(y) \right) = 0,
\end{equation}
\begin{equation}
    \phi^{(1)}(-\infty) =0 , \quad \quad \phi'^{(1)}(-\infty) =0,
    \tag{\theequation{a,b}}
\end{equation}
\begin{equation}
    \phi''^{(1)}(0) + \phi^{(1)}(0) = 0,
    \tag{\theequation{c}}
\end{equation}
\begin{align} \nonumber
  & \phi'''^{(1)}(0) - 3 \phi'^{(1)}(0) + i Re (c^{(0)} - 1) \phi'^{(0)}(0) \\ \nonumber &+ \dfrac{i G Re (-c^{(0)} \phi^{(1)}(0) + c^{(1)} \phi^{(0)}(0) + \phi^{(1)}(0))}{Bo (c^{(0)} - 1)^2} =0.
    \tag{\theequation{d}}
\end{align}

Similar to the long wave asymptotics, the solution must be free of a linear multiple of the $O(1)$ solution. After solving the $O(\epsilon)$ equation along with its boundary conditions, one obtains
\begin{equation}
    \phi^{(1)}(z) = \dfrac{G Re^2 \exp(z) (z-1) (2 z-3)}{16 Bo}, \quad \quad c^{(1)} = -\frac{3 i G^2 Re^3}{16 Bo^2}.
\end{equation}
The calculation is carried out up to $O(\epsilon^4)$, where the eigenvalue reveals an instability.

\bibliographystyle{jfm}
\bibliography{jfm}

\begin{thebibliography}{85}
\expandafter\ifx\csname natexlab\endcsname\relax\def\natexlab#1{#1}\fi
\def\au#1{#1} \def\ed#1{#1} \def\yr#1{#1}\def\at#1{#1}\def\jt#1{\textit{#1}}
  \def\bt#1{#1}\def\bvol#1{\textbf{#1}} \def\vol#1{#1} \def\pg#1{#1}
  \def\publ#1{#1}\def\arxiv#1{#1}\def\org#1{#1}\def\st#1{\textit{#1}}

\bibitem[Abdullah \& Park(2024)]{abdullah2024linear}
{\sc \au{Abdullah, M} \& \au{Park, GI}} \yr{2024}  \at{Linear stability
  analysis of oblique couette--poiseuille flows}.  \jt{Journal of Fluid
  Mechanics}  \bvol{998},  \pg{A25}.

\bibitem[Albert \& Charru(2000)]{albert2000small}
{\sc \au{Albert, F} \& \au{Charru, F}} \yr{2000}  \at{Small reynolds number
  instabilities in two-layer couette flow}.  \jt{European Journal of
  Mechanics-B/Fluids}  \bvol{19}~(2),  \pg{229--252}.

\bibitem[Ayet \& Chapron(2022)]{ayet2022dynamical}
{\sc \au{Ayet, A} \& \au{Chapron, B}} \yr{2022}  \at{The dynamical coupling of
  wind-waves and atmospheric turbulence: a review of theoretical and
  phenomenological models}.  \jt{Boundary-Layer Meteorology}  \bvol{183}~(1),
  \pg{1--33}.

\bibitem[Bai {\em et~al.\/}(1992)Bai, Chen \& Joseph]{bai1992lubricated}
{\sc \au{Bai, R}, \au{Chen, K} \& \au{Joseph, DD}} \yr{1992}  \at{Lubricated
  pipelining: stability of core—annular flow. part 5. experiments and
  comparison with theory}.  \jt{Journal of Fluid Mechanics}  \bvol{240},
  \pg{97--132}.

\bibitem[Barthelet {\em et~al.\/}(1995)Barthelet, Charru \&
  Fabre]{barthelet1995experimental}
{\sc \au{Barthelet, P}, \au{Charru, F} \& \au{Fabre, J}} \yr{1995}
  \at{Experimental study of interfacial long waves in a two-layer shear flow}.
  \jt{Journal of Fluid Mechanics}  \bvol{303},  \pg{23--53}.

\bibitem[Benjamin(1957)]{benjamin1957wave}
{\sc \au{Benjamin, T~Brooke}} \yr{1957}  \at{Wave formation in laminar flow
  down an inclined plane}.  \jt{Journal of Fluid Mechanics}  \bvol{2}~(6),
  \pg{554--573}.

\bibitem[Bonfils {\em et~al.\/}(2023)Bonfils, Mitra, Moon \&
  Wettlaufer]{bonfils2023flow}
{\sc \au{Bonfils, AF}, \au{Mitra, D}, \au{Moon, W} \& \au{Wettlaufer, JS}}
  \yr{2023}  \at{Flow-driven interfacial waves: an inviscid asymptotic study}.
  \jt{Journal of Fluid Mechanics}  \bvol{976},  \pg{A19}.

\bibitem[Boomkamp {\em et~al.\/}(1997)Boomkamp, Boersma, Miesen \&
  Beijnon]{boomkamp1997chebyshev}
{\sc \au{Boomkamp, PAM}, \au{Boersma, BJ}, \au{Miesen, RHM} \& \au{Beijnon,
  GV}} \yr{1997}  \at{A chebyshev collocation method for solving two-phase flow
  stability problems}.  \jt{Journal of computational Physics}  \bvol{132}~(2),
  \pg{191--200}.

\bibitem[Boomkamp \& Miesen(1996)]{boomkamp1996classification}
{\sc \au{Boomkamp, PAM} \& \au{Miesen, RHM}} \yr{1996}  \at{Classification of
  instabilities in parallel two-phase flow}.  \jt{International Journal of
  Multiphase Flow}  \bvol{22},  \pg{67--88}.

\bibitem[Buckley \& Veron(2016)]{buckley2016structure}
{\sc \au{Buckley, MP.} \& \au{Veron, F}} \yr{2016}  \at{Structure of the
  airflow above surface waves}.  \jt{Journal of Physical Oceanography}
  \bvol{46}~(5),  \pg{1377--1397}.

\bibitem[Burns(1953)]{burns1953long}
{\sc \au{Burns, JC}} \yr{1953} Long waves in running water.  \bt{In {\em
  Mathematical proceedings of the Cambridge philosophical society\/}}, ,
  \vol{vol.~49},  \pg{pp. 695--706}. Cambridge University Press.

\bibitem[Carpenter {\em et~al.\/}(2022)Carpenter, Buckley \&
  Veron]{carpenter2022evidence}
{\sc \au{Carpenter, JR}, \au{Buckley, MP} \& \au{Veron, F}} \yr{2022}
  \at{Evidence of the critical layer mechanism in growing wind waves}.
  \jt{Journal of Fluid Mechanics}  \bvol{948},  \pg{A26}.

\bibitem[Charles \& Lilleleht(1965)]{charles1965experimental}
{\sc \au{Charles, ME} \& \au{Lilleleht, LU}} \yr{1965}  \at{An experimental
  investigation of stability and interfacial waves in co-current flow of two
  liquids}.  \jt{Journal of Fluid Mechanics}  \bvol{22}~(2),  \pg{217--224}.

\bibitem[Charru \& Hinch(2000)]{charru2000phase}
{\sc \au{Charru, F} \& \au{Hinch, EJ}} \yr{2000}  \at{‘phase diagram’of
  interfacial instabilities in a two-layer couette flow and mechanism of the
  long-wave instability}.  \jt{Journal of Fluid Mechanics}  \bvol{414},
  \pg{195--223}.

\bibitem[Cheng {\em et~al.\/}(2024)Cheng, Ma, Pullin \& Luo]{cheng2024linear}
{\sc \au{Cheng, W}, \au{Ma, H}, \au{Pullin, DI} \& \au{Luo, X}} \yr{2024}
  \at{Linear stability analysis of generalized couette--poiseuille flow: the
  neutral surface and critical properties}.  \jt{Journal of Fluid Mechanics}
  \bvol{995},  \pg{R3}.

\bibitem[Cohen \& Hanratty(1965)]{cohen1965generation}
{\sc \au{Cohen, LS} \& \au{Hanratty, TJ}} \yr{1965}  \at{Generation of waves in
  the concurrent flow of air and a liquid}.  \jt{AIChE Journal}  \bvol{11}~(1),
   \pg{138--144}.

\bibitem[Constantin \& Ivanov(2019)]{constantin2019equatorial}
{\sc \au{Constantin, A} \& \au{Ivanov, RI}} \yr{2019}  \at{Equatorial
  wave--current interactions}.  \jt{Communications in Mathematical Physics}
  \bvol{370}~(1),  \pg{1--48}.

\bibitem[Constantin \& Johnson(2016)]{constantin2016exact}
{\sc \au{Constantin, A} \& \au{Johnson, RS}} \yr{2016}  \at{An exact, steady,
  purely azimuthal equatorial flow with a free surface}.  \jt{Journal of
  Physical Oceanography}  \bvol{46}~(6),  \pg{1935--1945}.

\bibitem[Constantin \& Johnson(2019)]{constantin2019ekman}
{\sc \au{Constantin, A} \& \au{Johnson, RS}} \yr{2019}  \at{Ekman-type
  solutions for shallow-water flows on a rotating sphere: a new perspective on
  a classical problem}.  \jt{Physics of Fluids}  \bvol{31}~(2).

\bibitem[Craik(1966)]{craik1966wind}
{\sc \au{Craik, ADD}} \yr{1966}  \at{Wind-generated waves in thin liquid
  films}.  \jt{Journal of Fluid Mechanics}  \bvol{26}~(2),  \pg{369--392}.

\bibitem[Dai {\em et~al.\/}(2010)Dai, Qiao, Sulisz, Han \&
  Babanin]{dai2010experiment}
{\sc \au{Dai, D}, \au{Qiao, F}, \au{Sulisz, W}, \au{Han, L} \& \au{Babanin, A}}
  \yr{2010}  \at{An experiment on the nonbreaking surface-wave-induced vertical
  mixing}.  \jt{Journal of Physical Oceanography}  \bvol{40}~(9),
  \pg{2180--2188}.

\bibitem[Deng {\em et~al.\/}(2025)Deng, Yang \& Shen]{deng2025mechanisms}
{\sc \au{Deng, BQ}, \au{Yang, Z} \& \au{Shen, L}} \yr{2025}  \at{Mechanisms
  underlying how free surfaces influence very-large-scale motions in turbulent
  plane open channel flows based on linear non-modal analysis}.  \jt{Journal of
  Fluid Mechanics}  \bvol{1002},  \pg{A9}.

\bibitem[Drazin \& Reid(2004)]{drazin2004hydrodynamic}
{\sc \au{Drazin, Philip~G} \& \au{Reid, William~Hill}} \yr{2004} {\em
  Hydrodynamic stability\/}.  \publ{Cambridge university press}.

\bibitem[Engevik(2000)]{engevik2000note}
{\sc \au{Engevik, L}} \yr{2000}  \at{A note on the instabilities of a
  horizontal shear flow with a free surface}.  \jt{Journal of Fluid Mechanics}
  \bvol{406},  \pg{337--346}.

\bibitem[Floryan {\em et~al.\/}(1987)Floryan, Davis \&
  Kelly]{floryan1987instabilities}
{\sc \au{Floryan, JM}, \au{Davis, SH} \& \au{Kelly, RE}} \yr{1987}
  \at{Instabilities of a liquid film flowing down a slightly inclined plane}.
  \jt{The Physics of fluids}  \bvol{30}~(4),  \pg{983--989}.

\bibitem[Francius \& Kharif(2017)]{francius2017two}
{\sc \au{Francius, M} \& \au{Kharif, C}} \yr{2017}  \at{Two-dimensional
  stability of finite-amplitude gravity waves on water of finite depth with
  constant vorticity}.  \jt{Journal of Fluid Mechanics}  \bvol{830},
  \pg{631--659}.

\bibitem[Govindarajan \& Sahu(2014)]{govindarajan2014instabilities}
{\sc \au{Govindarajan, R} \& \au{Sahu, KC}} \yr{2014}  \at{Instabilities in
  viscosity-stratified flow}.  \jt{Annual review of fluid mechanics}
  \bvol{46}~(1),  \pg{331--353}.

\bibitem[Guha \& Frigaard(2010)]{guha2010stability}
{\sc \au{Guha, A} \& \au{Frigaard, IA}} \yr{2010}  \at{On the stability of
  plane couette--poiseuille flow with uniform crossflow}.  \jt{Journal of Fluid
  Mechanics}  \bvol{656},  \pg{417--447}.

\bibitem[Hidy \& Plate(1966)]{hidy1966wind}
{\sc \au{Hidy, GM} \& \au{Plate, EJ}} \yr{1966}  \at{Wind action on water
  standing in a laboratory channel}.  \jt{Journal of Fluid Mechanics}
  \bvol{26}~(4),  \pg{651--687}.

\bibitem[Hinch(1984)]{hinch1984note}
{\sc \au{Hinch, EJ}} \yr{1984}  \at{A note on the mechanism of the instability
  at the interface between two shearing fluids}.  \jt{Journal of Fluid
  Mechanics}  \bvol{144},  \pg{463--465}.

\bibitem[Hooper(1989)]{hooper1989stability}
{\sc \au{Hooper, AP}} \yr{1989}  \at{The stability of two superposed viscous
  fluids in a channel}.  \jt{Physics of Fluids A: Fluid Dynamics}
  \bvol{1}~(7),  \pg{1133--1142}.

\bibitem[Hooper \& Boyd(1983)]{hooper1983shear}
{\sc \au{Hooper, AP} \& \au{Boyd, WGC}} \yr{1983}  \at{Shear-flow instability
  at the interface between two viscous fluids}.  \jt{Journal of Fluid
  Mechanics}  \bvol{128},  \pg{507--528}.

\bibitem[Hooper \& Boyd(1987)]{hooper1987shear}
{\sc \au{Hooper, AP} \& \au{Boyd, WGC}} \yr{1987}  \at{Shear-flow instability
  due to a wall and a viscosity discontinuity at the interface}.  \jt{Journal
  of Fluid Mechanics}  \bvol{179},  \pg{201--225}.

\bibitem[Ishkhanyan(2024)]{ishkhanyan2024quadratic}
{\sc \au{Ishkhanyan, AM}} \yr{2024}  \at{A quadratic transformation for a
  special confluent heun function}.  \jt{Heliyon}  \bvol{10}~(16).

\bibitem[Janssen(2004)]{janssen2004interaction}
{\sc \au{Janssen, P}} \yr{2004} {\em The interaction of ocean waves and
  wind\/}.  \publ{Cambridge University Press}.

\bibitem[Joseph \& Renardy(2013)]{joseph2013fundamentals}
{\sc \au{Joseph, DD} \& \au{Renardy, YY}} \yr{2013} {\em Fundamentals of
  two-fluid dynamics: Part i: Mathematical theory and applications\/}, ,
  \vol{vol.~3}.  \publ{Springer Science \& Business Media}.

\bibitem[Kadam {\em et~al.\/}(2023)Kadam, Patibandla \& Roy]{kadam2023wind}
{\sc \au{Kadam, Y}, \au{Patibandla, R} \& \au{Roy, A}} \yr{2023}
  \at{Wind-generated waves on a water layer of finite depth}.  \jt{Journal of
  Fluid Mechanics}  \bvol{967},  \pg{A12}.

\bibitem[Kaffel \& Renardy(2011)]{kaffel2011surface}
{\sc \au{Kaffel, Ahmed} \& \au{Renardy, Michael}} \yr{2011}  \at{Surface modes
  in inviscid free surface shear flows}.  \jt{ZAMM-Journal of Applied
  Mathematics and Mechanics/Zeitschrift f{\"u}r Angewandte Mathematik und
  Mechanik}  \bvol{91}~(8),  \pg{649--652}.

\bibitem[Kaffel \& Riaz(2015)]{kaffel2015eigenspectra}
{\sc \au{Kaffel, A} \& \au{Riaz, A}} \yr{2015}  \at{Eigenspectra and mode
  coalescence of temporal instability in two-phase channel flow}.  \jt{Physics
  of Fluids}  \bvol{27}~(4).

\bibitem[Kalliadasis {\em et~al.\/}(2011)Kalliadasis, Ruyer-Quil, Scheid \&
  Velarde]{kalliadasis2011falling}
{\sc \au{Kalliadasis, S}, \au{Ruyer-Quil, C}, \au{Scheid, B} \& \au{Velarde,
  MG}} \yr{2011} {\em Falling liquid films\/}, ,  \vol{vol. 176}.
  \publ{Springer Science \& Business Media}.

\bibitem[Kao \& Park(1972)]{kao1972experimental}
{\sc \au{Kao, TW} \& \au{Park, C}} \yr{1972}  \at{Experimental investigations
  of the stability of channel flows. part 2. two-layered co-current flow in a
  rectangular channel}.  \jt{Journal of Fluid Mechanics}  \bvol{52}~(3),
  \pg{401--423}.

\bibitem[Kapitza(1948)]{kapitza1948wave}
{\sc \au{Kapitza, PL}} \yr{1948}  \at{Wave flow of thin layer of a viscous
  fluid}.  \jt{Zhur. Eksperim. Teoret. Fiz.}  \bvol{18}~(3),  \pg{9}.

\bibitem[Kelly {\em et~al.\/}(1989)Kelly, Goussis, Lin \&
  Hsu]{kelly1989mechanism}
{\sc \au{Kelly, RE}, \au{Goussis, DA}, \au{Lin, SP} \& \au{Hsu, FK}} \yr{1989}
  \at{The mechanism for surface wave instability in film flow down an inclined
  plane}.  \jt{Physics of Fluids A: Fluid Dynamics}  \bvol{1}~(5),
  \pg{819--828}.

\bibitem[Kouris \& Tsamopoulos(2001)]{kouris2001core}
{\sc \au{Kouris, C} \& \au{Tsamopoulos, J}} \yr{2001}  \at{Core--annular flow
  in a periodically constricted circular tube. part 1. steady-state, linear
  stability and energy analysis}.  \jt{Journal of Fluid Mechanics}  \bvol{432},
   \pg{31--68}.

\bibitem[Lefebvre \& McDonell(2017)]{lefebvre2017atomization}
{\sc \au{Lefebvre, AH} \& \au{McDonell, VG}} \yr{2017} {\em Atomization and
  sprays\/}.  \publ{CRC press}.

\bibitem[Lin(1946)]{lin1946stability}
{\sc \au{Lin, CC}} \yr{1946}  \at{On the stability of two-dimensional parallel
  flows. iii. stability in a viscous fluid}.  \jt{Quarterly of Applied
  Mathematics}  \bvol{3}~(4),  \pg{277--301}.

\bibitem[Liu {\em et~al.\/}(2017)Liu, Chen, Bond \& Hu]{liu2017experimental}
{\sc \au{Liu, Y}, \au{Chen, WL}, \au{Bond, LJ} \& \au{Hu, H}} \yr{2017}  \at{An
  experimental study on the characteristics of wind-driven surface water film
  flows by using a multi-transducer ultrasonic pulse-echo technique}.
  \jt{Physics of Fluids}  \bvol{29}~(1).

\bibitem[Miesen \& Boersma(1995)]{miesen1995hydrodynamic}
{\sc \au{Miesen, R} \& \au{Boersma, BJ}} \yr{1995}  \at{Hydrodynamic stability
  of a sheared liquid film}.  \jt{Journal of fluid mechanics}  \bvol{301},
  \pg{175--202}.

\bibitem[Miles(1957)]{miles1957generation}
{\sc \au{Miles, JW}} \yr{1957}  \at{On the generation of surface waves by shear
  flows}.  \jt{Journal of Fluid Mechanics}  \bvol{3}~(2),  \pg{185--204}.

\bibitem[Miles(1959)]{miles1959generation}
{\sc \au{Miles, JW}} \yr{1959}  \at{On the generation of surface waves by shear
  flows. part 2}.  \jt{Journal of Fluid Mechanics}  \bvol{6}~(4),
  \pg{568--582}.

\bibitem[Miles(1960)]{miles1960hydrodynamic}
{\sc \au{Miles, JW}} \yr{1960}  \at{The hydrodynamic stability of a thin film
  of liquid in uniform shearing motion}.  \jt{Journal of Fluid Mechanics}
  \bvol{8}~(4),  \pg{593--610}.

\bibitem[Mohammadi \& Smits(2016)]{mohammadi2016stability}
{\sc \au{Mohammadi, A} \& \au{Smits, AJ}} \yr{2016}  \at{Stability of
  two-immiscible-fluid systems: a review of canonical plane parallel flows}.
  \jt{Journal of Fluids Engineering}  \bvol{138}~(10),  \pg{100803}.

\bibitem[Mohammadi \& Smits(2017)]{mohammadi2017linear}
{\sc \au{Mohammadi, A} \& \au{Smits, AJ}} \yr{2017}  \at{Linear stability of
  two-layer couette flows}.  \jt{Journal of Fluid Mechanics}  \bvol{826},
  \pg{128--157}.

\bibitem[Morland {\em et~al.\/}(1991)Morland, Saffman \&
  Yuen]{morland1991waves}
{\sc \au{Morland, LC}, \au{Saffman, PG} \& \au{Yuen, HC}} \yr{1991}  \at{Waves
  generated by shear layer instabilities}.  \jt{Proceedings of the Royal
  Society of London. Series A: Mathematical and Physical Sciences}
  \bvol{433}~(1888),  \pg{441--450}.

\bibitem[Motygin(2018)]{motygin2018evaluation}
{\sc \au{Motygin, OV}} \yr{2018} On evaluation of the confluent heun functions.
   \bt{In {\em 2018 Days on Diffraction (DD)\/}},  \pg{pp. 223--229}. IEEE.

\bibitem[N{\'a}raigh {\em et~al.\/}(2011)N{\'a}raigh, Spelt, Matar \&
  Zaki]{naraigh2011interfacial}
{\sc \au{N{\'a}raigh, L{\'O}}, \au{Spelt, PDM}, \au{Matar, OK} \& \au{Zaki,
  TA}} \yr{2011}  \at{Interfacial instability in turbulent flow over a liquid
  film in a channel}.  \jt{International Journal of Multiphase Flow}
  \bvol{37}~(7),  \pg{812--830}.

\bibitem[{\"O}zgen(2008)]{ozgen2008coalescence}
{\sc \au{{\"O}zgen, S}} \yr{2008}  \at{Coalescence of tollmien--schlichting and
  interfacial modes of instability in two-fluid flows}.  \jt{Physics of Fluids}
   \bvol{20}~(4).

\bibitem[{\"O}zgen {\em et~al.\/}(1998){\"O}zgen, Degrez \&
  Sarma]{ozgen1998two}
{\sc \au{{\"O}zgen, S}, \au{Degrez, G} \& \au{Sarma, GSR}} \yr{1998}
  \at{Two-fluid boundary layer stability}.  \jt{Physics of fluids}
  \bvol{10}~(11),  \pg{2746--2757}.

\bibitem[Paquier {\em et~al.\/}(2015)Paquier, Moisy \&
  Rabaud]{paquier2015surface}
{\sc \au{Paquier, A}, \au{Moisy, F} \& \au{Rabaud, M}} \yr{2015}  \at{Surface
  deformations and wave generation by wind blowing over a viscous liquid}.
  \jt{Physics of Fluids}  \bvol{27}~(12).

\bibitem[Paquier {\em et~al.\/}(2016)Paquier, Moisy \&
  Rabaud]{paquier2016viscosity}
{\sc \au{Paquier, A}, \au{Moisy, F} \& \au{Rabaud, M}} \yr{2016}  \at{Viscosity
  effects in wind wave generation}.  \jt{Physical Review Fluids}  \bvol{1}~(8),
   \pg{083901}.

\bibitem[Patibandla {\em et~al.\/}(2023)Patibandla, Basak, Dasgupta \&
  Roy]{patibandla2023surface}
{\sc \au{Patibandla, R}, \au{Basak, S}, \au{Dasgupta, R} \& \au{Roy, A}}
  \yr{2023}  \at{Surface and internal gravity waves on a viscous liquid layer:
  Initial-value problems}.  \jt{International Journal of Multiphase Flow}
  \bvol{169},  \pg{104592}.

\bibitem[Plant(1982)]{plant1982relationship}
{\sc \au{Plant, WJ}} \yr{1982}  \at{A relationship between wind stress and wave
  slope}.  \jt{Journal of Geophysical Research: Oceans}  \bvol{87}~(C3),
  \pg{1961--1967}.

\bibitem[Renardy \& Renardy(2013)]{renardy2013stability}
{\sc \au{Renardy, Y} \& \au{Renardy, M}} \yr{2013}  \at{On the stability of
  inviscid parallel shear flows with a free surface}.  \jt{Journal of
  Mathematical Fluid Mechanics}  \bvol{15}~(1),  \pg{129--137}.

\bibitem[Sangalli {\em et~al.\/}(1995)Sangalli, Gallagher, Leighton, Chang \&
  McCready]{sangalli1995finite}
{\sc \au{Sangalli, M}, \au{Gallagher, CT}, \au{Leighton, DT}, \au{Chang, H-C}
  \& \au{McCready, MJ}} \yr{1995}  \at{Finite-amplitude waves at the interface
  between fluids with different viscosity: theory and experiments}.
  \jt{Physical review letters}  \bvol{75}~(1),  \pg{77}.

\bibitem[Shapiro \& Timoshin(2005)]{shapiro2005patterns}
{\sc \au{Shapiro, E} \& \au{Timoshin, S}} \yr{2005}  \at{On the patterns of
  interaction between shear and interfacial modes in plane air--water
  poiseuille flow}.  \jt{Proceedings of the Royal Society A: Mathematical,
  Physical and Engineering Sciences}  \bvol{461}~(2057),  \pg{1583--1597}.

\bibitem[Shrira(1993)]{shrira1993surface}
{\sc \au{Shrira, VI}} \yr{1993}  \at{Surface waves on shear currents: solution
  of the boundary-value problem}.  \jt{Journal of Fluid Mechanics}  \bvol{252},
   \pg{565--584}.

\bibitem[Slavianov \& Lay(2000)]{slavianov2000special}
{\sc \au{Slavianov, SI} \& \au{Lay, W}} \yr{2000} {\em Special functions: a
  unified theory based on singularities\/}.  \publ{Oxford University Press}.

\bibitem[Smith(1990)]{smith1990mechanism}
{\sc \au{Smith, MK}} \yr{1990}  \at{The mechanism for the long-wave instability
  in thin liquid films}.  \jt{Journal of Fluid Mechanics}  \bvol{217},
  \pg{469--485}.

\bibitem[Smith \& Davis(1982)]{smith1982instability}
{\sc \au{Smith, MK} \& \au{Davis, SH}} \yr{1982}  \at{The instability of
  sheared liquid layers}.  \jt{Journal of Fluid Mechanics}  \bvol{121},
  \pg{187--206}.

\bibitem[Stern \& Adam(1973)]{stern1973capillary}
{\sc \au{Stern, ME} \& \au{Adam, YA}} \yr{1973}  \at{Capillary waves generated
  by a shear current in water}.  \jt{Mem. Soc. R. Sci. Liege}  \bvol{6}~(6),
  \pg{179}.

\bibitem[Stommel(1959)]{stommel1959wind}
{\sc \au{Stommel, H}} \yr{1959}  \at{Wind-drift near the equator}.  \jt{Deep
  Sea Research (1953)}  \bvol{6},  \pg{298--302}.

\bibitem[Sullivan \& McWilliams(2010)]{sullivan2010dynamics}
{\sc \au{Sullivan, PP} \& \au{McWilliams, JC}} \yr{2010}  \at{Dynamics of winds
  and currents coupled to surface waves}.  \jt{Annual Review of Fluid
  Mechanics}  \bvol{42}~(1),  \pg{19--42}.

\bibitem[Takamura {\em et~al.\/}(2012)Takamura, Fischer \&
  Morrow]{takamura2012physical}
{\sc \au{Takamura, K}, \au{Fischer, H} \& \au{Morrow, NR}} \yr{2012}
  \at{Physical properties of aqueous glycerol solutions}.  \jt{Journal of
  Petroleum Science and Engineering}  \bvol{98},  \pg{50--60}.

\bibitem[Timoshin(1997)]{timoshin1997instabilities}
{\sc \au{Timoshin, SN}} \yr{1997}  \at{Instabilities in a high-reynolds-number
  boundary layer on a film-coated surface}.  \jt{Journal of Fluid Mechanics}
  \bvol{353},  \pg{163--195}.

\bibitem[Timoshin \& Hooper(2000)]{timoshin2000mode}
{\sc \au{Timoshin, SN} \& \au{Hooper, AP}} \yr{2000}  \at{Mode coalescence in a
  two-fluid boundary-layer stability problem}.  \jt{Physics of Fluids}
  \bvol{12}~(8),  \pg{1969--1978}.

\bibitem[Trefethen(2000)]{trefethen2000spectral}
{\sc \au{Trefethen, LN}} \yr{2000} {\em Spectral methods in MATLAB\/}.
  \publ{SIAM}.

\bibitem[Weinstein \& Ruschak(2004)]{weinstein2004coating}
{\sc \au{Weinstein, SJ} \& \au{Ruschak, KJ}} \yr{2004}  \at{Coating flows}.
  \jt{Annu. Rev. Fluid Mech.}  \bvol{36}~(1),  \pg{29--53}.

\bibitem[Wolfram~Research(2024)]{Mathematica}
{\sc \au{Wolfram~Research, Inc.}} \yr{2024} Mathematica, version 14.2.
  Champaign, IL, 2024.

\bibitem[Yiantsios \& Higgins(1988)]{yiantsios1988linear}
{\sc \au{Yiantsios, Stergios~G} \& \au{Higgins, Brian~G}} \yr{1988}  \at{Linear
  stability of plane poiseuille flow of two superposed fluids}.  \jt{Physics of
  Fluids}  \bvol{31}~(11),  \pg{3225}.

\bibitem[Yih(1963)]{10.1063/1.1706737}
{\sc \au{Yih, C‐S}} \yr{1963}  \at{{Stability of Liquid Flow down an Inclined
  Plane}}.  \jt{The Physics of Fluids}  \bvol{6}~(3),  \pg{321--334}.

\bibitem[Yih(1967)]{yih1967instability}
{\sc \au{Yih, C-S}} \yr{1967}  \at{Instability due to viscosity
  stratification}.  \jt{Journal of Fluid Mechanics}  \bvol{27}~(2),
  \pg{337--352}.

\bibitem[Yih(1972)]{yih1972surface}
{\sc \au{Yih, C-S}} \yr{1972}  \at{Surface waves in flowing water}.
  \jt{Journal of Fluid Mechanics}  \bvol{51}~(2),  \pg{209--220}.

\bibitem[Young \& Van~Vledder(1993)]{young1993review}
{\sc \au{Young, IR} \& \au{Van~Vledder, GP}} \yr{1993}  \at{A review of the
  central role of nonlinear interactions in wind—wave evolution}.
  \jt{Philosophical Transactions of the Royal Society of London. Series A:
  Physical and Engineering Sciences}  \bvol{342}~(1666),  \pg{505--524}.

\bibitem[Young \& Wolfe(2014)]{young2014generation}
{\sc \au{Young, WR} \& \au{Wolfe, CL}} \yr{2014}  \at{Generation of surface
  waves by shear-flow instability}.  \jt{Journal of fluid mechanics}
  \bvol{739},  \pg{276--307}.

\bibitem[Zeisel {\em et~al.\/}(2008)Zeisel, Stiassnie \&
  Agnon]{zeisel2008viscous}
{\sc \au{Zeisel, A}, \au{Stiassnie, M} \& \au{Agnon, Y}} \yr{2008}  \at{Viscous
  effects on wave generation by strong winds}.  \jt{Journal of Fluid Mechanics}
   \bvol{597},  \pg{343--369}.

\end{thebibliography}
%Use of the above commands will create a bibliography using the .bib file. Shown below is a bibliography built from individual items.

%\bibliographystyle{jfm}
%\bibliography{jfm2esam}

%% End of file `jfm2esam.bib'.

\end{document}